\renewcommand{\theequation}{\thesection.\arabic{equation}}
\newcommand{\be}{\begin{theequation}}
\newcommand{\ee}{\end{theequation}}
\newcommand{\bea}{\begin{eqnarray}}
\newcommand{\eea}{\end{eqnarray}}
\newcommand{\nn}{\nonumber}
\newcommand{\p}[1]{(\ref{#1})}
\newcommand{\cJ}{{\cal J}}
\newcommand{\cH}{{\cal H}}
\newcommand{\cT}{{\cal T}}
\newcommand{\th}{\theta}
\newcommand{\bth}{\bar{\theta}}

\newcommand{\cD}{{\cal D}}
\newcommand{\cDb}{\bar{{\cal D}}}
\newcommand{\np}{Nucl.Phys.\ }

\newcommand{\cmp}{Comm.Math.Phys.\ }
\newcommand{\pl}{Phys.Lett.\ }
\newcommand{\mpl}{Mod.Phys.Lett.\ }

\newcommand{\tz}{\frac{\theta_{12}}{z_{12}}}
\newcommand{\zot}{\frac{1}{z_{12}}}

\newcommand{\tzb}{\frac{\bar{\theta}_{12}}{z_{12}}}
\newcommand{\tzzb}{\frac{\bar{\theta}_{12}}{z_{12}^2}}
\newcommand{\tzz}{\frac{{\theta}_{12}}{z_{12}^2}}
\newcommand{\tzzbb}{\frac{\theta_{12} \bar{\theta}_{12}}{z_{12}^2}}
\newcommand{\tzzzbb}{\frac{\theta_{12} \bar{\theta}_{12}}{z_{12}^3}}
\newcommand{\tzbb}{\frac{\theta_{12} \bar{\theta}_{12}}{z_{12}}}
\documentstyle[12pt]{article}
\def\theequation{\arabic{section}.\arabic{equation}}
\begin{document}
\renewcommand{\thefootnote}{\fnsymbol{footnote}}
\thispagestyle{empty}
\begin{flushright}
JINR E2-95-343\\
hep-th/9508005\\
July, 1995 \\
\end{flushright}
\vspace{2cm}
\begin{center}
$N=2$ AFFINE SUPERALGEBRAS  AND \\
HAMILTONIAN REDUCTION IN $N=2$ SUPERSPACE \\
\vspace{1cm}
{\sc Changhyun Ahn}\footnote{E-mail: ahn@thsun1.jinr.dubna.su
Address after Sept. 1, 1995: Dept. of Physics, Kyung Hee Univ.,\par
\hspace{1.45cm}
Seoul 130-701, Korea (cahn@photon.kyunghee.ac.kr)},
{\sc E. Ivanov}\footnote{E-mail: eivanov@thsun1.jinr.dubna.su} and
{\sc A. Sorin}\footnote{E-mail: sorin@thsun1.jinr.dubna.su} \\
\vspace{1cm}
{\it Bogoliubov Laboratory of Theoretical Physics, \\
 JINR, 141980, Dubna, Moscow Region, Russia }\\
\vspace{3cm}
{\bf {\sc Abstract}}
\end{center}

We construct $N=2$ affine current algebras for the superalgebras
$sl(n|n-1)^{(1)}$ in terms of $N=2$ supercurrents subjected to nonlinear
constraints and discuss the general procedure of the hamiltonian reduction
in $N=2$ superspace at the classical level. We consider in detail the
simplest case of $N=2$ $sl(2|1)^{(1)}$ and show how $N=2$ superconformal
algebra in $N=2$ superspace follows via the hamiltonian reduction.
Applying the hamiltonian reduction to the case of $N=2$ $sl(3|2)^{(1)}$,
we find two new extended $N=2$ superconformal algebras in a manifestly
supersymmetric $N=2$ superfield form. Decoupling of four component currents
of dimension $1/2$ in them yields, respectively, $ u(2|1) $ and $ u(3) $
Knizhnik-Bershadsky superconformal algebras. We also discuss how the $N=2$
superfield formulations of $N=2$ $W_{3}$ and $N=2$ $W_{3}^{(2)}$
superconformal algebras come out in this framework, as well as some unusual
extended $N=2$ superconformal algebras containing constrained $N=2$ stress
tensor and/or spin $0$ supercurrents.

\vspace{1cm}
\vfill
\setcounter{page}0
\setcounter{equation}{0}
\newpage

\section{\bf Introduction}
\setcounter{equation}{0}

For several last years an important progress has been achieved in
understanding the role of world-sheet superconformal symmetry and target
space symmetry of nonlinear $\sigma$-models in the context of string
theory and topological field theory \cite{W,Ler,Eg}. The BRST structure
of bosonic string  ( $W_{n}$ string ) generates a topologically twisted
$N=2$ superconformal algebra \cite{GaSe} ($N=2$ super-$W_{n}$ algebra
\cite{BLNW,IK} ). In obtaining these results, a heavy use of the
hamiltonian reduction  from WZNW models based on the superalgebra
$sl(n|n-1)$ has been made. Futhermore, any superstring theory possesses
$N=3$ twisted supersymmetry \cite{BLNW}. Recently, BRST structure has
been systematically constructed for superstrings with $N$ supersymmetries
by the hamiltonian reduction of affine extension of $osp(N+2|2)$ \cite{BLLS}.
$N=2$ analog for topological strings is the twisted $N=4$ $su(2)$
superconformal algebra (SCA) which has been obtained by the reduction of
affine extension of $sl(2|2)$ in \cite{LLS}.

As these and many other examples demonstrate, the hamiltonian reduction
is a powerful method of deducing new conformal \cite{DS,FRRTW,BTv} and
superconformal algebras and analysing the symmetry structure of the
conformal field theory and string theory models. Since a natural arena for
studying various superconformal symmetries and the related field theory
models is provided by superspace, it is tempting to have convenient
superspace generalizations of the hamiltonian reduction. $N=1$ superspace
version of this procedure in various aspects was discussed in Ref.
\cite{DRS}. On the other hand, a lot of interesting models (both in string
theory and topological field theory) reveal $N=2$ superconformal symmetries,
manifestly covariant formulations of which require $N=2$ superspace.
Motivated by this, in the present paper we generalize the hamiltonian
reduction procedure to $N=2$ superspace.

Let us recollect some well-known facts which are relevant to the problems
we address in the present paper.

Knizhnik \cite{K} and Bershadsky \cite{Ber} have proposed SCAs
with quadratic nonlinearity having as subalgebras $u(n)$ and
$so(n)$ affine algebras. It has been shown later \cite{GS} that the
nonlinear $so(3)$ and $so(4)$ Knizhnik-Bershadsky (KB) SCAs can be embedded
as subalgebras in usual linear $so(3)$ and $so(4)$ extended SCAs \cite{A}
after passing to some new basis for the currents of the latter (related to
the standard one by an invertible nonlinear transformation). By construction,
the usual $N=2$ and $N=4$ $su(2)$ SCAs \cite{A} are the same as $u(1)$ and
$u(2)$ KB SCAs, respectively.

Polyakov \cite{P} has found that there exist two types of classical
hamiltonian reductions for $sl(3)$: one yields $W_{3}$ algebra while the
other leads to $W_{3}^{(2)}$ which is a $u(1)$ "quasi" SCA in the sense
that dimension $3/2$ fields are bosonic ("wrong" statistics) and, besides,
it reveals a quadratic nonlinearity in the $u(1)$ current in its operator
product expansions (OPEs). Bershadsky \cite{B1} has further explained its
structure in detail. In Ref.\cite{Ro} new infinite
families of nonlinear extended conformal algebras, $u(n)$ and $sp(2 n)$
quasi SCAs, have been found. Independently it has been shown \cite{BTv,F}
that $u(n)$ quasi SCAs can be constructed by the hamiltonian reductions of
affine algebras $sl(n)^{(1)}$, based on non-principal embeddings of $sl(2)$
into $sl(n)$. A $N=2$ supersymmetric extension of
$W_3^{(2)}$ containing both $W_3^{(2)}$ and $N=2$ SCA as genuine subalgebras
have been constructed in \cite{{KrSo},{AKrSo}} by means of hamiltonian
reduction of the affine $sl(3|2)^{(1)}$ (at the level of component
currents). Recently, a formulation of this extended SCA in terms of
constrained $N=2$ superfields has been presented \cite{IKS}.

It was demonstrated in \cite{DTH,FL,Bo} that new SCAs
with quadratic nonlinearity, so-called $Z_{2}\times Z_{2}$ graded SCAs,
can be obtained by combining both fermionic and
bosonic spin-$3/2$ currents in the same $osp(m|2n)$ or $u(m|n)$
supermultiplet. The $u(n)$ KB SCAs and the algebra $W_{3}^{(2)}$ can be
identified with $Z_2\times Z_2$ graded SCAs associated with the
superalgebras $ u(n|0) $and $u(0|1)$, respectively
\footnote{There exist other conventions for these superalgebras, see,
e.g., Ref. \cite{FL}.}.

By applying the classical hamitonian reductuion to affine Lie superalgebra
$sl(n|2)^{(1)}$ and putting the constraint on the currents valued in its
bosonic $sl(2)$ part, in \cite{IM} the classical $u(n)$ KB SCAs has been
recovered in a new setting. In \cite{Ra}, this analysis was promoted to
$N=1$ superspace and a $N=1$ extension of $u(n)$ KB SCAs has been
constructed (at the classical level). However, an attempt to incorporate
$N=2$ supersymmetry has failed.  As we will show, this happened just because
nonlinear constraints on $N=2$ affine supercurrents have not been involved
into the game.

As was already said, the aim of this paper is to develop the hamiltonian
reduction at the classical level directly in $N=2$ superspace. In short,
its main steps are: (i) construction of $N=2$ affine current algebra for
some superalgebra admitting a complex structure (we limit our consideration
here to the superalgebras $sl(n|n-1)$); (ii) imposing appropriate constraints
on the relevant superalgebra valued $N=2$ supercurrents; (iii) deducing $N=2$
extended superconformal algebras in $N=2$ superfield formalism.
We would like to specially emphasize that we are always dealing with $N=2$
superfield approach in our scheme. To our knowledge, this was not done
before. Another point to be mentioned is that our construction here is
purely algebraic and does not resort to any specific field theory
realization of $N=2$ affine current superalgebras, e.g. to their WZNW
realizations. This is the difference from, e.g., Ref. \cite{DRS} where
$N=1$ superspace version of the hamiltonian reduction was discussed in the
WZNW context. Also, we will be mainly interested in such  extended $N=2$
SCAs which include as subalgebra the standard linear $N=2$ SCA, i.e.,
contain $N=2$ superconformal stress tensor among their defining
supercurrents.

The paper is organized as follows. In Section $2$ we construct
$N=2$ $sl(n|n-1)^{(1)}$ current algebra in terms of $N=2$ supercurrents
subjected to nonlinear constraints.
In Section $3$ we describe the general procedure of
the hamiltonian reduction in $N=2$ superspace and in Section 4 we exemplify
it by the simplest case of $N=2$
$sl(2|1)^{(1)}$ which gives rise to the standard $N=2$ SCA.
In Section $5$ we consider the case of $N=2$ $sl(3|2)^{(1)}$. We reproduce
the previously known $N=2$ $W_3$ and $N=2$ $W^{(2)}_3$ SCAs in
$N=2$ superfield formulation  and find two
new $N=2$ extended SCAs. We explain how the
factorization of the dimension $1/2$ component currents in these
superalgebras
works. And finally in Section $6$ we end with a few closing remarks.
In Appendices,
we give notations for $sl(n|n-1)$ superalgebras, $u(m|n)$ SCA and different
realization of $sl(n|n-1)$.

\section{\bf $N=2$ Current Algebra for $sl(n|n-1)^{(1)}$ }
\setcounter{equation}{0}

In \cite{HS} Hull and Spence have constructed
$N=2$ current algebra for bosonic algebra $g$ in terms of
$N=2$ superfield currents satisfying {\it nonlinear} constraints.
The only essential restriction on $g$ is that it is
even-dimensional and admits a complex structure.
The quadratic terms appearing in the r.h.s. of
superoperator product expansions (SOPEs) between the supercurrents
are necessary for the consistency between these SOPEs and aforementioned
nonlinear constraints.
The nonlinearity of $N=2$ current algebra while it is written in terms
of $N=2$ supercurrents is the price for manifest $N=2$ supersymmetry.
When formulated via ordinary currents or $N=1$ supercurrents,
the algebra can be put in a linear form (in an appropriate
basis).

If $g$ is an ordinary bosonic algebra, all the $N=2$ affine supercurrents are
fermionic and we cannot put them to be constants. On the other hand,
this kind of constraints imposed on bosonic (super)currents
is of common use in the standard hamiltonian reduction scheme.
We are going to generalize the latter to $N=2$ superspace,
expecting such a generalization to allow us
to deduce extended $N=2$ SCAs (both previously known and new)
in a manifestly supersymmetric $N=2$ superfield fashion.
To be able to impose the aforementioned constraints on the affine
supercurrents, we need to have bosonic ones among them. A natural
way to achieve this is to deal with $N=2$ affine extensions of
{\it superalgebras}. So we are led to generalize the approach
of Ref. \cite{HS} to the superalgebras admitting a complex structure.
In this paper we
confine our consideration to the superalgebras $sl(n|n-1)$.

Let $g$ be a classical simple Lie superalgebra $ g=g_0 \oplus g_1$, where
$g_0$ is the bosonic subalgebra and $g_1$ is the fermionic subspace,
with the generators $t_{A}$ satisfying graded commutation relations
$ [ t_{A}, t_{B} \} ={F_{A B}}^{C} t_{C} $.
Let us introduce new structure constants,
${f_{AB}}^{C}=
(-1)^{(d_{A}+1)d_{B}} {F_{AB}}^{C}, $ where
for $ t_{A} \in g_{\alpha},
\alpha \in 0,1 $ we used the grading
$d_{A}=\alpha+1$. Therefore, ${f_{AB}}^{C}$ are
antisymmetric in the indices A,B when A,B correspond to bosonic
generators and
symmetric otherwise.
It is convenient to choose a complex basis for $g$,
so that its generators are labelled by $a$ and $\bar{a}$,
$a=1,2,\ldots$, $\frac{1}{2} \dim \, g = {1\over 2} ((2n-1)^2-1)\;$.
In this basis the complex structure associated with the second
supersymmetry has eigenvalue $+i$ on the generators $t_{a}$ and
$-i$ on the conjugated ones $t_{\bar{a}} (= t_{a}^{\dagger} )$.
The Killing metric $g_{a \bar{b}} $ is given by $ Str(t_{a}t_{\bar{
b}})$, $g_{a \bar{b}} $ being symmetric for the indices related to
bosonic generators
and antisymmetric otherwise.
Any index can be raised and lowered with $g^{a \bar{b}}$
and $g_{a \bar{b}}$.

The affine superalgebra $\hat{g}=sl(n|n-1)^{(1)}$ we deal with
in this paper has the equal
number $2n(n-1)$ of fermionic and bosonic supercurrents. For example, in
the fermionic $g$ valued supercurrent in the fundamental representation
$\cJ \equiv \cJ_A t_B g^{AB}\;,$
top-left $n \times n$ and bottom-right $(n-1)
\times (n-1)$ matrix elements are fermionic,
so that $d_{a}, d_{\bar{a}}=1$.
Then the bosonic
supercurrents are entries of the top-right $n \times (n-1)$ and
bottom-left $(n-1)
\times n$ blocks in the supercurrent matrix, so for them
$d_{a}, d_{\bar{a}}=2$.
In the scheme of hamiltonian reduction which will be explained in the
next Section we impose non-zero constraints just on these supercurrents.

We refer the reader to Ref. \cite{HS} for details of how
the $N=2$ current algebra can be formulated in $N=2$ superspace.
The only new thing to be kept in mind in our case is that now
there are extra {\it bosonic} supercurrents besides the  fermionic ones.
The presence of supercurrents with different statistics will play
an important role in our construction. This property will manifest
itself in the appearance of some extra $(-1)$ factors
in the r.h.s. of SOPEs defining the $N=2$ affine superalgebra.

With all these remarks taken into account, we summarize the
$N=2$ affine current algebra corresponding to
$sl(n|n-1)^{(1)}$  with the level $k$ as the following set of SOPEs
between $N=2$ superfield currents satisfying the appropriate nonlinear
constraints \footnote{ By Z we denote the coordinates of $1D$ $N=2$
superspace, $Z=(z, \theta, \bar{\theta})$. From now on we do not write
down explicitly the regular parts of SOPEs.
All the supercurrents (currents) appearing in
the r.h.s. of SOPEs (OPEs) are evaluated at the point
$Z_{2}$ ($z_2$).}:
\bea
{\cJ}_{a} (Z_{1}) {\cJ}_{b} (Z_{2})   & = & -\tzb {f_{ab}}^{c} {\cJ}_{c}
-\tzbb \frac{1}{k} (-1)^{(d_{a}+1)d_{\bar{c}}} (-1)^{(d_{b}+1)d_{\bar{c}}}
{f_{a \bar{c}}}^{
d} {f_{b}}^{\bar{c}e}
 {\cJ}_{d} {\cJ}_{e}, \nn \\
 {\cJ}_{\bar{a}} (Z_{1}) {\cJ}_{\bar{b}} (Z_{2})  & = &
 \tz {f_{\bar{a} \bar{b}}}^{\bar{c}} {\cJ}_{\bar{c}}
 +\tzbb \frac{1}{k} (-1)^{(d_{\bar{a}}+1)d_{c}} (-1)^{(d_{\bar{b}}+1)d_{c}}
{f_{\bar{a}c}}^{\bar{d} }
 {f_{\bar{b} }}^{c\bar{e}} {\cJ}_{\bar{d}} {\cJ}_{\bar{e}}, \nn \\
 {\cJ}_{a} (Z_{1}) {\cJ}_{\bar{b}} (Z_{2})  & = &
 -\tzzbb \frac{1}{2} k g_
{a \bar{b}}    +\frac{1}{z_{12}} k g_{a \bar{b}}
  + \tz {f_{a \bar{b}}}^c {\cJ}_c  - \tzb
 {f_{a \bar{b}}}^{\bar{c}} {\cJ}_{\bar{c}} \nn \\
 & &  + \tzbb \left[ {f_{a \bar{b}}}^{c} \cDb {\cJ}_{c}
 - \frac{1}{k} (-1)^{(d_{a}+1)d_{\bar{c}}} (-1)^{(d_{\bar{b}}+1)d_{\bar{c}}}
{f_{a \bar{c}}}^{d} {f_{\bar{b}}}^{\bar{c} \bar{e}}
 {\cJ}_{d} {\cJ}_{\bar{e}}  \right],
\label{eq:qq}
\eea
where
\bea
\th_{12}=\th_{1}-\th_{2}, \; \bth_{12}=\bth_{1}-\bth_{2}, \;
z_{12}=z_{1}-z_{2}+\frac{1}{2}(\th_{1} \bth_{2} + \bth_{1} \th_{2})
\eea
and the constraints on the supercurrents read
\bea
\cD {\cJ}_{a} - \frac{1}{2k} (-1)^{d_{a}} {f_{a}}^{bc} {\cJ}_{b} {\cJ}_{c} =
0, \quad
\cDb {\cJ}_{\bar{a}} + \frac{1}{2k} (-1)^{d_{\bar{a}}}
{f_{\bar{a}}}^{\bar{b}
     \bar{c}} {\cJ}_{\bar{b}} {\cJ}_{\bar{c}} = 0
\label{eq:cons}
\eea
(the summation is assumed over repeated indices).
Here, we work with complex fermionic covariant derivatives
\bea
\cD=\frac{\partial}{\partial \theta}-\frac{1}{2} \bar{\theta}
\partial,\;\;\;\;\;
\cDb =\frac{\partial}{\partial \bar{\theta}}-\frac{1}{2}
\theta \partial \nn
\eea
satisfying the algebra
\bea
\{ \cD, \cDb \}=-\partial\;(=-\partial_{z}),
\label{eq:DDB}
\eea
all other anticommutators are vanishing.
If we restrict the indices in (\ref{eq:qq}) and (\ref{eq:cons})
to the fermionic supercurrents we reproduce the $N=2$ $sl(n)^{(1)} \oplus
sl(n-1)^{(1)} \oplus u(1)^{(1)}$
affine current algebra \cite{HS}.
We have checked that the whole $N=2$ current superalgebra
(\ref{eq:qq}) with the nonlinear
constraints (\ref{eq:cons}) satisfies the standard $Z_2$ graded
Jacobi identities and that SOPEs of the l.h.s. of (\ref{eq:cons})
with any affine supercurrent vanish on the shell of constraints (the
presence of nonlinear terms in the r.h.s. of (\ref{eq:qq}) is crucial
for this).
When we consider this superalgebra at the quantum level (to all orders
in contractions between the supercurrents), then there appears an extra term,
$\frac{1}{2} (-1)^{d_{a}+1} {f_{a \bar{c}}}^{d} {f_{\bar{b}}}^{\bar{c}
\bar{d}}$ in $\tzzbb$ in the r.h.s. of SOPE $\cJ_{a}(Z_{1})
\cJ_{\bar{b}}(Z_{2})$.
This is due to the fact that there exist additional contractions
between the supercurrents at the quantum level. In the remainder of
this paper we will deal with the classical relations (\ref{eq:qq}) and
(\ref{eq:cons}).

Generalizing the well-known
Sugawara construction to $N=2$ superspace yields the following
formula for the improved $N=2$ stress
tensor in terms of the affine supercurrents $\cJ_{a}, \cJ_{\bar{a}}$,
\bea
{\cT}_{sug}=\frac{1}{k} g^{a \bar{b}} \cJ_{a} \cJ_{\bar{b}}
+\alpha_{i} \cDb \cH_{i}+\alpha_{\bar{j}} \cD \cH_{\bar{j}}.
\label{eq:sug}
\eea
We denote by $\cH_{i}, \cH_{\bar{i}}$ $(i=1,2,..., n-1)$
the supercurrents associated with
Cartan generators of $sl(n|n-1)$. The $N=2$ super stress tensor
satisfies the following SOPE
\bea
{\cT}_{sug} (Z_{1}) {\cT}_{sug} (Z_{2})=\frac{c}{z_{12}^{2}} +
\left[ \tzzbb -\tz \cD+\tzb \cDb +\tzbb \partial  \right] {\cT}_{sug}
\label{eq:TT}
\eea
with
\bea \label{centrN2}
c= -2k\; \alpha_i \alpha_{\bar{j}} g_{i \bar{j}}\;.
\eea
With respect to this ${\cT}_{sug}$, the supercurrents
$\cH_{i}, \cH_{\bar{i}} $ are
quasi superprimary superfields of
the dimension $1/2$ with $u(1)$ charge $+1, -1$, respectively.
All other affine $N=2$ supercurrents are superprimary
\bea
{\cT}_{sug} (Z_{1}) {\cJ}_{a (\bar{a})} (Z_{2})=
\left[ s_{a (\bar{a})} \tzzbb -
\tz \cD + \tzb \cDb + \tzbb \partial +
 q_{a (\bar{a})} \frac{1}{z_{12}} \right] {\cJ}_{a (\bar{a})}.
\label{eq:JJ}
\eea
Their dimensions (superspins) $s_{a(\bar{a})}$ and $u(1)$ charges
$q_{a(\bar{a})}$ depend in a certain way
on  the parameters $\alpha_{i}, \alpha_{\bar{j}}$. The explicit formulas
for them  will be given later, for each specific example
we will consider.

Our last remark in this Section is that for the superalgebra
$sl(n|n-1)$ (and seemingly for other superalgebras admitting a complex
structure) the above $N=2$ extension actually coincides with the
$N=1$ extension given in \cite{DRS,FRS}. In other words, the latter
possesses a {\it hidden}
$N=2$ supersymmetry which becomes manifest in terms
of constrained $N=2$ supercurrents. Indeed, due to the fact that the
generators
of $sl(n|n-1)$ can be divided into the pairs of mutually conjugated ones,
the relevant
$N=1$ supercurrent for each pair is complex and its component content
(two real spins ${1\over 2}$ and two real spins $1$) is such that
these components can be combined into a $N=2$ supermultiplet
\footnote{Similar arguments for the case of bosonic
algebra $g$ were given in \cite{RASS}.}. To explicitly demonstrate
this, let us solve the
constraints (\ref{eq:cons}) via unconstrained
$N=1$ supercurrents $J_{a},J_{\bar{a}}$
\bea
{\cJ}_{a}=J_{a} +\theta^{1} \left[ iD J_{a}+ (-1)^{d_{a}} \frac{1}{k}
{f_{a}}^{bc} J_{b}J_{c} \right] , \nn \\
{\cJ}_{\bar{a}}=J_{\bar{a}} -\theta^{1} \left[
iD J_{\bar{a}}+ (-1)^{d_{\bar{a}}} \frac{1}{k} {f_{\bar{a}}}^{\bar{b}
\bar{c}}
J_{\bar{b}} J_{\bar{c}} \right].
\label{eq:jj}
\eea
Here the $N=1$ supercurrents are defined on a real $N=1$ superspace
$\widetilde{Z}=(z, \theta^2)$,
$\theta^2 \equiv {1\over 2} (\theta + \bar \theta)$,
$D=\frac{\partial}{\partial {\theta}^{2}} - {\theta}^{2}
\partial$ is a $N=1$ covariant fermionic derivative
and $\theta^{1} \equiv {1\over 2i} (\theta - \bar \theta)$ is an
extra fermionic coordinate.
The SOPEs between the $N=1$ $sl(n|n-1)^{(1)}$ affine supercurrents $J_A (
\widetilde{Z})$
\cite{DRS,FRS}
are given by
\bea
J_A (\widetilde{Z_1}) J_B (\widetilde{Z_2})=\frac{1}{\widetilde{z_{12}}}
k g_{AB}+\frac{\widetilde{\th_{12}} }{\widetilde{z_{12}}}
{f_{AB}}^{C} J_C
\label{eq:jj1}
\eea
where
\bea
\widetilde{\th_{12}}=\th_{1}^{2}-\th_{2}^{2}, \;
\widetilde{z_{12}}=z_{1}-z_{2}-\th_{1}^{2} \th_{2}^{2}.
\eea
In a complex basis, the indices $A, B, \cdots $ can be divided into
the two sets of the barred  and unbarred indices, thus
demonstrating that the number of $N=1$ supercurrents in the present case
coincides with the number of $N=2$ ones (of course, these complex
$N=1$ supercurrents are reducible, each containing two real $N=1$
supermultiplets). The superalgebra
(\ref{eq:jj1}) is equivalent to the superalgebra ( \ref{eq:qq}) supplemented
with the nonlinear constraints (\ref{eq:cons}). The $N=1$ superfield
formulation
clearly demonstrates that the nonlinearities in the r.h.s. of
eqs. (\ref{eq:qq}) are fake: they appear as the price for manifest
$N=2$ supersymmetry. In what follows the $N=1$ formulation will be
a useful guide of how to impose constraints on the relevant $N=2$
supercurrents corresponding to different embeddings of $sl(2|1)$ into
$sl(n|n-1)$ and to extract those preserving $N=2$ supersymmetry from their
general set.

In the next Sections, we will discuss different hamiltonian reductions
of the $N=2$ $sl(n|n-1)^{(1)}$ current algerbra in $N=2$ superspace for
the particular cases of $n=2, 3$. But before we will sketch the
basic peculiarities of the $N=2$ superspace version of the hamiltonian
reduction procedure.

\section{\bf Hamiltonian Reduction }
\setcounter{equation}{0}

To illustrate the basic idea of different reductions, we start by
considering how we can obtain
extended $N=2$ SCAs by imposing reduction constraints on
the $N=2$ affine supercurrents which we defined in Section $2$.

       From now on we will deal with the matrix elements
$\cJ_{mn}$ of the $sl(n|n-1)$ valued
affine $N=2$ supercurrent (with the $sl(n|n-1)$ generators in the
fundamental representation) rather than with its ajoint
representation components labelled by indices $a, \bar{a}$.
The explicit relation between them is given by
\bea
\cJ_{mn}=
\left( \begin{array}{cccccccccc}
\cH_{\bar{1}}& & & & & & & &    \\
             & \cH_{\overline{2}}+\cH_{1}&  & & & &
\mbox{unbarred} & &    \\
             & &. &  & & & \mbox{indices} & &    \\
             & & &  & & & & &   \\
             & & & & \cH_{\overline{n-1}}+\cH_{n-2} & & & &    \\
             & & & & &  \cH_{{n-1}} & & &    \\
             & & & & & & \cH_{\overline{1}}+\cH_{1}  & &    \\
             & \mbox{barred} & & & & & &. & &   \\
             &  \mbox{indices} & & & & & & &   \\
             & & & & & & & &   \cH_{\overline{n-1}}+\cH_{n-1}
\end{array} \right)
\label{eq:jmn}
\eea
(see also Appendices A and B).

We will consider only linear reduction constraints
like in \cite{DS,FRRTW,BTv}.
Then we are led to equate some of $\cJ_{mn}$ (we denote the corresponding
subset of indices by the symbol ``hat'') to constants
\bea
\Phi_{\hat{m} \hat{n}} \equiv \cJ_{\hat{m} \hat{n}}-c_{\hat{m} \hat{n}}=0.
\label{eq:constrgen}
\eea
The entries of the constant supermatrix $c_{\hat{m}\hat{n}}$ can be
either $0$, which is
possible both for bosonic and fermionic supercurrents, or $1$,  which  is
admissible only for bosonic supercurrents. In order to produce $N=2$
supersymmetric algebras these constraints should be invariant with respect
to $N=2$ superconformal transformations generated by improved $N=2$
stress tensor (\ref{eq:sug}), which means that the constrained supercurrents
with nonzero $c_{\hat{m} \hat{n}}$ should have zero spin and
$u(1)$ charge.

In Ref. \cite{FRS}, $W$ superalgebras which can be obtained by the
reductions associated with different embeddings
of $osp(1|2)$ into $sl(n|n-1)$ have been classified in $N=1$ superspace.
Once we know the constraints in $N=1$ superspace, the relation (\ref{eq:jj})
gives us constraints in $N=2$ superspace. Some of the constraints in $N=1$
superspace, being rewritten in $N=2$ superspace, explicitly
break $N=2$ supersymmetry. Meanwhile, we wish to deal with only those
reductions which preserve $N=2$ supersymmetry, because our eventual aim is to
get extended SCAs containing $N=2$ SCA
as a subalgebra. Only a subset of constraints in $N=1$ superspace
preserves $N=2$ supersymmetry, namely those which after substitution
into (\ref{eq:jj}) produce no explicit $\theta$'s in the r.h.s.,
i.e. lead to the $N=2$ constraints in the form (\ref{eq:constrgen}).
Thus, we can choose the approriate subset of constraints in $N=1$
superspace and then extract the constraints in $N=2$
superspace from (\ref{eq:jj}).

Then the first-class constraints, i.e. those which commute among
themselves on the constraints shell, generate a gauge invariance.
An infinitesimal gauge transformation of $\cJ_{kl}$ induced by
$\Phi_{\hat{m}\hat{n}}$ with a gauge parameter
${\Lambda}_{\hat{m} \hat{n}}$ can easily be calculated
\bea
\delta_{\Lambda} \cJ_{kl}(Z_{2})=\frac{1}{2\pi i} \oint d
Z_{2} {\Lambda}_{\hat{m} \hat{n}}
(Z_{1}) \left( \Phi_{\hat{m} \hat{n}}(Z_{1}) \cJ_{kl}(Z_{2}) \right)
\vert_{ \{\Phi_{\hat{m} \hat{n}}=0\} },
\label{eq:gauge}
\eea
where the symbol $\vert_{ \{ \Phi_{\hat{m} \hat{n}}=0 \} } $ means that
after computing the SOPEs we should pass on the constraints shell by
imposing the constraints (\ref{eq:constrgen}) on the resulting expression
and the gauge parameters $\Lambda_{\hat{m} \hat{n}}$ are general $N=2$
superfields which do not depend on $\cJ_{kl}$. It is clear that
the variation of the l.h.s. of (\ref{eq:cons}) vanishes identically
because the SOPEs of (\ref{eq:cons}) with any $\cJ_{kl}$, and,
in particular, with  $\Phi_{\hat{m} \hat{n}}$ are zero on the shell of
(\ref{eq:cons}) (see discussion in the paragraph below (\ref{eq:DDB})).

{\it By definition}, an extended $N=2$ SCA constructed
by the hamiltonian reduction based on the constraints
(\ref{eq:constrgen})
is a superalgebra generated by gauge invariant
differential - polynomial functionals of affine supercurrents $\cJ_{kl}$,
including some $N=2$ stress tensor.
It is possible to find these superalgebras by using
Dirac construction.
Let us remind its main steps.

At first, we should fix the gauge, which means that
we are led to enlarge the original set of first-class constraints by adding
the gauge-fixing conditions (standard gauge-fixing procedure),
such that the total set of constraints becomes second-class.
We denote this extended set of constraints by $\Psi_{\hat{m}\hat{n}}$.
The number of constraints $\Psi_{\hat{m}\hat{n}}$ is
exactly twice the number of $\Phi_{\hat{m}\hat{n}}$.
For the remaining unconstrained supercurrents we will use in this Section
Greek indices, $\alpha, \beta, \cdots$. Clearly, once a gauge freedom with
respect to the $\Lambda$ transformations has been somehow fixed,
the surviving supercurrents $\cJ_{\alpha \beta}$ are expressed as
some gauge invariant differential functionals of the original affine
supercurrents.

Secondly, we should construct Dirac brackets between these gauge
invariant supercurrents.
We generalize this procedure  to the $N=2$ supersymmetric case and represent
Dirac brackets in an
equivalent form of SOPEs. The new rules for
calculation of SOPEs of the gauge invariant supercurrents which we denote
by brackets with star,
$\left( \cJ_{\alpha \beta} (Z_{1}) \cJ_{\gamma \sigma} (Z_{2}) \right)^{*}$,
can be defined in terms of original SOPEs of the affine supercurrents
as follows
\vspace{1cm}
\bea
\left( \cJ_{\alpha \beta} (Z_{1}) \cJ_{\gamma \sigma} (Z_{2}) \right)^{*}
 \equiv
\left( \widetilde{\cJ_{
\alpha \beta}} (Z_{1}) \widetilde{\cJ_{\gamma \sigma}} (Z_{2}) \right)
\vert_{\{ \Psi_{\hat{m} \hat{n}}=0 \}} \nn
\eea
\bea
- \left[ \frac{1}{(2 \pi i)^2} \oint
\oint d \; Z_{3} d \; Z_{4} \left( \widetilde{\cJ_{\alpha \beta}} (Z_{1})
\Psi_{\hat{i} \hat{j}} (Z_{3}) \right) \triangle^{\hat{i} \hat{j},
\hat{k} \hat{l}} (Z_{3},
Z_{4}) \left( \Psi_{\hat{k} \hat{l}} (Z_{4}) \widetilde{\cJ_{\gamma \sigma}}
(Z_{2}) \right) \right]
\vert_{\{ \Psi_{\hat{m} \hat{n}}=0 \}},
\label{eq:dirac}
\eea
where $\widetilde{\cJ_{\alpha \beta}}$ are functionals of the original
supercurrents $\cJ_{kl}$ (including both unconstrained
$\cJ_{\alpha \beta}$ and constrained  $\cJ_{\hat{m} \hat{n}}$
supercurrents) which satisfy the following restrictions on the constraints
shell
\bea
\widetilde{\cJ_{\alpha \beta}} \vert_{\{ \Psi_{\hat{m}\hat{n}}=0 \}}=
\cJ_{\alpha \beta}
\label{eq:tilj}
\eea
and are arbitrary otherwise, the supermatrix
$\triangle^{\hat{i} \hat{j},\hat{k} \hat{l}}(Z_{1},Z_{2})$ is the
inverse of the supermatrix
\bea
\triangle_{\hat{i} \hat{j}, \hat{k} \hat{l}}(Z_{1},Z_{2})=
 \left( \Psi_{\hat{i} \hat{j}} (Z_{1}) \Psi_{\hat{k} \hat{l}} (Z_{2}) \right)
\vert_{\{ \Psi_{\hat{m} \hat{n}}=0 \}}
\eea
i.e.
\bea
\frac{1}{2 \pi i} \oint d \; Z_{2} \triangle^{\hat{i} \hat{j}, \hat{k}
\hat{l}} (Z_{1}, Z_{2})
\triangle_{
\hat{k} \hat{l}, \hat{m} \hat{n}} (Z_{2}, Z_{3})= \delta^{\hat{i}}_{\hat{m}}
\delta^{\hat{j}}_{\hat{n}}
\theta_{13} {\bar{\theta}}_{13}
\delta(z_{1}-z_{3}) \;.
\eea
Any gauge invariant supercurrent can be represented as some functional of
$\cJ_{\alpha \beta}$ and SOPEs between these functionals can be
calculated using SOPEs
(\ref{eq:dirac}) between $\cJ_{\alpha \beta}$.

It is a very complicated technical problem to calculate the inverse
supermatrix $\triangle^{\hat{i} \hat{j},\hat{k} \hat{l}}(Z_{1},Z_{2})$ in
the general case. To get round this difficulty,
we use the following trick. By looking at
(\ref{eq:dirac}), one can observe that for $\widetilde{\cJ_{\alpha \beta}}$
satisfying
\bea
\left( \widetilde{\cJ_{\alpha \beta}} (Z_{1}) \Psi_{\hat{m} \hat{n}} (Z_{2})
\right) \vert_{\{ \Psi_{\hat{m} \hat{n}}=0 \}} = 0
\label{eq:req}
\eea
the second term in the r.h.s. of (\ref{eq:dirac}) is vanishing.
We are at freedom to choose $\widetilde{\cJ_{\alpha \beta}}$ to satisfy
eq.(\ref{eq:req}) as these functionals are {\it a priori} arbitrary up to
the condition (\ref{eq:tilj}) which is obviously consistent
with (\ref{eq:req}). Then
the SOPEs with star between the gauge invariant supercurrents
coincide with ordinary SOPEs between
$\widetilde{\cJ_{\alpha \beta}}$ on the constraints shell and so
can be calculated using SOPEs (\ref{eq:qq}) for the original
affine supercurrents
\bea
\left( \cJ_{\alpha \beta} (Z_{1}) \cJ_{\gamma \sigma} (Z_{2}) \right)^{*}
& \equiv &
\left( \widetilde{\cJ_{
\alpha \beta}} (Z_{1}) \widetilde{\cJ_{\gamma \sigma}} (Z_{2})
\right) \vert_{\{ \Psi_{\hat{m} \hat{n}}=0 \}}.
\label{eq:req1}
\eea
In this way, the task of constructing $N=2$ extended superalgebras reduces
to that of constructing the functionals
$\widetilde{\cJ_{\alpha \beta}}$ satisfying
the restrictions (\ref{eq:tilj}), (\ref{eq:req}).

Now let us discuss the general structure of such functionals.
It is evident that only those of them which are {\it linear} in the
total set of constraints $\Psi_{\hat{m} \hat{n}}$ can actually
contribute to (\ref{eq:req1}), because the SOPEs including any higher order
monomial of $\Psi_{\hat{m} \hat{n}}$ are proportional to
$\Psi_{\hat{m} \hat{n}}$ and so obviously vanish on the constraints shell
$\{ \Psi_{\hat{m} \hat{n}} = 0 \}$. The coefficients in these linear
functionals can in general be nonlinear functionals of the remaining
unconstrained supercurrents $\cJ_{\alpha \beta}$.

Keeping this in mind, from now on we consider as a starting expression for
$\widetilde{\cJ_{\alpha \beta}}$
linear functionals of constraints $\Psi_{\hat{m} \hat{n}}$
(and derivatives of the latter) with nonlinear
in general coefficient-functions of $\cJ_{\alpha \beta}$. Taking for these
coefficients the
most general ansatz in terms of $\cJ_{\alpha \beta}$ with arbitrary
constant coefficients, such that it preserves superspins and $u(1)$ charges
with respect to the improved $N=2$ stress tensor (\ref{eq:sug}), and
substituting it into eqs.
(\ref{eq:tilj}), (\ref{eq:req}), one obtains the solution which proves
to be unique up to some unessential
coefficients which do not contribute to (\ref{eq:req1}).

In the next Section we will illustrate the formalism described above
by the simplest example of hamiltonian reduction of the
$N=2$ $sl(2|1)^{(1)}$ superalgebra.

\section{\bf Example: $N=2$ $sl(2|1)^{(1)}$ Affine Superalgebra}
\setcounter{equation}{0}

Let us apply the general procedure developed in the previous Section to
the superalgebra $N=2$ $sl(2|1)^{(1)}$. We will naturally come to the
$N=2$ superspace formulation of standard $N=2$ SCA in this way.

In Appendix A, for completeness we give the explicit form of generators,
structure constants and Killing metric for $sl(2|1)$ superalgebra in
the complex basis described in Section $2$, as well as the relations
between affine supercurrents $\cJ_{a}, \cJ_{\bar{a}}$ in this basis and
matrix elements $\cJ_{mn}$ introduced in Section $3$.
Substituting these formulas into (\ref{eq:qq}), (\ref{eq:cons}) and
(\ref{eq:sug}) one can obtain explicit expressions for the defining SOPEs
of $N=2$ affine extension of $sl(2|1)$,
for nonlinear constraints the relevant supercurrents satisfy, as well as
for the improved Sugawara $N=2$ stress tensor.
The last one has the following form
\bea
\cT_{sug}=\frac{1}{k} \left( \cJ_{12} \cJ_{21} - \cH_{1} \cH_{\bar{1}} -
\cJ_{13} \cJ_{31}-\cJ_{23} \cJ_{32} \right) +
\alpha_{1} \cDb \cH_{1}+\alpha_{\bar{1}}
\cD \cH_{\bar{1}},
\label{eq:sugsl21}
\eea
where two parameters, $\alpha_{1}$ and $\alpha_{\bar{1}}$, give rise to a
splitting of supercurrents into the grades with positive, zero and negative
dimensions and $u(1)$ charges (see Table 1).
Actually, this splitting is due to the existence of two
grading operators: $(\alpha_{1} t_{2} +\alpha_{\bar{1}} t_{\bar{2}} )/2 $ and
$\alpha_{1} t_{2}-\alpha_{\bar{1}} t_{\bar{2}}$, $t_i, t_{\bar{i}}$ being
Cartan generators of $sl(2|1)$ in the coadjoint representation.
The eigenvalues of the former are exactly
( ``dimension''- $1/2$)  in the Table 1 and those of the latter are
(``$u(1)$ charge'' $\pm 1$) where $+1$ is for barred supercurrents
and $-1$ for unbarred ones.
\bea
\begin{array}{ccc}
\mbox{Table 1} \nn \\ \hline
\mbox{scs} & \mbox{dim} & u(1)   \nn \\
\hline
\vspace{2mm}
\cH_{1}^F & 1/2 & 1  \nn \\
\vspace{2mm}
\cH_{\bar{1}}^F & 1/2 & -1  \nn \\
\vspace{2mm}
\cJ_{12}^F & (1+\alpha_{1}-\alpha_{\bar{1}})/2 &
(1+\alpha_{1}+\alpha_{\bar{1}})  \nn \\
\vspace{2mm}
\cJ_{21}^F & (1-\alpha_{1}+\alpha_{\bar{1}})/2 &
(-1-\alpha_{1}-\alpha_{\bar{1}})   \nn \\
\vspace{2mm}
\cJ_{13}^B & (1+\alpha_{1})/2 & (1+\alpha_{1})  \nn \\
\vspace{2mm}
\cJ_{31}^B & (1-\alpha_{1})/2 & (-1-\alpha_{1})  \nn \\
\vspace{2mm}
\cJ_{23}^B & (1+\alpha_{\bar{1}})/2 & (1-\alpha_{\bar{1}})  \nn \\
\vspace{2mm}
\cJ_{32}^B & (1-\alpha_{\bar{1}})/2 & (-1+\alpha_{\bar{1}})  \nn \\
\hline
\end{array}
\eea
In this and all subsequent Tables we use the following abbreviations:
``scs'' for supercurrents, ``dim'' for superconformal dimensions
and ``$u(1)$''
for $u(1)$ charges.

We also give the explicit form of the nonlinear constraints (\ref{eq:cons})
\bea
\cD \cH_{1}   = 0,  & \;\; & \cDb \cH_{\bar{1}}   =  0, \nn \\
\left( \cD  + \frac{1}{k} \cH_{1} \right) \cJ_{12}   =   0,  & \;\; &
\left( \cDb - \frac{1}{k} \cH_{\bar{1}} \right) \cJ_{21}   =   0, \nn \\
\cD \cJ_{13} - \frac{1}{k} \cJ_{12} \cJ_{23}   =  0,  & \;\; &
\left( \cDb -\frac{1}{k} \cH_{\bar{1}} \right) \cJ_{31} - \frac{1}{k}
\cJ_{21} \cJ_{32}   =  0, \nn \\
\left( \cD  - \frac{1}{k} \cH_{1} \right) \cJ_{23}   =  0, & \;\; &
\cDb \cJ_{32}   =   0.
\label{eq:constra}
\eea

Now we are ready to consider a hamiltonian reduction of $N=2$
$ sl(2|1)^{(1)}$
which produces $N=2$ SCA. To this end, we should first learn at which
values of parameters $\alpha_{1}$ and $\alpha_{\bar{1}}$ at least one
of the bosonic supercurrents could have the spin
and $u(1)$ charge characteristic of the $N=2$ stress tensor, i.e.
$1, 0$, respectively. It turns out possible with the following choice
\bea
\alpha_{1} = -1, \;\; \alpha_{\bar{1}} = 1.
\label{eq:alphab}
\eea
In this case, besides the fermionic supercurrents $\cH_{1},
\cH_{\bar{1}}$ with
the spin and $u(1)$ charge $1/2$ and $\pm 1$,
$N=2$ $ sl(2|1)^{(1)}$ superalgebra contains bosonic spin $0$
$( \cJ_{13}, \cJ_{32}) $  and spin $1$  $(\cJ_{31}, \cJ_{23} )$
ones with zero $u(1)$ charges, as well as the fermionic doublet
$\cJ_{12}, \cJ_{21}$ with spins $-1/2,3/2$ and $u(1)$ charges $1, -1$,
respectively.

Secondly, we should put first-class constraints on some supercurrents at
which at least one of two spin $1$ supercurrents
( $\cJ_{31} \; \mbox{or} \; \cJ_{23}$ )
is unconstrained in order to be able to identify it with
$N=2$ unconstrained stress tensor. At first sight, it seems
impossible to achieve this because from the
beginning all the supercurrents are constrained by the
conditions (\ref{eq:constra}). Nevertheless,  it can be done.
Let us briefly explain the basic idea of how unconstrained $N=2$
superfields can come out in this way.

By looking at the constraints (\ref{eq:constra}), one sees that they are
quadratically nonlinear and their number precisely matches with that of
supercurrrents. Moreover, in every constraint there is only one linear term
with spinor covariant derivative on some supercurrent, and different
constraints contain different linear terms, so they are in one-to-one
correspondence with the consistent set of standard chiral and
anti-chiral conditions. The last ones reduce the number of
independent superfield components by the factor two.
The same is evidently true for a nonlinear generalization of
these constraints  (\ref{eq:constra}): the only new point is that
the components which were forced to be zero in
the case of chiral constraints become some functions of the remaining
independent ones in the case of (\ref{eq:constra}). However, an
important difference of the latter from the linear constraints
is the following. If we replace
some bosonic supercurrents in (\ref{eq:constra}) by nonzero constants,
then in some constraints the nonlinear terms can produce {\it  a linear} one
without a spinor derivative on it. So, this constraint becomes algebraic
with respect to the supercurrent entering it linearly and can be solved
for the latter. Thus this supercurrent turns out to be eventually expressed
in terms of other ones and their spinor covariant derivatives. Now among
the remaining independent supercurrents one can find, in a number of cases,
unconstrained $N=2$ superfields. This is just what comes about in the
case at hand. An analogous resume
could be drawn from the analysis of solutions of $N=2$ constraints
(\ref{eq:constra}) in terms of unconstrained $N=1$ superfields (\ref{eq:jj}).

Keeping in mind the above remark, we choose first-class constraints as
follows
\bea
\cJ_{mn}^{constr}=
\left( \begin{array}{ccc}
\ast & 0 & 1 \\
\ast & \ast & \ast \\
\ast & 1 & \ast
\end{array} \right).
\label{eq:sl21constr}
\eea
They clearly preserve $N=2$ superconformal symmetry generated by
$\cT_{sug}$ (\ref{eq:sugsl21}), (\ref{eq:alphab}).
This set of constraints is also consistent
with eqs. (\ref{eq:constra}). Indeed, by substituting (\ref{eq:sl21constr})
into (\ref{eq:constra}) we find that those constraints from (\ref{eq:constra})
which include spinor derivative of the supercurrents
$\cJ_{13}, \cJ_{32}$ and $\cJ_{12}$ are satisfied identically while the
constraint containing  spinor derivative of $\cJ_{31}$ current becomes
algebraic and expresses $\cJ_{21}$ in terms of $\cJ_{31}$
\bea
\cJ_{21} = k \left( \cDb -\frac{1}{k} \cH_{\bar{1}} \right) \cJ_{31}.
\label{eq:constsol}
\eea
The remaining constraints from the set (\ref{eq:constra}) preserve
their form on the constraints
shell (\ref{eq:sl21constr}). Thus on the shell of
constraints (\ref{eq:sl21constr}) there arise no any restrictions on the
spin $1$, $u(1)$ charge $0$ bosonic supercurrent $\cJ_{31}$, so
the latter is an
unconstrained $N=2$ superfield and, as we will see soon, proves to be
directly related to the $N=2$ superconformal stress tensor.

Let us note that the constraints (\ref{eq:sl21constr}) actually amount
to the set of constraints imposed in  \cite{DRS}
in $N=1$ superspace. This latter set can be shown to
produce the above
constraints without breaking $N=2$ supersymmetry through the explicit
relation (\ref{eq:jj}) between $N=1$ and $N=2$ supercurrents.

Constraints (\ref{eq:sl21constr}) can easily be checked to have zero mutual
SOPEs on their shell, so they are first-class and give rise to a gauge
invariance which can be used to gauge away three more entries in the
supermatrix (\ref{eq:sl21constr}). Indeed, with respect to infinitesimal
gauge transformations
(\ref{eq:gauge}) generated by constraints (\ref{eq:sl21constr}) with the gauge
parameters $\Lambda_{12}, \Lambda_{13}$ and $\Lambda_{32}$ the currents
$\cJ_{23}, \cH_{1}$ and $\cH_{\bar{1}}$ are transformed inhomogeneously
\bea
\delta_{\Lambda_{12}} \cJ_{23} (Z_{2}) & = & -
\left( \cD -\frac{1}{k} \cH_{1} \right)\Lambda_{12},\nn\\
\delta_{\Lambda_{13}} \cH_{1} (Z_{2}) & = & \cD {\Lambda}_{13}, \nn \\
\delta_{\Lambda_{32}} \cH_{\bar{1}} (Z_{2}) & = & \cDb {\Lambda}_{32}.
\eea
One can explicitly check that these gauge transformations preserve
the constraints (\ref{eq:constra}).
As a result, we can consistently fix the gauge as\footnote{In this gauge
there remains a residual gauge freedom with chiral $\Lambda_{12}$ and
anti-chiral $\Lambda_{13}$, $\Lambda_{32}$. However, the final expression
for $N=2$ stress tensor turns out to be invariant under this
residual freedom.}
\bea
\cJ_{23} = 0, \;\;  \cH_{1} = 0, \;\; \cH_{\bar{1}} = 0.
\label{eq:gagfix}
\eea
It is easy to check that the total set of constraints, i.e. constraints
(\ref{eq:sl21constr}) and gauge fixing conditions (\ref{eq:gagfix}),
is second-class.
Substituting the gauge fixing conditions (\ref{eq:gagfix}) and
the expression (\ref{eq:constsol}) into supermatrix (\ref{eq:sl21constr})
we obtain the expression
for $N=2$ supercurrents $\cJ_{mn}$ in the highest
weight ( or Drinfeld-Sokolov \cite{DS} ) gauge
\bea
\cJ_{mn}^{DS}=
\left( \begin{array}{ccc}
0 & 0 & 1 \\
k \cDb \cJ_{31} & 0 & 0 \\
\cJ_{31} & 1 & 0
\end{array} \right).
\label{eq:ds}
\eea
So the superalgebra which is produced from $N=2$ $sl(2|1)^{(1)}$ by
hamiltonian reduction associated with the constraints (\ref{eq:sl21constr})
is generated by only one gauge invariant bosonic supercurrent
$\widetilde{\cJ_{31}}$
which coincides with $\cJ_{31}$ on the shell of total set of constraints
(see (\ref{eq:tilj})).

Our next task is to find $\widetilde{\cJ_{31}}$ from the
conditions (\ref{eq:tilj}), (\ref{eq:req}).
This can be easily done by making use of the general procedure described
in  Section 3. As a result we obtain the following expression for
$\widetilde{\cJ_{31}}$ up to unessential terms
\bea
\widetilde{\cJ_{31}}=\left( -\cJ_{12} \cJ_{21}+\cJ_{13}
\cJ_{31}+\cJ_{23} \right)+k \cDb \cH_{1}- k \cD \cH_{\bar{1}}+k^3
\cDb \cJ_{12}'+k^2 \cJ_{13}'-k^2 \cJ_{32}'.
\eea
Substituting this expression into (\ref{eq:req1}),  we get the
SOPE of superalgebra we are looking for. This SOPE coincides with
SOPE of $N=2$
SCA (\ref{eq:TT}) with central charge $-2 k$ after rescaling
\bea
\cJ_{31} \rightarrow -k \cJ_{31}.
\eea

Before closing this Section, we briefly mention that there
exists another choice for the gauge fixing, so called
diagonal gauge \cite{DS}. Namely,
\bea
\cJ_{23} = 0, \;\;  \cJ_{21} = 0, \;\; \cJ_{31} = 0.
\label{eq:gagfix1}
\eea
Repeating all the steps we have passed before, one can obtain the
following form for $N=2$ supercurrents $\cJ_{mn}$ in this case
\bea
\cJ_{mn}^{diag}=
\left( \begin{array}{ccc}
\cH_{\bar{1}} & 0 & 1 \\
0 & \cH_{1} & 0 \\
0 & 1 & \cH_{\bar{1}}+\cH_{1}
\end{array} \right),
\label{eq:dia}
\eea
where $\cH_{1}, \cH_{\bar{1}}$ are chiral and anti-chiral $N=2$ fermionic
superfields which form the $N=2$ $u(1)$ affine superalgebra
\bea
\cH_{1} (Z_{1}) \cH_{\bar{1}} (Z_{2}) = \tzzbb \frac{k}{2} - \zot k.
\eea

Two different gauge choices (\ref{eq:ds}), (\ref{eq:dia}) are
connected to each other by some gauge
transformation. If we would know this gauge transformation, then we could
obtain the standard Miura free field realization of $N=2$ SCA in terms of
chiral and
anti-chiral fermionic superfields. However, in our simple case it is
of no need to know this gauge transformation for deducing
Miura realization, if we observe that the total set of constraints
(\ref{eq:sl21constr}), (\ref{eq:gagfix1}) for the diagonal gauge on
the constraint shell is
invariant under transformations generated by the $N=2$ stress tensor
$\cT_{sug}$ (\ref{eq:sugsl21}), (\ref{eq:alphab}).
This means that $\cT_{sug}$ is gauge invariant and on the constraints shell
has the following form
\bea
\cT_{sug}= - \frac{1}{k} \cH_{1} \cH_{\bar{1}}  -
 \cDb \cH_{1} + \cD \cH_{\bar{1}}
\label{eq:sugsl21shell}
\eea
which coincides with the standard Miura free field form of
the $N=2$ stress tensor.

In the next Section we will discuss various reductions of
$N=2$ $sl(3|2)^{(1)}$ and deduce some new superfield extended $N=2$
SCAs in this way.

\section{\bf Hamiltonian Reductions of
$N=2$ $sl(3|2)^{(1)}$ Affine Superalgebra}
\setcounter{equation}{0}

The reductions of $N=2$ $sl(3|2)^{(1)}$ we will consider in this Section
give rise to four new types of extensions of $N=2$ SCA.
The first one is rather unusual in the sense that
the $N=2$ stress tensor is a constrained supercurrent. The second possesses
an unconstrained stress tensor, but contains spin $0$ supercurrents, such
that it turns out impossible to
decouple dimension $0$ component currents. We will concentrate
on the third and fourth cases corresponding to $N=2$ $u(2|1)$ and
$N=2$ $u(3)$ SCAs, respectively, because these are ``canonical'' in the sense
that the relevant $N=2$ stress tensor is unconstrained and there are
no spin $0$ supercurrents.  We will also illustrate
how the known $N=2$ $W_3$ \cite{LPRSW} and $N=2$ $W_{3}^{(2)}$ \cite{IKS}
SCAs reappear in the hamiltonian reduction approach in $N=2$ superspace.

It is rather straightforward to find the structure constants and Killing
metric in the complex basis for $sl(3|2)$, so we do not write down them
explicitly (see Appendix B). From the general expression for the
improved Sugawara $N=2$ stress tensor
(\ref{eq:sug}) we obtain it for $N=2$ $sl(3|2)^{(1)}$ in the following form
\bea
\cT_{sug} & = &
\frac{1}{k} \left( -\cH_{1} \cH_{\bar{1}}-\cH_{2} \cH_{\bar{2}}+
\cJ_{12} \cJ_{21}+\cJ_{13} \cJ_{31}+\cJ_{23} \cJ_{32}-\cJ_{14} \cJ_{41}
 \right. \nn \\
& & \left.
-\cJ_{15} \cJ_{51}-\cJ_{24} \cJ_{42}-\cJ_{25} \cJ_{52}-\cJ_{34} \cJ_{43}-
\cJ_{35} \cJ_{53}-\cJ_{45} \cJ_{54}-\cH_{2} \cH_{\bar{1}} \right) \nn \\
& & +\alpha_{1} \cDb \cH_{1}+\alpha_{\bar{1}} \cD \cH_{\bar{1}}+\alpha_{2}
\cDb \cH_{2}+\alpha_{\bar{2}} \cD \cH_{\bar{2}},
\label{eq:sugsl32}
\eea
where four parameters, $\alpha_{1}, \alpha_{\bar{1}}, \alpha_{2},
\alpha_{\bar{2}}$ split the supercurrents into the grades with positive,
zero and negative dimensions and $u(1)$ charges (see Table 2).

Let us stress that the nonlinear constraints (\ref{eq:cons}) for the case
of $sl(3|2)$ can be easily read off using the structure constants
of this superalgebra. They will play the important role in all the
calculations in the remainder of this paper.
Our main aim in this Section will be to find
extended $N=2$ SCAs which contain at least one
{\it bosonic unconstrained} supercurrent with dimension $1$ and vanishing
$u(1)$ charge by applying the general procedure of the hamiltonian
reduction to $N=2$ $sl(3|2)^{(1)}$ affine superalgebra.

In the next Subsections, we will present only the basic results and make
some comments without detailed explanations, because most
of technical points are a direct generalization of those
expounded in Section $4$ on the simpler example of $N=2$ $sl(2|1)^{(1)}$.

\bea
\begin{array}{ccc}
\mbox{Table 2} \nn \\
\hline
\mbox{scs} & \mbox{dim} & u(1) \nn \\ \hline
\vspace{2mm}
\cH_{\bar{1}}^F  & 1/2 & -1 \nn \\
\vspace{2mm}
\cH_{1}^F  & 1/2 & 1 \nn \\
\vspace{2mm}
\cH_{\bar{2}}^F  & 1/2 & -1 \nn \\
\vspace{2mm}
\cH_{2}^F  & 1/2 & 1 \nn \\
\vspace{2mm}
\cJ_{12}^F & (1+\alpha_{1}-\alpha_{\bar{1}}+\alpha_{\bar{2}})/2 &
(1+\alpha_{1}+\alpha_{\bar{1}}-
\alpha_{\bar{2}}) \nn \\
\vspace{2mm}
\cJ_{21}^F & (1-\alpha_{1}+\alpha_{\bar{1}}-\alpha_{\bar{2}})/2 &
(-1-\alpha_{1}-\alpha_{\bar{1}}+
\alpha_{\bar{2}}) \nn \\
\vspace{2mm}
\cJ_{13}^F & (1-\alpha_{\bar{1}}+\alpha_{2})/2 & (1+\alpha_{\bar{1}}+
\alpha_{2}) \nn \\
\vspace{2mm}
\cJ_{31}^F & (1+\alpha_{\bar{1}}-\alpha_{2})/2 & (-1-\alpha_{\bar{1}}-
\alpha_{2}) \nn \\
\vspace{2mm}
\cJ_{23}^F & (1-\alpha_{1}+\alpha_{2}-\alpha_{\bar{2}})/2 &
(1-\alpha_{1}+\alpha_{2}+
\alpha_{\bar{2}}) \nn \\
\vspace{2mm}
\cJ_{32}^F & (1+\alpha_{1}-\alpha_{2}+\alpha_{\bar{2}})/2 &
(-1+\alpha_{1}-
\alpha_{2}-\alpha_{\bar{2}}) \nn \\
\vspace{2mm}
\cJ_{14}^B & (1+\alpha_{1})/2 & (1+\alpha_{1}) \nn \\
\vspace{2mm}
\cJ_{41}^B & (1-\alpha_{1})/2 & (-1-\alpha_{1}) \nn \\
\vspace{2mm}
\cJ_{15}^B & (1-\alpha_{\bar{1}}+\alpha_{2}+\alpha_{\bar{2}})/2 &
(1+\alpha_{\bar{1}}+\alpha_{2}-
\alpha_{\bar{2}}) \nn \\
\vspace{2mm}
\cJ_{51}^B & (1+\alpha_{\bar{1}}-\alpha_{2}-\alpha_{\bar{2}})/2 &
(-1-\alpha_{\bar{1}}-\alpha_{2}+
\alpha_{\bar{2}}) \nn \\
\vspace{2mm}
\cJ_{24}^B & (1+\alpha_{\bar{1}}-\alpha_{\bar{2}})/2 & (1-\alpha_{\bar{1}}+
\alpha_{\bar{2}}) \nn \\
\vspace{2mm}
\cJ_{42}^B & (1-\alpha_{\bar{1}}+\alpha_{\bar{2}})/2 & (-1+\alpha_{\bar{1}}-
\alpha_{\bar{2}}) \nn \\
\vspace{2mm}
\cJ_{25}^B & (1-\alpha_{1}+\alpha_{2})/2 & (1-\alpha_{1}+\alpha_{2}) \nn \\
\vspace{2mm}
\cJ_{52}^B & (1+\alpha_{1}-\alpha_{2})/2 & (-1+\alpha_{1}-\alpha_{2}) \nn \\
\vspace{2mm}
\cJ_{34}^B & (1+\alpha_{1}+\alpha_{\bar{1}}-\alpha_{2})/2 & (1+\alpha_{1}-
\alpha_{\bar{1}}-\alpha_{2}) \nn \\
\vspace{2mm}
\cJ_{43}^B & (1-\alpha_{1}-\alpha_{\bar{1}}+\alpha_{2})/2 & (-1-\alpha_{1}+
\alpha_{\bar{1}}+
\alpha_{2}) \nn \\
\vspace{2mm}
\cJ_{35}^B & (1+\alpha_{\bar{2}})/2 & (1-\alpha_{\bar{2}}) \nn \\
\vspace{2mm}
\cJ_{53}^B & (1-\alpha_{\bar{2}})/2 & (-1+\alpha_{\bar{2}}) \nn \\
\vspace{2mm}
\cJ_{45}^F & (1-\alpha_{1}-\alpha_{\bar{1}}+\alpha_{2}+\alpha_{\bar{2}})/2 &
(1-\alpha_{1}+
\alpha_{\bar{1}}+\alpha_{2}-\alpha_{\bar{2}}) \nn \\
\vspace{2mm}
\cJ_{54}^F & (1+\alpha_{1}+\alpha_{\bar{1}}-\alpha_{2}-\alpha_{\bar{2}})/2
& (-1+\alpha_{1}-
\alpha_{\bar{1}}-\alpha_{2}+\alpha_{\bar{2}}) \nn \\
\hline
\end{array}
\eea

\subsection{\bf $N=2$ $W_{3}$ SCA}

In order to understand the reduction scheme in the case under consideration,
we take as a first example $N=2$ $ W_{3}$
SCA \cite{LPRSW} and study how it is reproduced in our method.

The algebra $N=2$ $W_{3}$ has one extra spin $2$ bosonic
supercurrent besides the spin $1$ $N=2$ stress tensor. This counting
suggests that we should impose ten constraints which is the ``maximal''
set. The point is that requiring the constraints to be first-class
restricts a possible number of such constraints. It can be
easily checked that this requirement cannot be met
if the number of constraints exceeds ten.

For the choice
\bea
\alpha_{1}=-1, \;\; \alpha_{\bar{1}}=2, \;\; \alpha_{2}=-2,
\;\; \alpha_{\bar{2}}=1
\label{eq:w3para}
\eea
in Table 3 we give the list of ``twisted'' dimensions and $u(1)$ charges of
those supercurrents which will be subjected to the reduction constraints
and corresponding gauge fixing conditions (we use for them, respectively,
the abbreviation ``constr. scs'' and ``g.f. scs'').

\bea
\begin{array}{ccccccccccc}
\mbox{Table 3} \nn \\
\hline
 u(1)  & 1 & 0 & 0 & 1 & 1 & 1 & 0 & 0 & 0 & 0  \nn \\
\mbox{dim} & -\frac{3}{2} & -1 & -1 & -\frac{1}{2} & -\frac{1}{2} &
 -\frac{1}{2} & 0 & 0 & 0 & 0 \nn \\
\hline
\mbox{constr. scs} & \cJ_{13}^F &  \cJ_{15}^B &  \cJ_{43}^F &
 \cJ_{12}^F & \cJ_{23}^F &
\cJ_{45}^F & \cJ_{14}^B & \cJ_{25}^B & \cJ_{42}^F &  \cJ_{53}^B  \nn \\
\hline
\hline
\mbox{g.f. scs} & \cJ_{34}^B &\cJ_{52}^B & \cJ_{32}^F &
 \cJ_{24}^B &
 \cJ_{35}^B & \cJ_{54}^F & \cH_{1}^F & \cH_{2}^F & \cH_{\bar{1}}^F &
 \cH_{\bar{2}}^F  \nn \\ \hline
\mbox{dim} & 2 & 1 & \frac{3}{2} & 1 & 1 & \frac{3}{2} & \frac{1}{2}
 & \frac{1}{2} & \frac{1}{2} & \frac{1}{2} \nn \\
 u(1)  & 0 & 0 & -1 &
0 & 0 & -1 & 1 & 1 & -1 & -1  \nn \\
\hline
\end{array}
\eea

We impose the constraints on all the negative and zero dimension
supercurrents as is summarized below
\bea
\cJ_{mn}^{constr}=
\left( \begin{array}{ccccc}
\ast & 0    & 0  & 1    & 0  \\
\ast & \ast & 0  & \ast & 1 \\
\ast & \ast & \ast & \ast & \ast \\
\ast & 1 & 0 & \ast & 0 \\
\ast & \ast & 1 & \ast & \ast
\end{array} \right).
\label{eq:w3cons}
\eea
As was repeatedly mentioned above, these first-class constraints generate
gauge invariances. In the upper line of Table 3 we place the supercurrents
which are subjected to the above constraints and are basically the
generators of these invariances according to the general formula
(\ref{eq:gauge}). The lower line collects the supercurrents which are
gauged away by these invariances. For example,
$\cJ_{34}$ can be
gauged away using the gauge transformation generated by
constraint $\cJ_{13}$ ($\cJ_{52}$ by $\cJ_{15}$ and so on).
Note that four constraints of units in (\ref{eq:w3cons}) are necessary
to gauge away four dimension $1/2$ supercurrents
corresponding to Cartan elements.

As we see, only four supercurrents $\cJ_{21}, \cJ_{31}, \cJ_{41}, \cJ_{51}$
eventually survive. Substituting (\ref{eq:w3cons})
into the nonlinear constraints (\ref{eq:cons}) we
find that $\cJ_{21}, \cJ_{31}$, before fixing the gauge, are expressed
as follows
\bea
\cJ_{21} & = & k \left( \cDb-\frac{1}{k} \cH_{\bar{1}} \right) \cJ_{41}, \nn \\
\cJ_{31} & = & k \left( \cDb-\frac{1}{k} \left( \cH_{\bar{1}}+\cH_{\bar{2}}
\right) \right) \cJ_{51}- \cJ_{21} \cJ_{52}- \cJ_{41} \cJ_{54}.
\label{eq:linear}
\eea
After gauging away the unphysical degrees of
freedom in accord with Table 3, we are left with the following supercurrent
matrix $\cJ_{mn}$ in the highest weight gauge
\bea
\cJ_{mn}^{DS}=
\left( \begin{array}{ccccc}
0 & 0    & 0  & 1    & 0  \\
k \cDb \cJ_{41} & 0 & 0  & 0 & 1 \\
k \cDb \cJ_{51} & 0 & 0 & 0 & 0 \\
\cJ_{41} & 1 & 0 & 0 & 0 \\
\cJ_{51} & 0 & 1 & 0 & 0
\end{array} \right),
\label{eq:w3ds}
\eea
Thus as an output we have two independent unconstrained supercurrents with
zero $u(1)$ charges: a dimension $1$ supercurrent $\cJ_{41}$
which is nothing but the $N=2$ stress tensor and a
dimension $2$ supercurrent $\cJ_{51}$.

We will not discuss here how to construct gauge invariant supercurrents
and which SOPEs they satisfy, because all these
formulas can be reproduced via a secondary
hamiltonian reduction from $N=2$ $W_{3}^{(2)}$ SCA which will be
discussed in the following
Subsection. Anticipating the result, the relevant set of SOPEs forms the
classical $N=2$ $W_3$ SCA \cite{LPRSW}.

\subsection{\bf $N=2$ $W_{3}^{(2)}$ SCA}

Let us now describe another reduction.

We wish to understand how $N=2$ $W_{3}^{(2)}$ SCA of Ref. \cite{IKS}
can be obtained within our procedure. Recall that
this algebra is described in $N=2$ superspace by
the spin $1/2, 2$ bosonic and $1/2, 2$ fermionic constrained supercurrents
in addition to the spin 1 bosonic unconstrained $N=2$ stress tensor.
To match this superfield content, we are led to impose nine constraints on
the $N=2$ affine supercurrents. One could try to proceed by relaxing one of
the constraints (\ref{eq:w3cons}), still with the same choice
of the splitting parameters (\ref{eq:w3para}).
However, in this basis one finds no spin 2 fermionic supercurrents required
by the superfield content of $N=2$ $W_3^{(2)}$ SCA.
So we are led to choose $\alpha_i, \alpha_{\bar i}$ in another way
(once again, the basic motivation for this choice is the presence of
at least one spin $1$ supercurrent with zero $u(1)$ charge after
splitting)
\bea
\alpha_{1}=-1, \alpha_{\bar{1}}=1, \alpha_{2}=-2, \alpha_{\bar{2}}=0.
\label{eq:w32para}
\eea
It turns out that this is the right choice to produce the $N=2$ $W_3^{(2)}$
SCA precisely in the form given in \cite{IKS}, one of the surviving
supercurrents being the corresponding unconstrained
$N=2$ stress tensor. Actually, the choices \p{eq:w3para}, \p{eq:w32para}
are closely related to each other: the relevant $N=2$ stress tensors
differ by an improving term containing a spin $1/2$ fermionic
supercurrent. We will come back to this point later, while discussing
the secondary reduction of $N=2$ $W_3^{(2)}$ SCA.

Proceeding as before, we list in Table 4 the dimensions and
$u(1)$ charges of the constrained and gauge fixed supercurrents,
and in Table 5 indicate the supercurrents surviving the whole set of the
hamiltonian reduction second class constraints to be defined below (we
denote these latter supercurrents as ``surv. scs'').
\bea
\begin{array}{cccccccccc}
\mbox{Table 4} \nn \\
\hline
 u(1)  & 0 & 0 & 1 & -1 & 1 & 0 & 0 & 0 & 0   \nn \\
\mbox{dim} & -1 & -1 & -\frac{1}{2} & -\frac{1}{2} & -\frac{1}{2}
 & 0 & 0 & 0 & 0  \nn \\
\hline
\mbox{constr. scs} & \cJ_{13}^F &  \cJ_{15}^B &  \cJ_{12}^F &
\cJ_{43}^B & \cJ_{45}^F &
\cJ_{23}^F & \cJ_{14}^B & \cJ_{25}^B & \cJ_{42}^B   \nn \\
\hline
\hline
\mbox{g.f. scs} & \cJ_{34}^B &\cJ_{52}^B & \cJ_{24}^B &
\cJ_{32}^F &
 \cJ_{54}^F & \cJ_{35}^B & \cH_{1}^F & \cH_{2}^F & \cH_{\bar{1}}^F \nn \\
\hline
\mbox{dim} & \frac{3}{2} & 1 & 1 & 1 & \frac{3}{2} & \frac{1}{2} &
\frac{1}{2} & \frac{1}{2} & \frac{1}{2}  \nn \\
 u(1)  & 1 & 0 & 0 &
0 & -1 & 1 & 1 & 1 & -1   \nn \\
\hline
\end{array}
\eea
\bea
\begin{array}{cccccc||c}
\mbox{Table 5} \nn \\
\hline
\mbox{surv. scs} & \cJ_{53}^B &  \cH_{\bar{2}}^F &  \cJ_{41}^B
&  \cJ_{31}^F & \cJ_{51}^B &
\cJ_{21}^F   \nn \\ \hline
\mbox{dim} & \frac{1}{2} & \frac{1}{2} & 1 & 2 & 2 & \frac{3}{2}  \nn \\
 u(1)  & -1 & -1 & 0 & 0 & 0 & -1   \nn \\
\hline
\end{array}
\eea

In Table 5 and in similar Tables for other cases studied in this
Section we adopt the following convention: to the right from the
double vertical line we place those of the surviving
supercurrents (actually the single current $\cJ_{21}$ in the case at hand)
which are expressed through other ones by the remnants of the nonlinear
constraints  (\ref{eq:cons}) after imposing the hamiltonian
reduction constraints.
These latter supercurrents themselves (they still can be constrained,
e.g., be chiral) are placed on the left.

                   From Table 4 we conclude that there are only three
bosonic affine supercurrents with both spin and $u(1)$ charge equal to zero,
namely, $\cJ_{14}$, $\cJ_{25}$ and $\cJ_{42}$. So we can put them
equal to $1$, while all the supercurrents with negative dimensions,
as in the previous examples, equal to zero. We also equate to zero
the fermionic supercurrent $\cJ_{23}$.
Thus the constraints we impose are of the form
\bea
\cJ_{mn}^{constr}=
\left( \begin{array}{ccccc}
\ast & 0    & 0  & 1    & 0  \\
\ast & \ast & 0  & \ast & 1 \\
\ast & \ast & \ast & \ast & \ast \\
\ast & 1 & 0 & \ast & 0 \\
\ast & \ast & \ast & \ast & \ast
\end{array} \right).
\label{eq:consw32}
\eea

By plugging (\ref{eq:consw32}) into the nonlinear constraints
(\ref{eq:cons}), we can solve
one of them for $\cJ_{21}$ and express the latter in terms of $\cJ_{41}$
\bea
\cJ_{21}=k \left( \cDb -\frac{1}{k} \cH_{\bar{1}} \right) \cJ_{41}.
\label{eq:j21}
\eea
As the next step we should fixe gauges. Gauge fixing procedure
can be performed using the same arguments as in the previous
examples and we eventually arrive at the following $\cJ_{mn}^{DS}$
\bea
\cJ_{mn}^{DS}=
\left( \begin{array}{ccccc}
0 & 0    & 0  & 1    & 0  \\
k \cDb \cJ_{41} & \cH_{\bar{2}} & 0  & 0  & 1 \\
\cJ_{31} & 0 & 0 & 0 & 0 \\
\cJ_{41} & 1 & 0 & 0 & 0 \\
\cJ_{51} & 0 & \cJ_{53} & 0 & \cH_{\bar{2}}
\end{array} \right).
\label{eq:dsw32}
\eea

Now we are ready to construct five independent gauge invariant
supercurrents  by exploiting the general
procedure expounded in Section 3. After a lengthy but straightforward
computation we find
\bea
\widetilde{\cJ_{53}} & = & \cJ_{53} - k \cDb \cJ_{23} - k^3 \cDb
\cJ_{13}' +
k \cH_{\bar{2}} \cD \cJ_{43} +
  k^2 \cJ_{13} \cH_{\bar{2}}' + k^2 \cJ_{15} \cJ_{53}' -
  k \cDb \cJ_{12} \cJ_{53}  \nn \\
& &  + k^2 \cDb \cJ_{13} \cD \cH_{\bar{2}} -
  k^2 \cDb \cJ_{15} \cD \cJ_{53}  - k \cDb \cJ_{45} \cJ_{53}
  - k \cD \cH_{\bar{2}} \cJ_{43} + k^2 \cJ_{43}', \nn \\
\widetilde{\cH_{\bar{2}}} & = & \cH_{\bar{1}} + \cH_{\bar{2}} +
k \cDb \cJ_{14} + k \cDb \cJ_{25} +
k^3 \cDb \cJ_{15}' +
  k^2 \cJ_{15} \cH_{\bar{2}}' \nn \\
& &  - k \cDb \cJ_{12} \cH_{\bar{2}} -
  k^2 \cDb \cJ_{15} \cD \cH_{\bar{2}} - k \cDb \cJ_{45} \cH_{\bar{2}}, \nn \\
\widetilde{\cJ_{41}} & = & \cJ_{24} + \cJ_{52} + k \cDb \cH_{1} + 2
k \cDb \cH_{2} +
k^3 \cDb \cJ_{12}' -
  k \cD \cH_{\bar{1}} + \cH_{2} \cH_{\bar{2}}  \nn \\
& & - \cJ_{12} \cJ_{21} - \cJ_{13} \cJ_{31} +
  \cJ_{14} \cJ_{41} + \cJ_{15} \cJ_{51} + \cJ_{35} \cJ_{53} +
  k^2 \cJ_{14}' - k^2 \cJ_{42}', \nn \\
\widetilde{\cJ_{31}} & = & -2 \cJ_{31} + k \cDb \cJ_{34} - k^3 \cDb
\cJ_{35}' -
  \cJ_{12} \cH_{\bar{2}} \cJ_{31} + \cJ_{14} \cJ_{31} - k \cJ_{15}
 \cD \cH_{\bar{2}} \cJ_{31}  \nn \\
& & + \cJ_{25} \cJ_{31} + \cJ_{31} \cJ_{42} - \cJ_{32} \cJ_{41} - \cJ_{34}
\cH_{\bar{2}} -
  \cJ_{35} \cJ_{21} + \cJ_{35} \cH_{\bar{2}} \cJ_{41}  \nn \\
& & + k \cJ_{35} \cH_{\bar{2}} \cD \cH_{\bar{2}} +
  k^2 \cJ_{35} \cH_{\bar{2}}' - \cJ_{45} \cH_{\bar{2}} \cJ_{31} -
  k \cDb \cJ_{35} \cJ_{41} - k^2 \cDb \cJ_{35} \cD \cH_{\bar{2}} \nn \\
& & +   k \cDb \cJ_{45} \cJ_{31}
 - k \cD \cH_{\bar{2}} \cJ_{32} +
  k^2 \cJ_{15}' \cJ_{31} +
 k^2 \cJ_{35}' \cH_{\bar{2}} -
  k^2 \cJ_{32}', \nn \\
\widetilde{\cJ_{51}} & = & -2 \cJ_{51} - k^3 \cDb \cH_{2}' - k \cD \cJ_{21} +
  k^3 \cD \cH_{\bar{1}}' - k \cH_{\bar{1}} \cD \cJ_{41} -
  k \cH_{\bar{2}} \cD \cJ_{52} \nn \\
& & - 2 k \cH_{\bar{2}} \cD \cJ_{43} \cJ_{53}' -
  2 k \cH_{\bar{2}} \cD \cJ_{43}' \cJ_{53} + \cH_{2} \cJ_{21} -
  \cH_{2} \cH_{\bar{2}} \cJ_{41} - k \cH_{2} \cH_{\bar{2}} \cD \cH_{\bar{2}}
\nn \\
& & + k \cJ_{13} \cH_{\bar{2}} \cD \cJ_{31} -
  2 k^2 \cJ_{13} \cH_{\bar{2}}' \cJ_{53}' -
  2 k^2 \cJ_{13} \cH_{\bar{2}}'' \cJ_{53} + \cJ_{14} \cJ_{51} +
  k \cJ_{14} \cD \cJ_{21}  \nn \\
& & - k \cJ_{15} \cH_{\bar{2}} \cD \cJ_{51}
 +
  k \cJ_{15} \cD \cJ_{31} \cJ_{53} -
  2 k^2 \cJ_{15} \cJ_{53}' \cJ_{53}' -
  2 k^2 \cJ_{15} \cJ_{53}'' \cJ_{53}
\nn \\
& &  - \cJ_{23} \cJ_{31} + \cJ_{25} \cJ_{51} + \cJ_{34} \cJ_{53} -
  k \cJ_{35} \cH_{\bar{2}} \cD \cJ_{53} - \cJ_{35} \cJ_{41} \cJ_{53} -
  k \cJ_{35} \cD \cH_{\bar{2}} \cJ_{53}  \nn \\
& & - k^2 \cJ_{35} \cJ_{53}' -
  \cJ_{41}' \cJ_{52} + \cJ_{42} \cJ_{51}
- k \cDb \cH_{2} \cJ_{41} -
  k \cDb \cJ_{12}  \cJ_{51} - k^2 \cDb \cJ_{12} \cD \cJ_{21}  \nn \\
& &   + 4 k \cDb \cJ_{12} \cJ_{53}' \cJ_{53} -
  k^2 \cDb \cJ_{13} \cD \cJ_{31} -
  2 k^2 \cDb \cJ_{13} \cD \cH_{\bar{2}} \cJ_{53}' -
  2 k^2 \cDb \cJ_{13} \cD \cH_{\bar{2}}' \cJ_{53}  \nn \\
& &   - k^2 \cDb \cJ_{14} \cD \cJ_{41} - k^2 \cDb \cJ_{15} \cD \cJ_{51} +
  2 k^2 \cDb \cJ_{15} \cD \cJ_{53}' \cJ_{53} +
  2 k^2 \cDb \cJ_{15} \cJ_{53}' \cD \cJ_{53}  \nn \\
& &   + 2 k \cDb \cJ_{23} \cJ_{53}'
+
  4 k \cDb \cJ_{45} \cJ_{53}' \cJ_{53} +
  2 k \cDb \cJ_{12}' \cJ_{53} \cJ_{53} -
  2 k^2 \cDb \cJ_{13}' \cD \cH_{\bar{2}} \cJ_{53}
\nn \\
& &   + 2 k^2 \cDb \cJ_{15}' \cJ_{53} \cD \cJ_{53} +
  2 k \cDb \cJ_{23}' \cJ_{53} +
  2 k \cDb \cJ_{45}' \cJ_{53} \cJ_{53} +
  2 k^3 \cDb \cJ_{13}'' \cJ_{53}
\nn \\
& &   + 2 k \cD \cH_{\bar{2}} \cJ_{43} \cJ_{53}' +
  2 k \cD \cH_{\bar{2}} \cJ_{43}' \cJ_{53} + k \cD \cJ_{21} \cJ_{42} +
  k \cD \cJ_{31} \cJ_{43} + k \cD \cJ_{32} \cJ_{53}  \nn \\
& &   + 2 k \cD \cH_{\bar{2}}' \cJ_{43}  \cJ_{53} -
  2 k \cH_{\bar{2}}' \cD \cJ_{43} \cJ_{53} -
  2 k^2 \cH_{2}' \cH_{\bar{2}} +
  k^2 \cJ_{12}' \cJ_{21}  -
  2 k^2 \cJ_{13}' \cH_{\bar{2}}' \cJ_{53}  \nn \\
& &   - k^2 \cJ_{14}' \cJ_{41} -
  2 k^2 \cJ_{15}' \cJ_{53}' \cJ_{53} -
  2 k^2 \cJ_{35}' \cJ_{53} -
  2 k^2 \cJ_{43}' \cJ_{53}' -
  2 \cJ_{53}' \cJ_{53}  \nn \\
& &  -k^2 \cH_{2} \cH_{\bar{2}}'
 -k \cJ_{21} \cD \cJ_{42}
+  2 k^3 \cDb \cJ_{13}' \cJ_{53}'
+ k \cD \cH_{\bar{1}}  \cJ_{41}
 - 2 k^2 \cJ_{43}'' \cJ_{53} \nn \\
& & - k^2 \cJ_{24}' -
  k^2 \cJ_{52}' - k^4 \cJ_{14}''
  - \frac{k^2}{2} [ \cD, \cDb ]
\widetilde{\cJ_{41}} + 2 \widetilde{\cJ_{53}'} \widetilde{\cJ_{53}} +
\frac{k^2}{2} \widetilde{\cJ_{41}}'.
\label{eq:tilw32}
\eea
It is easy to verify that these gauge invariant supercurrents
satisfy the condition (\ref{eq:tilj}).
Note that in the last three terms of $\widetilde{\cJ_{51}}$, in order
to shorten this expression, we kept
$\widetilde{\cJ_{41}}, \widetilde{\cJ_{53}}$ in their implicit form.

Now it is direct to calculate the star SOPEs between them using the rule
(3.9) and the explicit expressions (\ref{eq:tilw32}) and (\ref{eq:qq}).
The $N=2$ stress tensor is given by
\bea
\cT=  -\frac{1}{k} \widetilde{\cJ_{41}} = -\frac{1}{k} \cJ_{41},
\label{eq:Tw32}
\eea
where the second equality is fulfilled on the shell of constraints.
It has the central charge $-2 k$ and coincides
with $\cT_{sug}$ (\ref{eq:sugsl32}), (\ref{eq:w32para})
on the constraints shell.
After the redefinitions
\bea
\cJ_{53}  \rightarrow  \cJ_{53}, \;\;
\cH_{\bar{2}}  \rightarrow  \cH_{\bar{2}}, \;\;
\cJ_{31}  \rightarrow \frac{1}{k^3} \cJ_{31}, \;\;
\cJ_{51} \rightarrow \frac{1}{k^3} \cJ_{51}\;,
\eea
all the supercurrents except for $\cH_{\bar{2}}$ are superprimary with
respect to the stress tensor (\ref{eq:Tw32}) (see eq. (\ref{eq:JJ})),
and have the spins $1/2, 1/2, 2$ and $2$, respectively, while
$\cH_{\bar{2}}$ is quasi superprimary
\bea
\cT (Z_{1}) {\cH}_{\bar{2}} (Z_{2})=
 \frac{\theta_{12}}{z_{12}^2} 2 k +
\left[ \tzzbb \frac{1}{2} -
\tz \cD + \tzb \cDb + \tzbb \partial -
 \frac{1}{z_{12}} \right] {\cH}_{\bar{2}}.
\label{eq:Th2}
\eea

Finally, the remaining set of SOPEs is as follows (from now on, we omit the
index ``*'', keeping in mind that all such SOPEs are
computed according to the rule (3.9))
\bea
\cH_{\bar{2}}(Z_{1}) \cJ_{53} (Z_{2}) & = & \tz
\cJ_{53}, \nn \\
\cH_{\bar{2}}(Z_{1}) \cJ_{31} (Z_{2}) & = & -\tz
\cJ_{31},  \nn \\
\cH_{\bar{2}}(Z_{1}) \cJ_{51} (Z_{2}) & = &
-\frac{\theta_{12}}{z_{12}^3} 2 k + \frac{\theta_{12} {\bar{\theta}}_
{12}}{
z_{12}^3}  \cH_{\bar{2}}+
   \frac{1}{z_{12}^2} \cH_{\bar{2}}  + \frac{\theta_{12}}{z_{12}^2}
 \left[  \cT -
  \cD \cH_{\bar{2}} \right]  \nn \\
& &  +
   \tzzbb \left[ \frac{1}{2} \cDb \cT  +
    \frac{1}{2 k} \cH_{\bar{2}} \cD \cH_{\bar{2}} -
    \frac{1}{2 k} \cT \cH_{\bar{2}} +
    \frac{1}{2} \cH_{\bar{2}}'\right]  \nn \\
& &  +
   \frac{1}{z_{12}} \left[ \cDb \cT + \frac{1}{k} \cH_{\bar{2}}
\cD \cH_{\bar{2}} - \frac{1}{k}
\cT \cH_{\bar{2}} +
      \cH_{\bar{2}}' \right],  \nn \\
\cJ_{53}(Z_{1}) \cJ_{31}(Z_{2}) & = &
 \frac{\theta_{12}}{z_{12}^3} 2 k  - \frac{\theta_{12} {\bar{\theta}}_
{12}}{
z_{12}^3}  \cH_{\bar{2}}
   -\frac{1}{z_{12}^2} \cH_{\bar{2}}
 + \frac{\theta_{12}}{z_{12}^2} \left[-
 \cT +
 \cD \cH_{\bar{2}} \right]  \nn \\
& & + \tzzbb \left[ -
    \frac{1}{2}  \cDb \cT -
    \frac{1}{2 k} \cH_{\bar{2}} \cD \cH_{\bar{2}} +
    \frac{1}{2 k} \cT \cH_{\bar{2}} -
    \frac{1}{2}  \cH_{\bar{2}}' \right]  \nn \\
& & + \frac{1}{z_{12}} \left[
   - \cDb \cT - \frac{1}{k}  \cH_{\bar{2}} \cD \cH_{\bar{2}} +
\frac{1}{k}  \cT \cH_{\bar{2}}-
 \cH_{\bar{2}}'\right]-\tz \cJ_{51}, \nn \\
\cJ_{53}(Z_{1}) \cJ_{51}(Z_{2}) & = &
\frac{\theta_{12} {\bar{\theta}}_{12}}{z_{12}^3}  \cJ_{53}+
  \frac{1}{z_{12}^2}  \cJ_{53} - \frac{\theta_{12}}{z_{12}^2}
\cD \cJ_{53}  \nn \\
& &  + \tzzbb \left[
    \frac{1}{2 k} \cH_{\bar{2}} \cD \cJ_{53}  -
    \frac{1}{2 k}  \cT \cJ_{53} +
    \frac{1}{2 k} \cD \cH_{\bar{2}} \cJ_{53} +
    \frac{1}{2} \cJ_{53}' \right]  \nn \\
& & + \frac{1}{z_{12}} \left[
   \frac{1}{k}  \cH_{\bar{2}} \cD \cJ_{53} - \frac{1}{k}
\cT \cJ_{53} +
\frac{1}{k} \cD \cH_{\bar{2}} \cJ_{53} +
      \cJ_{53}' \right],    \nn \\
\cJ_{31}(Z_{1}) \cJ_{51}(Z_{2}) & = &
-\frac{1}{z_{12}^2} 2  \cJ_{31}  - \tzzb \frac{1}{k} \cH_{\bar{2}}
\cJ_{31} -
    \tzz  \cD \cJ_{31}  \nn \\
& & + \tzzbb \left[- \frac{1}{k} \cT \cJ_{31} +
    \frac{3}{k} \cD \cH_{\bar{2}} \cJ_{31} +
   \frac{3}{2}  \cJ_{31}' \right]
 - \zot  \cJ_{31}'  \nn \\
& & +
   \tzb \left[ \frac{1}{k^2} \cH_{\bar{2}} \cD \cH_{\bar{2}} \cJ_{31} -
    \frac{1}{k^2} \cT \cH_{\bar{2}} \cJ_{31} +
\frac{1}{k} \cDb \cT \cJ_{31} -
    \frac{1}{k} \cH_{\bar{2}}' \cJ_{31} \right]  \nn \\
& & + \tz \left[-
     \cD \cJ_{31}' + \frac{1}{k} \cT \cD \cJ_{31} -
    \frac{1}{k} \cD \cH_{\bar{2}} \cD \cJ_{31} - \frac{1}{k}
\cD \cT \cJ_{31} \right]  \nn \\
& & +
    \tzbb \left[ -\frac{1}{k}  \cT \cJ_{31}' +
    \frac{1}{k} \cDb \cT \cD \cJ_{31} +
    \frac{2}{k} \cD \cH_{\bar{2}} \cJ_{31}' +
    \frac{1}{k^2}  \cD \cT \cH_{\bar{2}} \cJ_{31}  \right. \nn \\
& & \left.
    + \frac{2}{k} \cD \cH_{\bar{2}}' \cJ_{31} +
     \cJ_{31}'' \right],  \nn \\
\cJ_{51} (Z_{1}) \cJ_{51} (Z_{2}) & = &
-\frac{1}{z_{12}^2} 2 \cJ_{51}  - \tzzb \left[ \frac{1}{k}
 \cH_{\bar{2}} \cJ_{51} +
    \frac{1}{k} \cJ_{31} \cJ_{53} \right] - \tzz \cD
\cJ_{51}
 - \zot \cJ_{51}' \nn \\
& & + \tzzbb \left[ -
    \frac{3}{k}  \cJ_{31} \cD \cJ_{53} -
    \frac{1}{k} \cT \cJ_{51} +
    \frac{3}{k} \cD \cH_{\bar{2}} \cJ_{51} +
    \frac{3}{2}  \cJ_{51}' \right]
\nn \\
& &     +
   \tzb \left[ -\frac{1}{k^2} \cH_{\bar{2}} \cJ_{31} \cD
\cJ_{53} +
    \frac{1}{k^2} \cH_{\bar{2}} \cD \cH_{\bar{2}}
\cJ_{51} -
    \frac{1}{k} \cJ_{31} \cJ_{53}'
 \right. \nn \\
& &   - \frac{1}{k^2} \cT \cJ_{31} \cJ_{53} +
    \frac{1}{k} \cDb \cT \cJ_{51}  +
    \frac{1}{k^2} \cD \cH_{\bar{2}} \cJ_{31} \cJ_{53} -
    \frac{1}{k} \cH_{\bar{2}}' \cJ_{51}   \nn \\
& & \left. -    \frac{1}{k^2} \cT \cH_{\bar{2}} \cJ_{51} \right] \nn \\
& & + \tz \left[-
    \cD \cJ_{51}' + \frac{1}{k} \cT \cD \cJ_{51} -
    \frac{1}{k} \cD \cH_{\bar{2}} \cD \cJ_{51}
 +
    \frac{1}{k} \cD \cJ_{31} \cD \cJ_{53} \right. \nn \\
& & \left. - \frac{1}{k} \cD \cT \cJ_{51}
\right]  \nn \\
& &  +
 \tzbb \left[-
    \frac{2}{k} \cJ_{31} \cD \cJ_{53}' -
    \frac{1}{k}  \cT \cJ_{51}' +
    \frac{1}{k} \cDb \cT \cD \cJ_{51}  +
    \frac{2}{k} \cD \cH_{\bar{2}} \cJ_{51}'
      \right. \nn \\
& &
   + \frac{1}{k^2}  \cD \cT \cH_{\bar{2}} \cJ_{51} +
    \frac{1}{k^2} \cD \cT \cJ_{31} \cJ_{53} +
    \frac{2}{k} \cD \cH_{\bar{2}}' \cJ_{51} +\cJ_{51}'' \nn \\
& & \left. -
    \frac{2}{k} \cJ_{31}' \cD \cJ_{53} \right].
\eea

So we end up with the following five $N=2$ supercurrents:
a general spin $1$
$\cT$, spin $1/2$ antichiral fermionic $\cH_{\bar{2}}$ and bosonic
$\cJ_{53}$, constrained spin $2$ fermionic $\cJ_{31}$ and
bosonic $\cJ_{51}$ ones.

We also write down the constraints which stem from the original
nonlinear constraints on the affine supercurrents
\bea
\cDb \cH_{\bar{2}}  =  0, & \;\; & \cDb \cJ_{53}  =  0, \nn \\
\left( \cDb - \frac{1}{k} \cH_{\bar{2}} \right)
\cJ_{31}  =  0, & \;\; & \left( \cDb  -
\frac{1}{k} \cH_{\bar{2}} \right) \cJ_{51} -
\frac{1}{k} \cJ_{31} \cJ_{53}  =  0.
\label{eq:w32con}
\eea
By construction, all the above SOPEs are compatible with these constraints.
These SOPEs and constraints constitute the superfield description of $N=2$
$W_3^{(2)}$ superalgebra given in \cite{IKS}.

As shown in \cite{IKS}, we can obtain $N=2$ $W_{3}$ SCA from $N=2$
$W_{3}^{(2)}$ SCA by means of secondary hamiltonian
reduction \cite{DFRS} (by the primary hamiltonian reduction we
mean the one which proceeds directly from affine (super)algebra).

With respect to the new stress tensor $\cT_{\mbox{new}}$,
\bea
\cT_{\mbox{new}}= \cT+ \cD \cH_{\bar{2}}
\label{eq:w32newT}
\eea
$\cJ_{53}$ has vanishing spin and $u(1)$ charge.
Then we can put nonzero constraint on $\cJ_{53}$
\bea
\cJ_{53}-1 =0
\eea
and gauge away $\cH_{\bar{2}}$
\bea
\cH_{\bar{2}}=0.
\eea
These additional constraints are consistent with the first and
second of eqs. (\ref{eq:w32con}), respectively.
One observes that now $\cJ_{31}$ is expressed from the fourth of
eqs. (\ref{eq:w32con}) as
\bea
\cJ_{31}=  k \left( \cDb -\frac{1}{k} \cH_{
\bar{2}} \right) \cJ_{51}\;,
\eea
after which the third of eqs. (\ref{eq:w32con}) is satisfied identically.
Then we are left with the same $\cJ_{mn}^{DS}$ as in (\ref{eq:w3ds}).
Thus the surviving independent supercurrents are $\cT_{\mbox{new}}$ and $
\cJ_{51}$ and it remains to construct the appropriate gauge invariant
supercurrents and to compute their SOPEs using the rule (3.9).
The dimension of $\cJ_{51}$ and its $u(1)$ charge
with respect to $\cT_{\mbox{new}}$ are the same as in Table 5,
i.e. 2 and 0, which are characteristic of the second supercurrent of
$N=2$ $W_3$ SCA.
According to \cite{IKS}, the resulting superalgebra is precisely
the $N=2$ $W_{3}$ SCA \cite{LPRSW}. Of course, we could arrive
at the same SOPEs directly in the framework of the primary hamiltonian
reduction procedure described in the previous Subsection.

Let us remark that $\cT_{\mbox{new}}$ \p{eq:w32newT}
exactly corresponds to the previous choice of the
splitting parameters (\ref{eq:w3para}), in the sense that the dimensions
and $u(1)$ charges of all the supercurrents with respect to it
are the same as in Subsect. 5.1.
This implies that the bases (\ref{eq:w32para})
and (\ref{eq:w3para}) are related through the shift
$\sim \cD \cH_{\bar{2}}$ of the respective $N=2$ stress tensors.
We could equally derive $N=2$ $W_3^{(2)}$ SCA sticking
to the choice (\ref{eq:w3para}) and relaxing one of the constraints
of units (on $\cJ_{53}$) directly in the supermatrix (\ref{eq:w3cons}).
However, in the corresponding
basis the $N=2$ $W_3^{(2)}$ supercurrents are even not quasi superprimary.
To put this superalgebra in the standard form given in
\cite{IKS} one should pass to the stress tensor $\cT$ (\ref{eq:Tw32})
by the relation \p{eq:w32newT}.

It is worth to notice that one can produce eight reduction constraints by
relaxing of the constraint on $\cJ_{23}$ in the supermatrix (5.7). Then the
surviving supercurrents have extra two supercurrents $\cJ_{23}, \cJ_{35}$
in addition to the superfield contents of $N=2$ $W_{3}^{(2)}$ SCA.

In next Subsection we will consider more examples of extended $N=2$
conformal superalgebras obtained from $N=2$ $sl(3|2)^{(1)}$ via
$N=2$ superfield hamiltonian reduction.

\subsection{\bf New Extended $N=2$ SCAs}

           From now on we will concentrate on those examples of
hamiltonian reduction in $N=2$ superspace which generate
new extended $N=2$ SCAs.

Next natural step is to consider the cases in which the number of
the reduction constraints is less than nine. Let us first
describe the case with five constraints (this number is the minimal one
at which the constraints can still be chosen to be first-class).
As before, the reason why we choose the specific values for
splitting parameters as below stems from the demand that among
the surviving supercurrents there is at least one bosonic supercurrent
with spin $1$ and $u(1)$ charge zero.

For the choice
\bea
\alpha_{1}=-1, \;\; \alpha_{\bar{1}}=0, \;\; \alpha_{2}=0,
\;\; \alpha_{\bar{2}}=1
\eea
we list the dimensions and $u(1)$ charges of supercurrents in
Tables 6 and 7.
\bea
\begin{array}{cccccc}
\mbox{Table 6} \nn \\
\hline
 u(1)  & -1 & 0 & 2 & 0 & 0    \nn \\
\mbox{dim} & -\frac{1}{2} & 0 & 0 & 0 & 0   \nn \\
\hline
\mbox{constr. scs} & \cJ_{54}^F & \cJ_{14}^B &  \cJ_{24}^B &
  \cJ_{34}^B &
\cJ_{53}^B  \nn \\
\hline
\hline
\mbox{g.f. scs}  & \cJ_{43}^B & \cH_{1}^F &\cJ_{12}^F &
 \cJ_{13}^F &
\cH_{\bar{2}}^F  \nn \\  \hline
\mbox{dim} & 1 & \frac{1}{2} & \frac{1}{2} & \frac{1}{2} & \frac{1}{
2}  \nn \\
 u(1)  & 0 & 1 & -1 & 1 &-1   \nn \\
\hline
\end{array}
\eea

\bea
\begin{array}{cccccccccccc||ccc}
\mbox{Table 7} \nn \\
\hline
\mbox{surv. scs}&\cJ_{51}^B&\cJ_{52}^B&\cJ_{21}^F&\cJ_{23}^F&
\cH_{\bar{1}}^F&\cH_{2}^F&\cJ_{15}^B&\cJ_{35}^B&\cJ_{41}^B&\cJ_{42}^B&
\cJ_{25}^B&\cJ_{31}^F&\cJ_{32}^F&\cJ_{45}^F\nn \\ \hline
\mbox{dim} & 0 & 0 & \frac{1}{2} & \frac{1}{2} & \frac{1}{2}
 & \frac{1}{2} & 1 & 1 & 1 & 1 & 1
& \frac{1}{2} & \frac{1}{2} & \frac{3}{2}  \nn \\
 u(1)  & 0 & -2 & 1 & -1 & -1 & 1 & 0 & 0 & 0 & -2 & 2 & -1
& -3 & 1   \nn \\
\hline
\end{array}
\eea

We can choose the appropriate constraints as follows
\bea
\cJ_{mn}^{constr}=
\left( \begin{array}{ccccc}
\ast & \ast    & \ast  & 1    & \ast  \\
\ast & \ast & \ast  & 0 & \ast \\
\ast & \ast & \ast & 0 & \ast \\
\ast & \ast & \ast & \ast & \ast \\
\ast & \ast & 1 & 0 & \ast
\end{array} \right).
\label{eq:conso5}
\eea
One observes that it is not a subset of constraints discussed in
previous Subsections, (\ref{eq:w3cons}), (\ref{eq:consw32}).
Further we fix the gauge according to Table 6 and
quote the surviving supercurrents in Table 7.
There are three supercurrents expressible at expense of the
remaining ones, which can be seen by substituting \p{eq:conso5}
into the nonlinear constraints
(\ref{eq:cons})
\bea
\cJ_{31} & = & k \left( \cDb -\frac{1}{k} \left( \cH_{\bar{1}} +\cH_{\bar{2}}
\right) \right) \cJ_{51} - \cJ_{21} \cJ_{52}, \nn \\
\cJ_{32} & = & k \left( \cDb -\frac{1}{k} \cH_{\bar{2}} \right) \cJ_{52},
\nn \\
\cJ_{45} & = & k \left( \cD +\frac{1}{k} \cH_{1} \right) \cJ_{15} -
\cJ_{12} \cJ_{25}-\cJ_{13} \cJ_{35}.
\eea
Thus we come to the following $\cJ_{mn}^{DS}$
\bea
\cJ_{mn}^{DS}=
\left( \begin{array}{ccccc}
\cH_{\bar{1}} & 0    & 0  & 1    & \cJ_{15}  \\
\cJ_{21} & 0 & \cJ_{23}  & 0 & \cJ_{25} \\
k \left( \cDb -\frac{1}{k}  \cH_{\bar{1}} \right) \cJ_{51} -
\cJ_{21} \cJ_{52}  & k \cDb \cJ_{52} & \cH_{2} & 0 & \cJ_{35} \\
\cJ_{41} & \cJ_{42} & 0 & \cH_{\bar{1}} & k \cD \cJ_{15} \\
\cJ_{51} & \cJ_{52} & 1 & 0 & \cH_{2}  \\
\end{array} \right).
\label{eq:DS5}
\eea
The supercurrents $\cJ_{51}, \cJ_{52},$ and $\cJ_{15}$ remain unconstrained.

Once we know the gauge invariant supercurrents,
it is straightforward to deduce their algebra.
The consruction of these gauge invariant quantities is the crucial
(and most difficult) step of our approach.
With the above choice of five constraints, it is rather lengthy and
cumbersome to find correct ansatz for gauge invariant supercurrents
because two spin $0$ supercurrents $\cJ_{51}, \cJ_{52}$ are present
(let us remind that $\widetilde{\cJ_{\alpha \beta}}$ are some
nonlinear functionals of $\cJ_{\alpha \beta}$ and their derivatives).
As the first step we write down $\widetilde{\cJ_{\alpha
\beta}}$ as a lowest order monomial in $\cJ_{51}, \cJ_{52}$,
and check whether it
satisfies the conditions
(\ref{eq:tilj}), (\ref{eq:req}). If this is not the case, we include
next order terms in $\cJ_{51}, \cJ_{52}$, etc., until the
conditions (\ref{eq:tilj}), (\ref{eq:req}) are satisfied.

Finally we obtain the gauge invariant
supercurrents $\widetilde{
\cJ_{ \alpha \beta}}$ in the following form
\bea
\widetilde{\cJ_{51}} & = &
\cJ_{34} + 2 \cJ_{51} - k \cD \cJ_{54} + \cH_{2} \cJ_{54}+
\cJ_{23} \cJ_{52} \cJ_{54} +\cJ_{24} \cJ_{52} - \cJ_{51} \cJ_{53}, \nn \\
\widetilde{\cJ_{52}} & = &
2 \cJ_{52} - \cJ_{14} \cJ_{52}, \nn \\
\widetilde{\cJ_{21}} & = &
\cJ_{21} + k \cDb \cJ_{24} + \cJ_{14} \cJ_{21} -
\cJ_{21} \cJ_{53} + \cJ_{23} \cH_{\bar{1}} \cJ_{54} -
\cJ_{24} \cH_{\bar{1}} + k \cDb \cJ_{23} \cJ_{54}, \nn \\
\widetilde{\cJ_{23}} & = &
\cJ_{23} + \cJ_{23} \cJ_{14} - \cJ_{23} \cJ_{53},
\nn \\
\widetilde{\cH_{\bar{1}}}  & = &
\cH_{\bar{1}} + k \cDb \cJ_{14}, \nn \\
\widetilde{\cH_{2}}  & = &
\cH_{2} + k \cD \cJ_{53}, \nn \\
\widetilde{\cJ_{15}} & = &
\cJ_{43} - k \cDb \cJ_{13} + \cJ_{15} \cJ_{53}, \nn \\
\widetilde{\cJ_{35}} & = &
-k \cD \cH_{\bar{2}} + \cH_{2} \cH_{\bar{2}} - k \cH_{2}
\cDb \cJ_{14} -
  \cH_{2} \cJ_{15} \cJ_{54}  \nn \\
& &  - \cJ_{15} \cJ_{34} + k \cJ_{15} \cD \cJ_{54} +
  \cJ_{35} \cJ_{53} + \cJ_{45} \cJ_{54} - k^2 \cJ_{14}', \nn \\
\widetilde{\cJ_{41}} & = & k \cDb \cH_{1} + k \cH_{\bar{1}} \cD \cJ_{53} +
\cH_{1} \cH_{\bar{1}} - \cJ_{12} \cJ_{21} -
  \cJ_{13} \cH_{\bar{2}} \cJ_{51} - \cJ_{13} \cJ_{31} \cJ_{53}
  \nn \\
& & + \cJ_{14} \cJ_{41} + \cJ_{23} \cJ_{42} \cJ_{
54} +
  \cJ_{24} \cJ_{42} - \cJ_{43} \cJ_{51}  + k \cDb \cJ_{13} \cJ_{51} +
  k^2 \cJ_{53}', \nn \\
& &  -  \cJ_{13} \cJ_{41} \cJ_{54}  \nn \\
\widetilde{\cJ_{42}} & = &
-k \cDb \cJ_{12} - \cJ_{13} \cH_{\bar{2}} \cJ_{52} -
\cJ_{13} \cJ_{32} \cJ_{53} -
  \cJ_{13} \cJ_{42} \cJ_{54} + \cJ_{42} \cJ_{53} - \cJ_{43} \cJ_{52} \nn \\
& & +
  k \cDb \cJ_{13} \cJ_{52}, \nn \\
\widetilde{\cJ_{25}} & = &
 \cJ_{14} \cJ_{25} - \cJ_{15} \cJ_{24} + \cJ_{23}
\cH_{\bar{2}} - k \cJ_{23} \cDb \cJ_{14} +
  k \cJ_{23} \cDb \cJ_{53} - \cJ_{23} \cJ_{15} \cJ_{54}.
\eea

We also construct $N=2$ stress tensor
\bea
{\cal T} =-\frac{1}{k} \left[ \cJ_{35} +\cJ_{41} +
\cH_{2} \cH_{\bar{1}} +
 \cJ_{15} \cJ_{51} -k
\cJ_{23} \cDb \cJ_{52} + \cJ_{25}
\cJ_{52} \right]
\label{eq:T}
\eea
with central charge $2k$. The remnants of nonlinear irreducibility
constraints are given by
\bea
\cDb \cH_{\bar{1}} =   0, & \;\; & \cD \cH_{2}  =  0, \nn \\
\left( \cDb  -\frac{1}{k} \cH_{\bar{1}} \right)
\cJ_{21}  =  0, & \;\; & \left( \cD  +\frac{1}{k} \cH_{2} \right)
\cJ_{23}  = 0, \nn \\
\left( \cD - \frac{1}{k} \cH_{2} \right)
\cJ_{35}  =  0, & \;\; & \left( \cDb - \frac{1}{k} \cH_{\bar{1}} \right)
\cJ_{41}+
\frac{1}{k} \cJ_{21} \cJ_{42}  =  0, \nn \\
\cDb \cJ_{42}  =  0, & \;\; & \cD \cJ_{25} -\frac{1}{k} \cJ_{23}
\cJ_{35}  =  0.
\eea

The above gauge invariant supercurrents form some extended
$N=2$ SCA, in particular, the stress tensor \p{eq:T} generates
the standard $N=2$ SCA. Here we do not present this superalgebra
explicitly and leave its study to the future. The reason is that
it does not meet one of the criterions by which we limited from the
beginning our study in this paper. Namely, the
$N=2$ stress tensor (\ref{eq:T}) is constrained
because the linear terms in (\ref{eq:T}) $\cJ_{35}, \cJ_{41}$
are constrained. The only unconstrained bosonic supercurrent with
the spin and $u(1)$ charge appropriate for $N=2$ stress tensor,
$\cJ_{15}$, enters nonlinearly into $\cT$ (\ref{eq:T}),
so eq. (\ref{eq:T}) does not imply an invertible relation between
$\cT$ and $\cJ_{15}$.

We would like to mention that the supercurrents
\bea
\cH_{2}, \;\;\
 \cH_{\bar{1}}, \;\;\
\cJ_{35}-k \cDb \cH_{2}
\eea
are quasi superprimary, and
\bea
\cJ_{51}, \;\;
\cJ_{52}, \;\;
\cJ_{21}, \;\;
\cJ_{23}, \;\;
\cJ_{15}, \;\;
\cJ_{42}, \;\;
\cJ_{25}-\frac{k}{2} \cDb \cJ_{23}
\eea
are superprimary with respect to $\cT$ (\ref{eq:T}).

Now we turn to another choice of five constraints
which leads to an unconstrained $N=2$ stress tensor and so seems to be
more interesting. For
\bea
\alpha_{1}=-1, \alpha_{\bar{1}}=1, \alpha_{2}=0, \alpha_{\bar{2}}=0
\label{eq:new5para}
\eea
we have the spins and $u(1)$ charges as is given in Tables 8, 9.
\bea
\begin{array}{cccccc}
\mbox{Table 8} \nn \\
\hline
 u(1)  & 1 & 2 & 0 & 0 & 1  \nn \\
\mbox{dim} & -\frac{1}{2} & 0 & 0 & 0 & \frac{1}{2}  \nn \\
\hline
\mbox{constr. scs} & \cJ_{12}^F &  \cJ_{13}^F &  \cJ_{42}^B &
\cJ_{14}^B & \cJ_{43}^B \nn \\
\hline
\hline
\mbox{g.f. scs}  & \cJ_{24}^B &\cJ_{34}^B & \cH_{\bar{1}}^F
 & \cH_{1}^F &
 \cJ_{32}^F  \nn \\ \hline
\mbox{dim} & 1 & \frac{1}{2} & \frac{1}{2} & \frac{1}{2} & 0 \nn \\
 u(1)  & 0 & -1 & -1 & 1 & -2  \nn \\
\hline
\end{array}
\eea
\bea
\begin{array}{cccccccccccc||ccc}
\mbox{Table 9} \nn \\
\hline
\mbox{surv. scs} & \cJ_{15}^B &  \cJ_{52}^B &  \cJ_{53}^B
&  \cJ_{35}^B &
\cH_{2}^F & \cH_{\bar{2}}^F & \cJ_{25}^B & \cJ_{51}^B & \cJ_{23}^F &
\cJ_{41}^B &
\cJ_{31}^F & \cJ_{54}^F & \cJ_{45}^F & \cJ_{21}^F  \nn \\  \hline
\mbox{dim} & 0 & 0 & \frac{1}{2} & \frac{1}{2} & \frac{1}{2} & \frac{1}{2}
& 1 & 1 &
 1 & 1 & 1 &
\frac{1}{2} & \frac{1}{2} & \frac{3}{2}  \nn \\
 u(1) & 2 & -2 & -1 & 1 & 1 & -1 & 2 & -2 & 2 & 0 & -2 & -3
& 3 & -1   \nn \\
\hline
\end{array}
\eea

With this choice of parameters, we impose the following constraints:
\bea
\cJ_{mn}^{constr}=
\left( \begin{array}{ccccc}
\ast & 0    & 0  & 1    & \ast  \\
\ast & \ast & \ast  & \ast & \ast \\
\ast & \ast & \ast & \ast & \ast \\
\ast & 1 & 0 & \ast & \ast \\
\ast & \ast & \ast & \ast & \ast
\end{array} \right).
\label{eq:consn5}
\eea

Then, fixing gauges according to Table 8, for surviving supercurrents
we have Table 9.
In Table 9, last three supercurrents are expressed through
the remaining ones by the relations
\vspace{2cm}
\bea
\cJ_{45} & = & k \left( \cD +\frac{1}{k} \cH_{1} \right) \cJ_{15}, \nn \\
\cJ_{21} & = & k \left( \cDb -\frac{1}{k} \cH_{\bar{1}} \right) \cJ_{41},
\nn \\
\cJ_{54} & = & k \left( \cDb -\frac{1}{k} \cH_{\bar{2}} \right) \cJ_{52} -
\cJ_{32} \cJ_{53}.
\eea
The supercurrents $\cJ_{15}, \cJ_{41},$ and $\cJ_{52}$ are unconstrained.
Then one gets the following $\cJ_{mn}^{DS}$
\bea
\cJ_{mn}^{DS}=
\left( \begin{array}{ccccc}
0 & 0    & 0  & 1    & \cJ_{15}  \\
k \cDb \cJ_{41} & \cH_{\bar{2}} & \cJ_{23}  & 0 & \cJ_{25} \\
\cJ_{31} & 0 & \cH_{2} & 0 & \cJ_{35} \\
\cJ_{41} & 1 & 0 & 0 & k \cD \cJ_{15} \\
\cJ_{51} & \cJ_{52} & \cJ_{53} & k \left( \cDb -\frac{1}{k} \cH_{\bar{2}}
\right)
\cJ_{52} & \cH_{2}+\cH_{\bar{2}}
\end{array} \right).
\label{eq:dsds}
\eea

The gauge invariant supercurrents are given by the following
expressions
\bea
\widetilde{\cJ_{15}} & = &
2 \cJ_{15} - \cJ_{15} \cJ_{42} + k \cDb \cJ_{12} \cJ_{15}, \nn \\
\widetilde{\cJ_{52}} & = &
2 \cJ_{52} -\cJ_{12} \cJ_{54} - \cJ_{14}  \cJ_{52},
\nn \\
\widetilde{\cJ_{53}} & = &
\cJ_{53} - \cJ_{13} \cJ_{54} -\cJ_{13} \cH_{\bar{2}} \cJ_{52}
 - \cJ_{43} \cJ_{52} +  k \cDb \cJ_{13} \cJ_{52}, \nn \\
\widetilde{\cJ_{35}} & = & \cJ_{35} + \cH_{2} \cJ_{15} \cJ_{32} -
\cJ_{15} \cJ_{34} - k \cJ_{15} \cD \cJ_{32} -
  \cJ_{45} \cJ_{32}, \nn \\
\widetilde{\cH_{2}} & = & \cH_{2}, \nn \\
\widetilde{\cH_{\bar{2}}} & = & \cH_{\bar{1}} + \cH_{\bar{2}}, \nn \\
\widetilde{\cJ_{25}} & = &
 \cJ_{13} \cJ_{35} \cH_{\bar{2}} + \cJ_{14} \cJ_{25} -
\cJ_{15} \cJ_{24} +
  k \cJ_{15} \cD \cH_{\bar{1}} + \cJ_{23} \cJ_{15} \cJ_{32} +
\cJ_{35} \cJ_{43}  \nn \\
& & +
  \cJ_{45} \cH_{\bar{1}} - k \cDb \cJ_{13} \cJ_{35} -
k \cDb \cJ_{14} \cJ_{45} -  k^2 \cJ_{14}' \cJ_{15}, \nn \\
\widetilde{\cJ_{51}} & = &
\cH_{1} \cJ_{54} - \cH_{2} \cJ_{32}  \cJ_{53} -
\cJ_{23} \cJ_{32} \cJ_{52} +
  \cJ_{24} \cJ_{52} + \cJ_{34} \cJ_{53} + \cJ_{42} \cJ_{51}
  \nn \\
& & -
  k \cD \cH_{\bar{1}} \cJ_{52} + k \cD \cJ_{32} \cJ_{53} + k \cD \cJ_{42}
\cJ_{54} +
  k^2 \cJ_{12}' \cJ_{54} + k^2 \cJ_{14}' \cJ_{52}, \nn \\
& & - k \cDb \cJ_{12} \cJ_{51}   \nn \\
\widetilde{\cJ_{23}} & = &
k \cD \cJ_{43} + \cH_{2} \cJ_{43} - k
\cJ_{12} \cDb \cJ_{23} +
  \cJ_{12} \cJ_{23} \cH_{\bar{2}} - k \cJ_{13} \cD \cH_{\bar{2}}  \nn \\
& & + \cJ_{23} \cJ_{14} -
  k \cDb \cH_{2} \cJ_{13} + k \cDb \cJ_{12} \cJ_{23} + k^2
\cJ_{13}', \nn \\
\widetilde{\cJ_{41}} & = &
\cJ_{24} + k \cDb \cH_{1} + k^3 \cDb \cJ_{12}' -
 k \cD \cH_{\bar{1}} -
  \cJ_{12} \cJ_{21}  \nn \\
& &  - \cJ_{13} \cJ_{31} + \cJ_{14} \cJ_{41} -
\cJ_{23} \cJ_{32} +
  k^2 \cJ_{14}' - k^2 \cJ_{42}', \nn \\
\widetilde{\cJ_{31}} & = &
k \cDb \cJ_{34} + \cJ_{31} \cJ_{42} - \cJ_{32} \cJ_{41}-
\cJ_{34} \cH_{\bar{2}} -
  k \cDb \cH_{2} \cJ_{32}  \nn \\
& & - k \cDb \cJ_{12} \cJ_{31} - k \cD \cH_{\bar{2}} \cJ_{
32} -  k^2 \cJ_{32}'.
\eea

The whole set of irreducibility constraints for surviving supercurrents
is as follows
\bea
& & \cD \cH_{2}  =  0, \; \cDb \cH_{\bar{2}}  =  0, \;
\cDb \cJ_{53}  =  0, \nn \\
& & \left( \cD - \frac{1}{k} \cH_{2} \right)
 \cJ_{35}  =  0, \;
\left( \cD  + \frac{1}{k} \cH_{2} \right)
\cJ_{23}  =  0, \nn \\
& & \cD \cJ_{25} - \frac{1}{k} \cJ_{23}
\cJ_{35}  =  0, \;
\left( \cDb  - \frac{1}{k} \cH_{\bar{2}} \right)
\cJ_{31}  =  0, \nn \\
& & \left( \cDb  - \frac{1}{k} \cH_{\bar{2}} \right)
\cJ_{51}+ \frac{1}{k} \left(
\cJ_{21} \cJ_{52}+
\cJ_{31} \cJ_{53}+ \cJ_{41} \cJ_{54} \right)  =  0.
\label{eq:5cons}
\eea

The stress tensor has the following form:
\bea
\cT=-\frac{1}{k} \left[ \cH_{2} \cH_{\bar{2}}+
\cJ_{41}+
\cJ_{15} \cJ_{51}+ \cJ_{25}
\cJ_{52}+
\cJ_{35} \cJ_{53}+
k \cD \cJ_{15} ( k \cDb \cJ_{52}- \cH_{\bar{2}}
\cJ_{52} ) \right]
\label{eq:str}
\eea
and possesses the central charge $-2 k$.
On the constraints shell the Sugawara $N=2$ stress tensor coincides with
$\cT$. With respect to $\cT$, the following combinations of supercurrents
\bea
\cJ_{15}, \;\;
\cJ_{52}, \;\;
\cJ_{53}, \;\;
\cJ_{23}, \;\;
\cH_{2},  \;\;
\cH_{\bar{2}}, \;\;
\cJ_{35}, \;\;
\cJ_{31}, \;\;
\cJ_{25}-\frac{k^2}{2} ( [ \cD, \cDb ] \cJ_{15}-
 \cJ_{15}'), \nn \\
\cJ_{51}-\frac{k^2}{2} [ \cD, \cDb ] \cJ_{52}-
\frac{k^2}{2} \cJ_{52}'+k \cD \cH_{\bar{2}} \cJ_{52}+
k \cJ_{52} \cT
\label{eq:primar}
\eea
are superprimary.

It is straightforward to derive the complete set of SOPEs
between the above supercurrents $
\cJ_{\alpha \beta}$'s (\ref{eq:primar}). The $N=2$ stress tensor
(\ref{eq:str}) entering into this $N=2$ SCA is {\it unconstrained}
since the linear term $\cJ_{41}$ in
(\ref{eq:str}) is unconstrained. We do not give here the SOPEs between
the surviving supercurrents because these are
very complicated due to the presence of dimension zero supercurrents
$\cJ_{15}, \cJ_{52}$. Let us only point out
that in the present case one cannot decouple
two fields of dimension $0$ after passing to the component form of
the superalgebra.

In the next Subsection we will show that the above
unpleasant features of SCA under consideration disappear after
the appropriate secondary hamiltonian reduction of it.
The resulting SCA does not contain any spin $0$ supercurrents; all the
involved supercurrents are superprimary with respect to the corresponding
$N=2$ stress tensor. This reduction is accomplished by
adding two more constraints to the set \p{eq:consn5} and so corresponds
to imposing some seven constraints on the original supermatrix of
$N=2$ $sl(3|2)^{(1)}$ affine supercurrents.

\subsection{\bf $N=2$ $u(2|1)$ SCA}

In this Subsection we show that there exists a natural reduction
of the second of extended $N=2$ SCAs considered in the previous
Subsection, such that it yields a $N=2$ extension of the $u(2|1)$ SCA of
Ref. \cite{DTH}.

The $u(2|1)$ SCA is some graded version of the $u(3)$
KB SCA and is generated by $16$ component currents,
the number of bosonic and fermionic ones being the same.
The spins of them are greater than $1/2$. The details of this algebra
will be given later, the only point we wish to mention at once is
that there is no standard supersymmetry subalgebra in this SCA.
Anticipating our results, the $N=2$ supersymmetric extension of this SCA,
$N=2$ $u(2|1)$, contains four extra spin $1/2$ currents: two of them are
bosonic, others fermionic.
This current content immediately implies that the number of
the hamiltonian reduction constraints should be seven. One could start
directly from $N=2$ $sl(3|2)^{(1)}$ current algebra, i.e. make use of the
primary hamiltonian reduction procedure. However, it is simpler to
deduce the same results in an equivalent way, applying a secondary
reduction to the extended $N=2$ SCA described in the end of
previous Subsection.

Thus we start with the same choice of splitting parameters
(\ref{eq:new5para}) and wish to strengthen the set of constraints
\p{eq:consn5} by adding two more ones. A natural desire
is to get rid of the unwanted spin $0$ supercurrents, viz.
$\cJ_{15}$, $\cJ_{25}$ (see Table 9). It turns out that they both are
eliminated by enforcing the constraint
\bea
\cJ_{15} = 0.
\label{eq:constru21}
\eea
Then we can gauge away $\cJ_{52}$ using the gauge
transformation generated by this new constraint:
\bea
\cJ_{52} = 0.
\label{eq:fixu21}
\eea
We also note that \p{eq:constru21}, via the nonlinear
constraints (\ref{eq:cons}), automatically implies
\bea
\cJ_{45} = 0\;.
\eea
So the final supermatrix of constraints is given by
\bea
\cJ_{mn}^{constr}=
\left( \begin{array}{ccccc}
\ast & 0    & 0  & 1    & 0  \\
\ast & \ast & \ast  & \ast & \ast \\
\ast & \ast & \ast & \ast & \ast \\
\ast & 1 & 0 & \ast & 0 \\
\ast & \ast & \ast & \ast & \ast
\end{array} \right).
\label{eq:consu2}
\eea

As in previous examples, we list the constrained, gauge fixed and
surviving supercurrents in Tables 10 and 11.
\bea
\begin{array}{cccccccc}
\mbox{Table 10} \nn \\
\hline
 u(1)  & 1 & 2 & 0 & 0 & 1 & 2 & 3  \nn \\
\mbox{dim} & -\frac{1}{2} & 0 & 0 & 0 & \frac{1}{2} & 0 &
\frac{1}{2} \nn \\
\hline
\mbox{constr. scs} & \cJ_{12}^F &  \cJ_{13}^F &  \cJ_{42}^B &
\cJ_{14}^B & \cJ_{43}^B & \cJ_{15}^B & \cJ_{45}^F \nn \\
\hline
\hline
\mbox{g.f. scs}  & \cJ_{24}^B &\cJ_{34}^B &
\cH_{\bar{1}}^F & \cH_{1}^F &
 \cJ_{32}^F & \cJ_{52}^B & \cJ_{54}^F \nn \\ \hline
\mbox{dim} & 1 & \frac{1}{2} & \frac{1}{2} & \frac{1}{2} & 0 & 0 &
\frac{1}{2} \nn \\
 u(1)  & 0 & -1 & -1 & 1 & -2 & -2 & -3 \nn \\
\hline
\end{array}
\eea
\bea
\begin{array}{cccccccccc||c}
\mbox{Table 11} \nn \\
\hline
\mbox{surv. scs} & \cJ_{53}^B &  \cJ_{35}^B &
\cH_{2}^F & \cH_{\bar{2}}^F & \cJ_{25}^B & \cJ_{51}^B & \cJ_{23}^F
 & \cJ_{41}^B &
\cJ_{31}^F & \cJ_{21}^F  \nn \\  \hline
\mbox{dim} & \frac{1}{2} & \frac{1}{2} & \frac{1}{2} &
\frac{1}{2} & 1 & 1 & 1 & 1 & 1 &
 \frac{3}{2}  \nn \\
 u(1)  & -1 & 1 & 1 & -1 & 2 & -2 & 2 & 0 & -2 & -1   \nn \\
\hline
\end{array}
\eea

After substituting (\ref{eq:constru21}), (\ref{eq:fixu21})
into (\ref{eq:dsds}), the relevant $\cJ_{mn}^{DS}$ takes the following form
\bea
\cJ_{mn}^{DS}=
\left( \begin{array}{ccccc}
0 & 0    & 0  & 1    & 0  \\
k \cDb \cJ_{41} & \cH_{\bar{2}} & \cJ_{23}  & 0 & \cJ_{25} \\
\cJ_{31} & 0 & \cH_{2} & 0 & \cJ_{35} \\
\cJ_{41} & 1 & 0 & 0 & 0 \\
\cJ_{51} & 0 & \cJ_{53} & 0 & \cH_{2}+\cH_{\bar{2}}
\end{array} \right).
\eea
All the elementary supercurrents here, except for $\cJ_{41}$,
are still subjected to the constraints which are obtained
by substituting (\ref{eq:constru21}), (\ref{eq:fixu21})
into (\ref{eq:5cons}):
\bea
\cD {\cH_{2}} =  0, & \;\; &
\cDb \cH_{\bar{2}}  =  0, \nn \\
\left( \cD  - \frac{1}{k} {\cH_{2}} \right)
\cJ_{35}  =  0, & \;\; &
\cDb \cJ_{53}  =  0,  \nn \\
\left( \cD + \frac{1}{k} {\cH_{2}} \right)
{\cJ_{23}}  =  0, & \;\; &
\left( \cDb  - \frac{1}{k} \cH_{\bar{2}} \right)
\cJ_{31}  =  0, \nn \\
\cD {\cJ_{25}} - \frac{1}{k} {\cJ_{23}}
\cJ_{35}  =  0, & \;\; &
\left( \cDb  - \frac{1}{k} \cH_{\bar{2}} \right)
\cJ_{51}-\frac{1}{k} \cJ_{31} \cJ_{53}
  =  0.
\label{eq:ucons}
\eea

Using the same techniques as before, we get the following expressions for
gauge invariant supercurrents in terms of the original ones (forming the
previous SCA with five constraints)
\bea
\widetilde{\cJ_{35}} & = & \cJ_{35}, \nn \\
\widetilde{\cJ_{53}} & = & \cJ_{53}, \nn \\
\widetilde{\cH_{\bar{2}}} & = & \cH_{\bar{2}}, \nn \\
\widetilde{\cH_{2}} & = & \cH_{2}, \nn \\
\widetilde{\cJ_{25}} & = & \cJ_{25} - \frac{k^2}{2} [\cD, \cDb ]
\cJ_{15} - k \cJ_{15} \cD \cH_{\bar{2}} +
\cJ_{15}
\cJ_{35} \cJ_{53} -
  k \cD \cJ_{15} \cH_{\bar{2}} + \frac{k^2}{2}
\cJ_{15}',  \nn \\
\widetilde{\cJ_{23}} & = & \cJ_{23} + \cH_{2}
\cJ_{15}
\cJ_{53} + k \cJ_{15} \cD \cJ_{53} + k
\cD \cJ_{15} \cJ_{53},  \nn \\
\widetilde{\cJ_{41}} & = & \cJ_{41} + \cJ_{15}
\cJ_{51} +
\cJ_{25} \cJ_{52}, \nn \\
\widetilde{\cJ_{31}} & = & \cJ_{31} + k \cJ_{35} \cDb
\cJ_{
52} - \cJ_{35} \cH_{\bar{2}} \cJ_{52} +
  k \cDb \cJ_{35} \cJ_{52},  \nn \\
\widetilde{\cJ_{51}} & = & \cJ_{51} - \frac{k^2}{2} [\cD, \cDb ]
\cJ_{52} - k \cH_{\bar{2}} \cD \cJ_{52} +
\cJ_{35} \cJ_{52} \cJ_{53} -
  \cJ_{41} \cJ_{52} - \frac{k^2}{2}
\cJ_{52}'.
\eea
We would like to stress that the SOPEs between $\cJ_{kl}$
appearing in the r.h.s. of above equations can be found by
using the SOPEs of second superalgebra presented in the Subsection $5.3$.
The $N=2$ stress tensor is given by
\bea
\cT=-\frac{1}{k} \left[ \cH_{2} \cH_{\bar{2}}+
\cJ_{41}+
\cJ_{35} \cJ_{53} \right]
\label{eq:u21T}
\eea
with central charge $-2 k$. On the shell of constraints Sugawara
$N=2$ stress tensor coincides with this stress tensor and
contains linearly supercurrent $\cJ_{41}$, so $\cT$
(\ref{eq:u21T}) is unconstrained.

Then $N=2$ $u(2|1)$ SCA (the reason why we call it this way will be
soon clear) besides general spin $1$
supercurrent $\cT$, $N=2$ stress tensor, contains the following
eight constrained $N=2$ supercurrents: spin $1/2$ $\cH_{2}$,
$\cJ_{53}$, $\cJ_{35}$ and $\cH_{\bar{2}}$, spin $1$ $\cJ_{51},
\cJ_{25},$ $\cJ_{23},$ and $\cJ_{31}$. All these supercurrents are
superprimary with respect to $\cT$. After rescaling
\bea
\cJ_{25} \rightarrow \frac{1}{k} \cJ_{25}, \;\;
\cJ_{23} \rightarrow \frac{1}{k} \cJ_{23}, \;\;
\cJ_{31} \rightarrow \frac{1}{k} \cJ_{31}, \;\;
\cJ_{51} \rightarrow \frac{1}{k} \cJ_{51},
\eea
the rest of nonvanishing SOPEs are as follows
\vspace{2cm}
\bea
 {\cH_{2}} (Z_{1}) \cH_{\bar{2}} (Z_{2})  & = &
\tzzbb \frac{k}{2} -\zot k, \nn \\
 \cH_{\bar{2}} (Z_{1}) \cJ_{35} (Z_{2}) & = &
-\tz  \cJ_{35},  \nn \\
\cH_{\bar{2}} (Z_{1}) \cJ_{53} (Z_{2}) & = &
\tz \cJ_{53}, \nn \\
\cJ_{35} (Z_{1}) \cJ_{53} (Z_{2}) & = &
-\tzzbb \frac{k}{2} +\zot k + \tz {\cH_{2}} + \tzbb \cDb {\cH_{2}},
\label{eq:u11}
\eea
\bea
& & {\cH_{2}} (Z_{1})
\left\{\begin{array}{ccl}
{\cJ_{23}} (Z_{2}) & = &
\tzb {\cJ_{23}}, \nn \\
{\cJ_{25}} (Z_{2}) & = &
\tzb {\cJ_{25}}, \nn \\
 \cJ_{31} (Z_{2}) & = &
-\tzb \cJ_{31}, \nn \\
\cJ_{51} (Z_{2}) & = &
-\tzb \cJ_{51}, \nn \\
\end{array} \right. \nn \\
& & \cH_{\bar{2}} (Z_{1})
\left\{\begin{array}{ccl}
{\cJ_{23}} (Z_{2}) & = &
\tz {\cJ_{23}},  \nn \\
\cJ_{31} (Z_{2}) & = &
-\tz \cJ_{31}, \nn \\
\end{array} \right. \nn \\
& & \cJ_{35} (Z_{1})
\left\{\begin{array}{ccl}
{\cJ_{23}} (Z_{2}) & = &
-\tzb {\cJ_{25}} + \tzbb \frac{1}{k} \left[
{\cH_{2}} {\cJ_{25}}+\cJ_{35}
{\cJ_{23}} \right], \nn \\
{\cJ_{25}} (Z_{2}) & = &
\tzbb \frac{1}{k} \cJ_{35} {\cJ_{25}}, \nn \\
\cJ_{31} (Z_{2}) & =&
-\tzbb \frac{1}{k} \cJ_{35} \cJ_{31},
\nn \\
\cJ_{51} (Z_{2}) & = &
-\tzb \cJ_{31} + \tzbb
\frac{1}{k} \left[ {\cH_{2}} \cJ_{31} -
\cJ_{35} \cJ_{51} \right], \nn \\
\end{array} \right. \nn \\
& & \cJ_{53} (Z_{1})
\left\{\begin{array}{ccl}
{\cJ_{25}} (Z_{2}) & = &
\tz  {\cJ_{23}},  \nn \\
\cJ_{31} (Z_{2}) & = &
-\tz \cJ_{51}, \nn \\
\end{array} \right.
\eea
\bea
 {\cJ_{23}} (Z_{1}) {\cJ_{25}} (Z_{2}) & = &
-\tzbb \frac{1}{k} {\cJ_{23}} {\cJ_{25}},
\nn \\
{\cJ_{23}} (Z_{1}) \cJ_{31} (Z_{2}) & = &
\tzzzbb k-
   \frac{1}{z_{12}^{2}} k + \tzzb  \cH_{\bar{2}} +
\tzz  {\cH_{2}}  \nn \\
& &   +
   \tzzbb \left[-\frac{1}{2} \cT + \frac{3}{2}  \cDb {\cH_{2}} +
    \frac{1}{2}  \cD \cH_{\bar{2}} + \frac{1}{2 k}
{\cH_{2}}
\cH_{\bar{2}} -
    \frac{1}{k} \cJ_{35} \cJ_{53} \right]  \nn \\
& &    +
   \zot \left[  \cT- \cD \cH_{\bar{2}} +\frac{1}{k}
{\cH_{2}} \cH_{\bar{2}}+
\frac{2}{k} \cJ_{35}
\cJ_{53} - \cDb {\cH_{2}} \right]  \nn \\
& & + \tzb \left[  \cDb \cT-\frac{1}{k} \cH_{\bar{2}} \cT+
\frac{1}{k} \cH_{\bar{2}}
\cD \cH_{\bar{2}}-\frac{2}{k^2}
\cJ_{35} \cH_{\bar{2}} \cJ_{53}+
\frac{2}{k} \cDb {\cH_{2}}
\cH_{\bar{2}}  \right. \nn \\
& & \left.  + \frac{2}{k} \cDb \cJ_{35} \cJ_{53}+
\cH_{\bar{2}}'\right]  \nn \\
& & +
\tz \left[ -\frac{1}{k} {\cH_{2}} \cT+\frac{1}{k}
 {\cH_{2}} \cDb {\cH_{2}} +\frac{1}{k}
{\cH_{2}}
\cD \cH_{\bar{2}}-\frac{2}{k^2}
 {\cH_{2}} \cJ_{35} \cJ_{53}+
{\cH_{2}}' \right]
\nn \\
& & + \tzbb \left[
\cDb {\cH_{2}}'+\frac{1}{k} {\cH_{2}} \cDb \cT-
\frac{1}{k^2} {\cH_{2}} \cH_{\bar{2}} \cT+
\frac{1}{k^2} {\cH_{2}} \cH_{\bar{2}} \cD
\cH_{\bar{2}} \right. \nn \\
& & - \frac{2}{k^3} {\cH_{2}} \cJ_{35}
\cH_{\bar{2}} \cJ_{53}+
\frac{1}{k^2} {\cH_{2}} \cDb {\cH_{2}}
\cH_{\bar{2}}+\frac{2}{k^2} {\cH_{2}} \cDb
\cJ_{35}
\cJ_{53}+
\frac{1}{k} {\cJ_{23}} \cJ_{31}  \nn \\
& &  - \frac{1}{k} \cDb {\cH_{2}} \cT+\frac{1}{k}
 \cDb {\cH_{2}} \cDb \cH_{2}+\frac{1}{k} \cDb {\cH_{2}} \cD \cH_{\bar{2}}-
\frac{2}{k^2} \cDb {\cH_{2}} \cJ_{35}
\cJ_{53} \nn \\
& & \left. +
\frac{1}{k} \left( {\cH_{2}}
\cH_{\bar{2}} \right)' \right], \nn \\
{\cJ_{23}} (Z_{1}) \cJ_{51} (Z_{2}) & = &
- \tzzb  \cJ_{53} + \tzzbb \left[- \frac{1}{2}
\cD \cJ_{53} -
    \frac{3}{2 k} {\cH_{2}} \cJ_{53} \right]  \nn \\
& & +
 \zot \left[  \cD \cJ_{53}+\frac{1}{k}
 {\cH_{2}} \cJ_{53} \right]  \nn \\
& &  +
 \tzb \left[-\frac{1}{k} \cH_{\bar{2}} \cD \cJ_{53}+
\frac{2}{k^2} {\cH_{2}}
\cH_{\bar{2}} \cJ_{53}+
\frac{2}{k^2} \cJ_{35} \cJ_{53} \cJ_{53}+
\frac{1}{k} \cT \cJ_{53}
\right. \nn \\
& & \left. - \frac{1}{k} \cD \cH_{\bar{2}} \cJ_{53}-
\cJ_{53}' \right] - \tz \frac{1}{k}
 {\cH_{2}} \cD \cJ_{53}  \nn \\
& & + \tzbb \left[-\frac{1}{k^2} {\cH_{2}}
\cH_{\bar{2}} \cD \cJ_{53}+\frac{2}{k^3}
{\cH_{2}} \cJ_{35}
\cJ_{53} \cJ_{53}
+ \frac{1}{k^2} {\cH_{2}} \cT \cJ_{53}  \right. \nn \\
& &  - \frac{1}{k^2}
{\cH_{2}} \cDb {\cH_{2}} \cJ_{53} -
\frac{1}{k^2} {\cH_{2}} \cD \cH_{\bar{2}} \cJ_{
53}
 -\frac{1}{k} \left( {\cH_{2}} \cJ_{53} \right)'+
\frac{1}{k} {\cJ_{23}} \cJ_{51} \nn \\
& & \left. -
\frac{1}{k} \cDb {\cH_{2}} \cD
\cJ_{53} \right] \nn \\
{\cJ_{25}} (Z_{1}) \cJ_{31} (Z_{2}) & = &
-\tzz   \cJ_{35} + \tzzbb \left[ -\frac{3}{2} \cDb
\cJ_{35} +
    \frac{1}{k} \cJ_{35} \cH_{\bar{2}} \right]  \nn \\
& & +
\zot \cDb \cJ_{35}+ \tzb \frac{1}{k} \cDb
\cJ_{35} \cH_{\bar{2}}  \nn \\
& & + \tz \left[ \frac{1}{k^2} {\cH_{2}}
\cJ_{35} \cH_{\bar{2}}+\frac{1}{k}
 \cJ_{35} \cT- \frac{1}{k} \cJ_{35} \cD
\cH_{\bar{2}}+
\frac{2}{k^2} \cJ_{35} \cJ_{35}
\cJ_{53}  \right. \nn \\
& & \left. - \frac{1}{k} \cDb {\cH_{2}} \cJ_{35}-
\cJ_{35}' \right]  \nn \\
& & + \tzbb \left[-
\cDb \cJ_{35}'-\frac{1}{k^2} {\cH_{2}} \cDb
\cJ_{35} \cH_{\bar{2}}+\frac{1}{k}
\cJ_{35}
\cDb \cT-\frac{1}{k^2} \cJ_{35}
\cH_{\bar{2}} \cT  \right. \nn \\
& & + \frac{1}{k^2} \cJ_{35} \cH_{\bar{2}} \cD
\cH_{\bar{2}}-\frac{2}{k^3}
\cJ_{35} \cJ_{35}
\cH_{\bar{2}} \cJ_{53}+\frac{4}{k^2}
\cJ_{35} \cDb \cJ_{35} \cJ_{53}+
\frac{1}{k} \left( \cJ_{35} \cH_{\bar{2}} \right)'
\nn \\
& & \left.
- \frac{1}{k} \cDb {\cH_{2}} \cDb \cJ_{35}+\frac{2}{k^2}
 \cDb {\cH_{2}} \cJ_{35} \cH_{\bar{2}}+
\frac{1}{k} \cDb \cJ_{35} \cT
 -\frac{1}{k}
 \cDb \cJ_{35} \cD \cH_{\bar{2}} \right],
\nn \\
{\cJ_{25}} (Z_{1}) \cJ_{51} (Z_{2}) & = &
\tzzzbb k
-\frac{1}{z_{12}^2} k + \tzzb  \cH_{\bar{2}} \nn \\
& & +
\tzzbb \left[ -\frac{1}{2} \cT +
    \frac{1}{2} \cD \cH_{\bar{2}} - \frac{1}{2 k}
{\cH_{2}} \cH_{\bar{2}}-
    \frac{2}{k} \cJ_{35}  \cJ_{53} \right]  \nn \\
& & +
\zot \left[  \cT - \cD \cH_{\bar{2}}+\frac{1}{k}
 {\cH_{2}} \cH_{\bar{2}}+
\frac{2}{k} \cJ_{35}
\cJ_{53} \right]  \nn \\
& &  + \tzb \left[ \cDb \cT-\frac{1}{k} \cH_{\bar{2}} \cT+
\frac{1}{k} \cH_{\bar{2}}
\cD \cH_{\bar{2}}-
\frac{2}{k^2} \cJ_{35} \cH_{\bar{2}}
\cJ_{53}  \right. \nn \\
& & \left. + \frac{1}{k} \cDb {\cH_{2}} \cH_{\bar{2}}+
\frac{1}{k} \cDb \cJ_{35} \cJ_{53}+
\cH_{\bar{2}}' \right] + \nn \\
& &  \tz \left[ \frac{1}{k^2} {\cH_{2}} \cJ_{35}
\cJ_{53}
+\frac{1}{k} \cJ_{35} \cD \cJ_{53} \right]  \nn \\
& & + \tzbb \left[\frac{2}{k^3} {\cH_{2}} \cJ_{35}
\cH_{\bar{2}}
\cJ_{53}-
\frac{1}{k^2} {\cH_{2}} \cDb \cJ_{35}
\cJ_{53} + \frac{1}{k} {\cJ_{23}} \cJ_{31}
  \right. \nn \\
& & -
\frac{1}{k^2} \cJ_{35} \cH_{\bar{2}} \cD
\cJ_{53}
+\frac{2}{k^3} \cJ_{35} \cJ_{35}
\cJ_{53} \cJ_{53}+ \frac{1}{k^2}
\cJ_{35} \cT \cJ_{53} \nn \\
 & & \left. + \frac{1}{k} \cDb \cJ_{35} \cD \cJ_{53}-
\frac{1}{k}
\left( \cJ_{35} \cJ_{53} \right)'
-\frac{1}{k^2} \cJ_{35} \cD \cH_{\bar{2}} \cJ_{53}
\right], \nn \\
\eea

It can be checked that our algebra satisfies all the Jacobi identities.
The supercurrents $\cH_{2}, \cH_{\bar{2}}, \cJ_{35}, \cJ_{53}$ form $N=2$
$u(1|1)^{(1)}$ current algebra as a subalgebra (their SOPEs form a closed
set as is seen from eq. (\ref{eq:u11})).

Let us show that $N=2$ $W_{3}^{(2)}$ SCA constructed in Subsection
$5.2$ using primary hamiltonian reduction can be equally obtained
via a secondary hamiltonian reduction from $N=2$ $u(2|1)$ SCA. The
existence of such a possibility follows already from the fact that
the constraints \p{eq:consu2} form a subclass of the $N=2$ $W_3^{(2)}$
constraints \p{eq:consw32}.

With respect to the new stress tensor $\cT_{\mbox{new}}$
\bea
\cT_{\mbox{new}}= \cT - 2 \cDb \cH_{2}
\eea
the spins ($u(1)$ charges) of $\cJ_{25}$ and $\cJ_{23}$
are $0(0)$. In this new basis,
we can add two extra constraints such that
\bea
\cJ_{25}=1, \;\; \cJ_{23}=0\;,
\label{eq:cons25}
\eea
and, as usual, make use of the gauge freedom associated with
these constraints for gauging away two more supercurrents
\bea
\cH_{2}=0, \;\; \cJ_{35}=0\;.
\label{eq:cons2}
\eea
Using (\ref{eq:cons25}), (\ref{eq:cons2}) one can check
that (\ref{eq:ucons}) precisely reduces to
(\ref{eq:w32con}) and $\cJ_{mn}^{DS}$ coincides with (\ref{eq:dsw32}).
It can be easily checked that the dimensions and
$u(1)$ charges of the surviving supercurrents $\cJ_{53}, \cH_{\bar{2}},
\cJ_{51}, \cJ_{31}$ with respect to $\cT_{\mbox{new}}$
change and take the same values as in Table 5.
After finding gauge invariant supercurrents which we did not write down
explicitly, the reduced algebra becomes the algebra $N=2$ $W_{3}^{(2)}$
SCA elaborated in Subsection $5.2$.

Let us analyze in some detail the component structure of the extended
$N=2$ SCA constructed here.

After solving the constraints (\ref{eq:ucons})
for the involved supercurrents
we are left with the following set of $(10 + 10)$ currents: one Virasoro
spin $2$ stress tensor, two bosonic and four fermionic spin $3/2$ currents,
five bosonic and four fermionic spin $1$ currents, two bosonic and two
fermionic spin $1/2$ currents. For the time being we do not give the
precise relation of these currents to the components of supercurrents,
we only note that four spin $1/2$ currents appear as
the $\theta, \bar{\theta}$ independent parts of $\cJ_{35}, \cJ_{53},
\cH_{\bar{2}}, \cH_{\bar{2}}$. The Virasoro stress tensor,
pair of fermionic spin
$3/2$ currents and one bosonic spin $1$ current form $N=2$ SCA as
a subalgebra, while the remainder of currents are spread over
$N=2$ multiplets.

It is not too enlightening to present the OPEs
between these latter currents. For a better understanding
what we have obtained, it is more appropriate to pass, by means of
some nonlinear invertible
transformation, to another basis of the constituent currents
in which the $N=2$ multiplet structure becomes implicit but the
spin $1/2$ currents commute with all other ones and so can be factored out.
The possibility of such a factorization agrees with the general
statement of Ref. \cite{GS}. Below we give the explicit correspondence
between the modified currents (commuting with the spin $1/2$ ones) and
the initial supercurrents
\bea
k {J^{2}}_{1} & = & \cJ_{25} \vert, \nn \\
k {J^{1}}_{2} & = & \cJ_{51} \vert,  \nn \\
- k {J^{3}}_{1} & = & \cJ_{23} \vert,  \nn \\
k {J^{1}}_{3} & = & \cJ_{31} \vert,   \nn \\
k \left( {J^{1}}_{1}-{J^{2}}_{2}-{J^{3}}_{3} \right) & = &
\left(-k \cT-\cH_{2} \cH_{\bar{2}}-
\cJ_{35} \cJ_{53} \right) \vert, \nn \\
{J^{2}}_{3} & = & \cDb \cJ_{35} \vert,  \nn \\
-{J^{3}}_{2} & = & \left( \cD \cJ_{53}+\frac{1}{k}
\cJ_{53} \cH_{2}
\right) \vert, \nn \\
{J^{3}}_{3} & = & \left( \cD \cH_{\bar{2}}-\frac{1}{k}
\cJ_{35} \cJ_{53}
\right) \vert, \nn \\
{J^{2}}_{2}+{J^{3}}_{3} & = & \cDb \cH_{2} \vert, \nn \\
T & = & \left( -\frac{1}{2k} \left[ k [ \cD, \cDb ] \cT-
\cH_{2}' \cH_{\bar{2}}+
\cH_{2} \cH_{\bar{2}}'-
\cJ_{35}' \cJ_{53}+\cJ_{35}
\cJ_{53}' \right]
\right) \vert, \nn \\
-i \frac{k}{\sqrt{2}} G^{2} & = &
\left( \cDb \cJ_{25}+\frac{1}{k} \cJ_{25}
\cH_{\bar{2}} \right) \vert, \nn \\
-i \frac{k}{\sqrt{2}} {\bar{G}}_{2} & = &
\cD \cJ_{51} \vert,  \nn \\
-i \frac{k}{\sqrt{2}} G^{3} & = &
\left( \cDb \cJ_{23}+\frac{1}{k} \cH_{\bar{2}}
\cJ_{23}-\frac{1}{k}
\cJ_{53} \cJ_{25} \right) \vert, \nn \\
-i \frac{k}{\sqrt{2}} {\bar{G}}_{3} & = &
 \left( \cD \cJ_{31}-\frac{1}{k} \cH_{2}
\cJ_{31}+
\frac{1}{k} \cJ_{51} \cJ_{35} \right) \vert, \nn \\
-i \frac{k}{\sqrt{2}} {\bar{G}}_{1} & = &
\left( -k \cD \cT+\cH_{2} \cD \cH_{\bar{2}}-
\frac{1}{k} \cH_{2} \cJ_{35} \cJ_{53}-
\cJ_{35} \cD \cJ_{53} \right) \vert, \nn \\
-i \frac{k}{\sqrt{2}} G^{1} & = &
\left( -k \cDb \cT-\cDb \cH_{2} \cH_{\bar{2}}-
\cDb \cJ_{35} \cJ_{53}
\right) \vert,
\eea
where $|$ means the $\theta, \bar{\theta}$ independent part of corresponding
supercurrents. After decoupling spin $1/2$ currents the
quotient algebra includes the Virasoro stress tensor $T$, two
bosonic and four fermionic spin $3/2$ currents,
respectively, $G^{3}, {\bar{G}}_{3}$ and $G^{a}, {\bar{G}}_{a},
(a=1,2) $, five bosonic and four fermionic spin $1$ currents,
respectively ${J^{a}}_{b}, (a,b=1,2), {J^{3}}_{3}$
and ${J^{a}}_{3}, {J^{3}}_{a}, a=1,2 $.
Nine spin $1$ currents turn out to generate $u(2|1)$
current algebra\footnote{We give the relations of
$u(m|n)$ current algebra in Appendix C.}.
Spin $3/2$ currents transform under fundamental
and conjugate representaions of $u(2|1)$ for upper and lower positions
of indices. Their OPEs contain a quadratic nonlinearity in the
$u(2|1)$ currents. All the currents are primary with respect to $T$.

A simple inspection shows that this quotient algebra is none other than
the $Z_2\times Z_2$ graded extension of $u(2|1)$ current superalgebra,
$u(2|1)$ SCA \cite{DTH},
which is some graded version of the $u(3)$ KB SCA (the
precise correspondence comes out with the choice $k= -\kappa,
m = 2, n=1$ in the general formulas of \cite{DTH}). In contrast to
the original $N=2$ algebra with the spin $1/2$ currents added, the
quotient algebra does not contain the standard linear $N=2$ SCA
as a subalgebra; respectively, the $N=2$ multiplet structure of the
currents turns out to be lost. Thus we see that the adding of the
spin $1/2$ currents to the $u(2|1)$ SCA makes it possible to extend
it to some extended $N=2$ SCA, and this is why we call the latter
$N=2$ $u(2|1)$ SCA. The relation between this SCA
and its quotient by the spin $1/2$ currents strongly resembles, say,
the relation between linear $N=3$ SCA and nonlinear $so(3)$ KB
SCA \cite{GS}.
The essential difference consists, however, in that both
$N=2$ $u(2|1)$ SCA and its quotient are {\it nonlinear} algebras.
Nonetheless, we can say that the first algebra is still ``more linear''
compared to the second one, because passing to it linearizes two of
four nonlinear supersymmetries of $u(2|1)$ SCA.

Let us also remind that in the component version of hamiltonian reduction
of $sl(3|2)^{(1)}$, when we constrain both $sl(3)$ and $sl(2)$
blocks, $N=2$ $W_{3}$ or $N=2$  $W_{3}^{(2)}$ SCAs come out.
It is also known that we can obtain $u(3)$ KB SCA by imposing
constraints only on the $sl(2)$ block \cite{Ro,IM}. In terms of
component currents, $u(2|1)$ SCA corresponds to the
reduction when constraints are placed
only on the $sl(3)$ block of the $5\times 5$ $sl(3|2)^{(1)}$ supermatrix of
currents.

In the next Subsection we show that there exists another kind of
hamiltonian reduction of $N=2$ $sl(3|2)^{(1)}$ with the same number
$7$ of constraints. It yields some nonlinear extended $N=2$ SCA which by the
same reasoning as above can be called $N=2$ $u(3)$ SCA.

\subsection{\bf $N=2$ $u(3)$ SCA}

Using exactly the same arguments as given in previous Subsections, we can
continue our reduction procedure. We want to construct an $N=2$ extension
of $u(3)$ KB SCA which has $16$ component currents:
that is, $10$ bosonic currents and $6$ fermionic ones. The minimal way to
equalize the number of bosonic and fermionic currents is to add $4$
extra fermionic currents. This implies that the number of
the relevant reduction constraints should be again equal to $7$.

We choose
\bea
\alpha_{1}=1, \alpha_{\bar{1}}=0, \alpha_{2}=0, \alpha_{\bar{2}}=-1
\eea
and list the dimensions and $u(1)$ charges of supercurrents in Tables 12
and 13.
\bea
\begin{array}{cccccccc}
\mbox{Table 12} \nn \\
\hline
 u(1)  & 1 & 0 & -2 & 0 & 2 & -1 & 3   \nn \\
\mbox{dim} & -\frac{1}{2} & 0 & 0 & 0 & 0 & \frac{1}{2} &
\frac{1}{2}  \nn \\
 \hline
\mbox{constr. scs} & \cJ_{45}^F &  \cJ_{42}^B &  \cJ_{43}^B &
\cJ_{25}^B & \cJ_{15}^B &
\cJ_{23}^F & \cJ_{12}^F  \nn \\
\hline
\hline
\mbox{g.f. scs} & \cJ_{24}^B &\cH_{\bar{1}}^F & \cJ_{32}^F &
\cH_{2}^F &
 \cJ_{41}^B & \cJ_{35}^B & \cJ_{21}^F   \nn \\  \hline
\mbox{dim} & 1 & \frac{1}{2} & \frac{1}{2} & \frac{1}{2} & 0 & 0 &
\frac{1}{2}  \nn \\
 u(1)  & 0 & -1 & 1 &
1 & -2 & 2 & -3   \nn \\
\hline
\end{array}
\eea
\bea
\begin{array}{cccccccccc||c}
\mbox{Table 13} \nn \\
\hline
\mbox{surv. scs} & \cJ_{13}^F &  \cJ_{31}^F &  \cH_{1}^F &
\cH_{\bar{2}}^F & \cJ_{52}^B &
\cJ_{14}^B & \cJ_{34}^B & \cJ_{51}^B & \cJ_{53}^B & \cJ_{54}^F \nn \\ \hline
\mbox{dim} & \frac{1}{2} & \frac{1}{2} & \frac{1}{2} & \frac{1}{
2} & 1 & 1 & 1 & 1 & 1  & \frac{3}{2}
 \nn \\
 u(1)  & 1 & -1 & 1 & -1 & 0 & 2 & 2 & -2 & -2 & -1   \nn \\
\hline
\end {array}
\eea

We impose the following reduction constraints
\bea
\cJ_{mn}^{constr}=
\left( \begin{array}{ccccc}
\ast & 0    & \ast  & \ast    & 0  \\
\ast & \ast & 0  & \ast & 1 \\
\ast & \ast & \ast & \ast & \ast \\
\ast & 1 & 0 & \ast & 0 \\
\ast & \ast & \ast & \ast & \ast
\end{array} \right).
\label{eq:cou3}
\eea
These constraints are a subset of those we imposed in $N=2$
$W_{3}^{(2)}$ case. This implies, by the way, that
we can produce $N=2$ $W_{3}^{(2)}$ (or $N=2$ $W_{3}$) SCA by
secondary hamiltonian reduction starting with these
seven constraints and imposing two (three) more constraints.
As usual, the gauge fixing procedure goes in accord with the Table 12
and, as the result, we are left with the set of surviving currents
indicated in the Table 13.

Using the nonlinear irreducibility constraints, we may express $\cJ_{54}$
through the other supercurrents
\bea
\cJ_{54}= k \left( \cDb -\frac{1}{k} \cH_{\bar{2}} \right) \cJ_{52}-
\cJ_{32} \cJ_{53}\;,
\eea
and finally arrive at the following $\cJ_{mn}^{DS}$
\bea
\cJ_{mn}^{DS}=
\left( \begin{array}{ccccc}
0 & 0    & \cJ_{13}  & \cJ_{14}    & 0  \\
0 & \cH_{\bar{2}}+\cH_{1} & 0  & 0 & 1 \\
\cJ_{31} & 0 & 0 & \cJ_{34} & 0 \\
0 & 1 & 0 & \cH_{1} & 0 \\
\cJ_{51} & \cJ_{52} & \cJ_{53} &
 k \left( \cDb -\frac{1}{k} \cH_{\bar{2}} \right) \cJ_{52}
 & \cH_{\bar{2}}
\end{array} \right).
\eea

The remnants of the irreducibility constraints read
\vspace{3cm}
\bea
\cD \cH_{1}  =  0, & \;\; &
\cDb \cH_{\bar{2}}  =  0, \nn \\
\left( \cD  + \frac{1}{k} \cH_{1} \right)
\cJ_{13}  =  0, & \;\; &
\left( \cDb - \frac{1}{k} \cH_{\bar{2}} \right)
\cJ_{31}  =  0, \nn \\
\cDb \cJ_{53}  =  0, & \;\; &
\left( \cD  - \frac{1}{k} \cH_{1} \right)
\cJ_{34}  =  0, \nn \\
\cD \cJ_{14} - \frac{1}{k} \cJ_{13}
\cJ_{34}  =  0, & \;\; &
\left( \cDb - \frac{1}{k} \cH_{\bar{2}} \right)
\cJ_{51}-\frac{1}{k} \cJ_{31} \cJ_{53}
 =  0.
\label{eq:u3cons}
\eea

The computation of gauge invariant supercurrents is not very hard
due to the absence of dimension $0$ supercurrents among the surviving
currents. We give the results without entering into details
\bea
\widetilde{\cJ_{13}} & = &
 3 \cJ_{13} - \cJ_{13} \cJ_{25} -
\cJ_{13} \cJ_{42} - k \cJ_{13} \cDb \cJ_{45}, \nn \\
\widetilde{\cJ_{31}} & = &
 -\cJ_{31} + \cJ_{25} \cJ_{31} + \cJ_{31} \cJ_{42} +
 k \cDb \cJ_{45} \cJ_{31}, \nn \\
\widetilde{\cH_{1}} & = &
\cH_{1} + \cH_{2} + k \cD \cJ_{42} - k
\cH_{1} \cDb \cJ_{45} + k \cDb \cH_{1} \cJ_{45} -
  k^2 \cJ_{45}', \nn \\
\widetilde{\cH_{\bar{2}}} & = &
\cH_{\bar{1}} + \cH_{\bar{2}} + k \cDb \cJ_{25}, \nn \\
\widetilde{\cJ_{52}} & = &
\cJ_{24} + k \cDb \cH_{2} - k \cD \cH_{\bar{1}} +
\cH_{1} \cH_{\bar{1}} + \cH_{2} \cH_{\bar{2}} +
  \cJ_{14} \cJ_{41} + \nn \\
& & \cJ_{15} \cJ_{51} + \cJ_{25} \cJ_{52} + \cJ_{34}
\cJ_{43} +
  \cJ_{35} \cJ_{53}  + \cJ_{45} \cJ_{54}, \nn \\
\widetilde{\cJ_{14}} & = &
\cJ_{14} - k \cDb \cJ_{12} - k \cH_{1} \cDb \cJ_{
15} - \cJ_{13} \cJ_{32} -
  \cJ_{13} \cJ_{15} \cJ_{31} + \nn \\
& &  \cJ_{13} \cJ_{35} \cH_{\bar{2}} -
  k \cJ_{14} \cDb \cJ_{45} + k \cDb \cH_{1} \cJ_{15} - k
\cDb \cJ_{13} \cJ_{35} -
  k \cDb \cJ_{14} \cJ_{45} - k^2 \cJ_{15}', \nn \\
\widetilde{\cJ_{34}} & = &
-\cJ_{34} + k \cD \cJ_{32} - \cH_{1} \cJ_{32} -
k \cH_{1} \cDb \cJ_{35} -
  2 \cH_{1} \cJ_{15} \cJ_{31} + \nn \\
& & \cH_{1} \cJ_{35} \cH_{\bar{2}} +
\cJ_{12} \cJ_{31} +
  \cJ_{13} \cJ_{35} \cJ_{31} + \cJ_{14} \cJ_{45} \cJ_{31} +
  k \cJ_{15} \cD \cJ_{31} + \nn \\
& & \cJ_{25} \cJ_{34} +\cJ_{34} \cJ_{42} -
  \cJ_{34} \cJ_{45} \cH_{\bar{2}} - k \cJ_{35}  \cD  \cH_{\bar{2}} - k
\cDb \cJ_{34} \cJ_{45} -
  k^2 \cJ_{35}', \nn \\
\widetilde{\cJ_{51}} & = &
-\cJ_{51} + k \cD \cJ_{21} -\cH_{1} \cJ_{21} +
\cH_{1} \cH_{\bar{2}} \cJ_{41} -
  \cH_{1} \cJ_{31} \cJ_{43} + \nn \\
& & k \cH_{\bar{2}}  \cD \cJ_{41} -
\cJ_{13} \cJ_{31} \cJ_{41} -
  \cJ_{23} \cJ_{31} + \cJ_{25} \cJ_{51} - \cJ_{41} \cJ_{52} + \cJ_{42}
\cJ_{51} + \nn \\
& &
  k \cDb \cH_{1} \cJ_{41} + k \cDb \cJ_{45} \cJ_{51} + k \cD \cJ_{31}
\cJ_{43} +
  k^2 \cJ_{41}', \nn \\
\widetilde{\cJ_{53}} & = &
\cJ_{53} - k \cDb \cJ_{23} + k \cH_{\bar{2}} \cD \cJ_{
43} + \cJ_{13} \cJ_{21} +
  \cJ_{13} \cJ_{31} \cJ_{43} - \nn \\
& &  \cJ_{43} \cJ_{52} - k \cDb \cJ_{13}
\cJ_{41} -
  k \cD \cH_{\bar{2}} \cJ_{43} + k^2 \cJ_{43}',
\eea

The unconstrained $N=2$ stress tensor is given by
\bea
\cT=\frac{1}{k} \cJ_{13} \cJ_{31}-
\frac{1}{k}  \cJ_{52}+\cDb \cH_{1}-
\cD \cH_{\bar{2}}
\eea
with central charge $2 k$. All the supercurrents are superprimary with
respect to $\cT$.

After rescaling
\bea
\cJ_{14} \rightarrow \frac{1}{k} \cJ_{14}, \;\;
\cJ_{34} \rightarrow \frac{1}{k} \cJ_{34}, \;\;
\cJ_{51} \rightarrow \frac{1}{k} \cJ_{51}, \;\;
\cJ_{53} \rightarrow \frac{1}{k} \cJ_{53},
\eea
we can write down the remaining SOPEs in the following form
\vspace{2cm}
\bea
\cH_{1}(Z_{1}) \cH_{\bar{2}}(Z_{2}) & = &
\tzzbb \frac{k}{2}-
\frac{1}{z_{12}} k, \nn \\
\cH_{1}(Z_{1}) \cJ_{13}(Z_{2}) & = & \tzb
\cJ_{13}, \nn \\
\cH_{1}(Z_{1}) \cJ_{31}(Z_{2}) & = & -\tzb \cJ_{31}, \nn \\
\cH_{\bar{2}}(Z_{1})
\cJ_{13} (Z_{2}) & = &
\tz \cJ_{13}, \nn \\
\cH_{\bar{2}} (Z_{1}) \cJ_{31}(Z_{2}) & = & -\tz
\cJ_{31}, \nn \\
\cJ_{13} (Z_{1}) \cJ_{31}(Z_{2}) & = & -
\tzzbb \frac{k}{2} +
\frac{1}{z_{12}} k-
\tzb \cH_{\bar{2}}-\tz \cH_{1}+ \nn \\
& & \tzbb \left[ -\cDb \cH_{1}-\frac{1}{k}
\cH_{1} {\cH_{\bar{2}}} + \frac{1}{k}
\cJ_{13} \cJ_{31} \right],
\label{eq:u2}
\eea
\bea
& & \cH_{1} (Z_{1})
\left\{ \begin{array}{ccl}
\cJ_{14} (Z_{2}) & = & \tzb \cJ_{14}, \nn \\
\cJ_{51} (Z_{2}) & = & -\tzb \cJ_{51}, \nn \\
\end{array} \right. \nn \\
& & \cH_{\bar{2}} (Z_{1})
\left\{\begin{array}{ccl}
\cJ_{34} (Z_{2}) & = &
-\tz \cJ_{34},  \nn \\
 \cJ_{53} (Z_{2}) & = &
\tz \cJ_{53}, \nn \\
\end{array} \right. \nn \\
& & \cJ_{13} (Z_{1})
\left\{\begin{array}{ccl}
\cJ_{14} (Z_{2}) & = &
-\tzbb \frac{1}{k} \cJ_{13} \cJ_{14}, \nn \\
\cJ_{34} (Z_{2}) & = &
-\tzb \cJ_{14} - \tzbb \frac{1}{k} \cH_{1}
\cJ_{14}, \nn \\
\cJ_{51} (Z_{2}) & = &
\tzb \cJ_{53} + \tzbb \frac{1}{k} \left[ \cH_{1}
\cJ_{53} +
    \cJ_{13} \cJ_{51} \right], \nn \\
\end{array} \right. \nn \\
& & \cJ_{31} (Z_{1})
\left\{\begin{array}{ccl}
\cJ_{14} (Z_{2}) & = &
\tz \cJ_{34} + \tzbb  \frac{1}{k}
 \cJ_{34} \cH_{\bar{2}},  \nn \\
\cJ_{34} (Z_{2}) & = &
\tzbb \frac{1}{k} \cJ_{31} \cJ_{34}, \nn \\
 \cJ_{53} (Z_{2}) & = &
-\tz \cJ_{51} - \tzbb \frac{1}{k} \left[
\cH_{\bar{2}}
\cJ_{51}+
\cJ_{31} \cJ_{53} \right], \nn \\
\end{array} \right.
\eea
\bea
\cJ_{14} (Z_{1}) \cJ_{34} (Z_{2}) & = &
-\tzbb \frac{1}{k} \cJ_{14} \cJ_{34}, \nn \\
\cJ_{14} (Z_{1}) \cJ_{51} (Z_{2}) & = &
-\tzzzbb k+
   \frac{1}{z_{12}^2} k - \tzzb  \cH_{\bar{2}}  \nn \\
& & + \tzzbb  \left[
-\frac{1}{2} \cT -    \frac{1}{2} \cD \cH_{\bar{2}} -
    \frac{1}{2 k} \cH_{1} \cH_{\bar{2}} +
    \frac{2}{k} \cJ_{13} \cJ_{31} \right]  \nn \\
& & +
   \zot \left[ \cT +  \cD \cH_{\bar{2}} + \frac{1}{k}
\cH_{1}
\cH_{\bar{2}} - \frac{2}{k} \cJ_{13}
 \cJ_{31} \right]  \nn \\
& &  +
    \tzb \left[ \cDb \cT - \frac{1}{k} \cH_{\bar{2}} \cT -
    \frac{1}{k} \cH_{\bar{2}} \cD \cH_{\bar{2}} -
\frac{1}{k^2} \cJ_{13}
\cH_{\bar{2}} \cJ_{31} \right. \nn \\
& & \left.
   + \frac{1}{k} \cDb \cH_{1} \cH_{\bar{2}} -
\frac{1}{k} \cDb \cJ_{13}
{\bar{
\cJ_{13}}} - \cH_{\bar{2}}' \right]
- \tz \frac{1}{k} \cD \left( \cJ_{13}
\cJ_{31} \right)  \nn \\
& & + \tzbb \left[
    -\frac{1}{k^3} \cH_{1} \cJ_{13}
\cH_{\bar{2}}
\cJ_{31} -
    \frac{1}{k^2} \cH_{1} \cDb \cJ_{13}
\cJ_{31} +
    \frac{1}{k^2} \cJ_{13} \cJ_{31}
\cT  \right. \nn \\
& &
    \left. + \frac{1}{k^2} \cJ_{13} \cD \cH_{\bar{2}}
 \cJ_{31} +
    \frac{1}{k} \left( \cJ_{13} \cJ_{31} \right)' +
    \frac{1}{k} \cJ_{34} \cJ_{53} +
    \frac{1}{k} \cDb \cJ_{13} \cD \cJ_{31}
\right], \nn \\
\cJ_{14} (Z_{1}) \cJ_{53} (Z_{2}) & = &
\tzz  \cJ_{13} + \tzzbb \left[ \frac{3}{2} \cDb
\cJ_{13} -
    \frac{1}{2 k} \cJ_{13} \cH_{\bar{2}}
\right]  \nn \\
& & + \zot
   \left[- \cDb \cJ_{13} + \frac{1}{k} \cJ_{13}
\cH_{\bar{2}}
\right] + \tzb \frac{1}{k}
\cDb \cJ_{13} \cH_{\bar{2}}  \nn \\
& & + \tz \left[
    -\frac{1}{k^2} \cH_{1} \cJ_{13}
\cH_{\bar{2}} + \frac{1}{k}
 \cJ_{13}  \cT -
    \frac{1}{k} \cDb \cH_{1} \cJ_{13} +
\cJ_{13}' \right]  \nn \\
& &  +
    \tzbb \left[ \cDb \cJ_{13}'
     -\frac{1}{k} \cDb \cH_{1} \cDb \cJ_{13}  -
    \frac{1}{k^2} \cDb \left( \cH_{1} \cJ_{13}
\cH_{\bar{2}} \right) +
    \frac{1}{k} \cDb \left( \cJ_{13} \cT \right) \right],  \nn \\
\cJ_{34} (Z_{1}) \cJ_{51} (Z_{2}) & = &
-\tzzb  \cJ_{31} + \tzzbb \left[ -
\frac{1}{2} \cD \cJ_{31} +
    \frac{3}{2 k}  \cH_{1} \cJ_{31} \right]  \nn \\
& & +
   \zot \left[  \cD \cJ_{31} - \frac{1}{k} \cH_{1}
\cJ_{31} \right]
+ \tz \frac{1}{k} \cH_{1}
\cD \cJ_{31}  \nn \\
& &  +
\tzb \left[- \frac{1}{k^2} \cH_{1} \cH_{\bar{2}}
\cJ_{31} -
    \frac{1}{k} \cJ_{31} \cT -
 \frac{1}{k} \cD
\cH_{\bar{2}}
\cJ_{31} -
    \cJ_{31}'  \right]  \nn \\
& &  +
    \tzbb \left[  \frac{1}{k^2} \cH_{1}
\cJ_{31} \cT -
     \frac{1}{k^2} \cH_{1} \cDb \cH_{1}
\cJ_{31} +
     \frac{1}{k^2} \cH_{1} \cD \cH_{\bar{2}}
\cJ_{31}  \right. \nn \\
& & \left.
   + \frac{1}{k} \left( \cH_{1} \cJ_{31} \right)'
 -
     \frac{1}{k} \cJ_{34} \cJ_{51} +
    \frac{1}{k}  \cDb \cH_{1} \cD \cJ_{31} \right],     \nn \\
\cJ_{34} (Z_{1}) \cJ_{53} (Z_{2}) & = &
-\tzzzbb k +
   \frac{1}{z_{12}^2} k + \tzz  \cH_{1} \nn \\
& & + \tzzbb \left[
-\frac{1}{2} \cT +
    \frac{3}{2} \cDb \cH_{1} -
    \frac{1}{2 k} \cH_{1} \cH_{\bar{2}}
\right]  \nn \\
& & +
   \zot \left[ \cT - \cDb \cH_{1} + \frac{1}{k}
\cH_{1}
\cH_{\bar{2}} \right] +
\tzb \left[   \cDb \cT +
    \frac{1}{k} \cDb \cH_{1} \cH_{\bar{2}}
\right]  \nn \\
& &   + \tz \left[ \frac{1}{k} \cH_{1} \cT -
    \frac{1}{k} \cH_{1} \cDb \cH_{1} +
\cH_{1}' \right]  \nn \\
& &   +
    \tzbb \left[  \cDb \cH_{1}' -
    \frac{1}{k} \cH_{1} \cDb \cT +
    \frac{1}{k} \cDb \cH_{1} \cT -
    \frac{1}{k} \cDb \cH_{1} \cDb \cH_{1} \right].
\eea

Let us summarize the $N=2$ $u(3)$ SCA.
It contains unconstrained spin $1$ $N=2$ stress tensor
$\cT$, the spin $1/2$ chiral and anti-chiral supercurernts
$\cH_{1}$ and $
\cH_{\bar{2}}$, the spin $1/2$ supercurrents $\cJ_{
13}$ and $\cJ_{31}$ subjected to the nonlinear chirality constraints,
the spin $1$ anti-chiral supercurrent $\cJ_{53}$
and the spin $1$ constrained supercurrents $\cJ_{14}, \cJ_{51},
\cJ_{34}$. All these supercurrents are bosonic (fermionic) for integer
(half-integer) spin.
The supercurrents $\cH_{1},\cH_{\bar{2}}, \cJ_{13}, \cJ_{31}$ possess
a closed set of SOPEs (see eqs. (\ref{eq:u2})) and form $N=2$ $u(2)=u(2|0)$
current subalgebra.

We would like to note that in \cite{Ra} an $N=1$ superfield
extension of $u(3)$ KB SCA has been found.
The field content of both $N=1$ $u(3)$ SCA of Ref. \cite{Ra} and our
superalgebra is the same (modulo different choices of the basis for the
constituent currents), but the novel point is that
we have succeeded in arranging the relevant currents into $N=2$
supermultiplets
(by putting them into properly constrained $N=2$ supercurrents)
and thereby revealed $N=2$ supersymmetry of this superlagebra which was
hidden in the formulation of Ref. \cite{Ra}.

Let us now consider a secondary Hamiltonian reduction of $N=2$ $u(3)$
SCA to $N=2$ $W_{3}^{(2)}$ SCA. It goes as follows.
With respect to the new stress tensor $\cT_{\mbox{new}}$
\bea
\cT_{\mbox{new}}=\cT - 2 \cDb \cH_{1}- \cD
\cH_{\bar{2}}
\eea
the supercurrent $\cJ_{14}$ has zero spin and $u(1)$ charge,
while the spin and $u(1)$ charge of $\cJ_{13}$ are equal, respectively,
to $-1/2$ and $-1$. Thus we can impose two first-class constraints
\bea
\cJ_{14}=1, \;\; \cJ_{13}=0.
\label{eq:u314}
\eea
Gauge fixing procedure for either constraints can be done as usual.
So we fix the gauge by
\bea
\cH_{1}=0, \;\; \cJ_{34}=0 \;.
\label{eq:u3h1}
\eea
Using (\ref{eq:u314}), (\ref{eq:u3h1}) we see that (\ref{eq:u3cons})
is reduced to (\ref{eq:w32con}).
The dimensions and $u(1)$ charges of the surviving
supercurrents $\cJ_{31}, \cH_{\bar{2}},
\cJ_{51}, \cJ_{53}$ with respect to $\cT_{\mbox{new}}$ coincide with
those in Table.5.

Let us come back to discussion of $N=2$ $u(3)$ SCA. A simple
inspection of its current content shows that there are four
spin $1/2$ currents in it besides the
set of $16$ currents with higher spins. Like in the case
of $N=2$ $u(2|1)$ SCA, they can be factored out by passing to
a new basis where they (anti)commute with the remainder of the currents.
After decoupling of these spin $1/2$ currents our $N=2$ $u(3)$ SCA
reproduces $u(3)$ KB SCA \cite{K,Ber}.

Let us remind the current content of $u(3)$ KB SCA.
It is generated by $16$ currents:  Virasoro stress tensor $
T_{KB}$,  six spin $3/2$ currents $G^{a}_{KB}$ and
${\bar{G}}_{a\;KB}$,
and nine spin $1$ currents forming the $u(3)$ affine current algebra,
namely,  $u(1)$ current $H_{KB}$ and eight $su(3)$ currents
${J^{a}}_{b\;KB}$ with zero trace $ ( {J^{a}}_{a\;KB} =0 ) $.
Indices $a, b$ are running from $1$ to $3$ and
correspond to the fundamental $3$ and its conjugate $\bar{3}$
representations of $su(3)$ (for upper and lower positions, respectively).

Below we give the precise correspondence
between these $u(3)$ KB SCA currents and components of the original set of
$N=2$ $u(3)$ SCA supercurrents
\bea
{J^{3}}_{2,KB}
 & = & \left( \cDb \cJ_{13}-\frac{1}{k} \cJ_{13}
\cH_{\bar{2}} \right) \vert, \nn \\
-{J^{2}}_{3,KB}
& = & \left( \cD \cJ_{31}-\frac{1}{k} \cH_{1}
\cJ_{31} \right) \vert, \nn \\
-\frac{1}{3} H_{KB}-{J^{2}}_{2,KB} & = &
\left( \cDb \cH_{1}-\frac{1}{k} \cJ_{13}
\cJ_{31} \right) \vert, \nn \\
\frac{1}{3} H_{KB}-{J^{1}}_{1,KB}-{J^{2}}_{2,KB} & = &
\left( \cD \cH_{\bar{2}} -\frac{1}{k} \cJ_{13}
\cJ_{31} \right) \vert, \nn \\
-k {J^{1}}_{2,KB} & = & \cJ_{14} \vert, \nn \\
\frac{k}{\sqrt{2}} {\bar{G}}_{2,KB} & = &
\left( \cDb \cJ_{14}+\frac{1}{k} \cH_{\bar{2}}
\cJ_{14} \right) \vert, \nn \\
-k {J^{1}}_{3,KB} & = & \cJ_{34} \vert, \nn \\
\frac{k}{\sqrt{2}} {\bar{G}}_{3,KB} & = &
\left( \cDb \cJ_{34} +\frac{1}{k} \cJ_{31}
\cJ_{14} \right) \vert, \nn \\
-k {J^{2}}_{1,KB} & = & \cJ_{51} \vert, \nn \\
\frac{k}{\sqrt{2}} G^{2}_{KB} & = &
\cD \cJ_{51} \vert, \nn \\
-\frac{k}{3} {H}_{KB}-k {J^{1}}_{1,KB} & = &
\left( -k \cT+\cJ_{13} \cJ_{31}+k \cDb
\cH_{1}-k \cD \cH_{\bar{2}}-
\cH_{1} \cH_{\bar{2}} \right) \vert, \nn \\
\frac{k}{\sqrt{2}} G^{1}_{KB} & = &
\left( -k \cD \cT-\frac{1}{k} \cH_{1} \cJ_{13}
\cJ_{31}-\cJ_{13} \cD
\cJ_{31}+\cH_{1} \cD \cH_{\bar{2}}
\right) \vert, \nn \\
\frac{k}{\sqrt{2}} {\bar{G}}_{1,KB} & = &
\left( -k \cDb \cT+\cDb \cJ_{13} \cJ_{31}-
\frac{1}{k} \cJ_{13} \cH_{\bar{2}}
\cJ_{31}-\cDb \cH_{1} \cH_{\bar{2}}
\right) \vert, \nn \\
T_{KB} & = &
\left( -\frac{1}{2k} \left[ k [ \cD, \cDb ] \cT+ \cH_{1}
\cH_{\bar{2}}'-
\cH_{1}' \cH_{\bar{2}}+ \cJ_{13}'
\cJ_{31}-
\cJ_{13} \cJ_{31}' \right] \right) \vert, \nn \\
-k {J^{3}}_{1,KB} & = & \cJ_{53} \vert, \nn \\
\frac{k}{\sqrt{2}} {G^{3}}_{KB} & = &
\left( \cD \cJ_{53}+\frac{1}{k} \cH_{1}
\cJ_{53}+\frac{1}{k}
\cJ_{13} \cJ_{51} \right)
\vert.
\eea
The OPEs of these currents are a particular case of OPEs of
$u(m|n)$ SCA given in Appendix C, eqs. (\ref{eq:comp}),
with the following correspondence
\bea
k=\kappa, \;\; T_{KB}=T, \;\; H_{KB}={J^{a}}_{a}, \nn \\
{J^{a}}_{b, KB}={J^{a}}_{b}-\frac{1}{3} \delta^{a}_{b} {J^{c}}_{c}, \;\;
G^{a}_{KB}=i G^{a}, \;\; \bar{G}_{a,KB}=i {\bar{G}}_{a},
\eea
and $m=3, n=0$.

It is worth to notice that $G^{1}_{KB}, \bar{G}_{1,KB}$ are
related to the two fermionic components of linear $N=2$
superconformal stress tensor, $\cT$, through nonlinear transformations.
So, two of six supersymmetries of $u(3)$ KB SCA are linearized
by passing to $N=2$ $u(3)$ SCA(viz., by adding four spin $1/2$ fermionic
currents), but four of them remain nonlinear.

\section{\bf Conclusion and outlook}
\setcounter{equation}{0}

In this paper we constructed $N=2$ $sl(n|n-1)^{(1)}$ current superalgebras
and developed a general scheme of classical
hamiltonian reduction in $N=2$ superspace. We applied it to
$N=2$ extension of affine superalgebra $sl(3|2)^{(1)}$.
As the main result, we deduced some new extensions
of $N=2$ SCA, $N=2$ $u(2|1)$ and $N=2$ $u(3)$ SCAs.
Within our scheme, these two new algebras turn out to be more
fundamental than the previously explored
$N=2$ $W_{3}^{(2)},$ $N=2$ $ W_{3}$ SCAs in the sense that the
latter can be generated by secondary hamiltonian reductions from the former.
The following diagram depicts basic points of our reduction procedure.

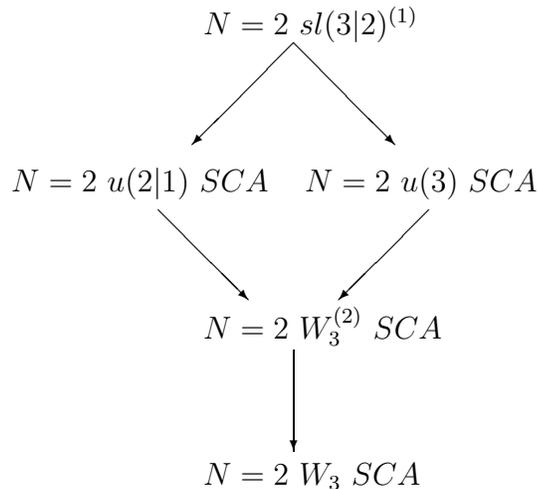
\begin{picture}(500,200)(-200,-170)
\setlength{\unitlength}{0.3mm}
\put(0,0){$N=2$ $sl(3|2)^{(1)}$}
\put(40,-4){\vector(-1,-1){45}}
\put(40,-4){\vector(1,-1){45}}
\put(-85,-70) {$N=2$ $u(2|1)$ $SCA$}
\put(45,-70) {$N=2$ $u(3)$ $SCA$}
\put(0,-135) {$N=2$ $W_{3}^{(2)}$ $SCA$}
\put(-20,-78) {\vector(1,-1){40}}
\put(100,-78) {\vector(-1,-1){40}}
\put(0,-200) {$N=2$ $W_{3}$ $SCA$}
\put(40,-140) {\vector(0,-1){45}}
\end{picture}
There are several problems to be worked out and questions
which at present are open.

Quantizing $W$ algebras associated with arbitray
embeddings of $sl(2)$ into (super)algebras has been studied
in \cite{ST}. These results were extended to $N=1$ affine Lie
superalgebras in superspace formalism \cite{MR}.
It is interesting to see whether the quantization of our superconformal
algebras can be carried out in $N=2$ superfield formalism.

It would be also interesting to study how $N=2$ $W_{4}$
\cite{YW}, and $N=2$ extensions (yet to be constructed)
of some other reductions of $sl(4)$ could come out in the
framework of hamiltonian reduction applied to $N=2$
$sl(4|3)^{(1)}$ superalgebra.

There exist some other superalgebras
which have completely fermionic simple root
system and admit $osp(1|2)$ principal embedding: $osp(2n \pm 1|2n),
osp(2n|2n), osp(2n+2|2n)$ $n \geq 1$ and $D(2,1; \alpha)$
$ \alpha \neq0,-1$ \cite{LSS}.
It is natural to apply our general procedure to these
superalgebras and see whether they admit $N=2$
superfield extensions.

It is also rather straightforward to construct free superfield
realizations for $N=2$ $u(2|1)$ and $N=2$ $u(3)$ SCAs. An interesting
related problem is to understand how these latter algebras reappear
in the $N=2$ superfield Toda and WZNW setting \footnote{For $N=2$ $W_n$
this is discussed in \cite{DM}.}.

It is rather exciting task to extend the techniques developed here
to the $N=4$ case, and, as a first step, to regain ``small''
$N=4$ SCA within the hamiltonian reduction framework in a
manifestly supersymmetric $N=4$ superfield fashion.

\vspace{1cm}

{\Large \bf Acknowledgments}
\vspace{0.5cm}

We would like to thank M. Magro, E. Ragoucy, A. Semikhatov,
P. Sorba, F. Toppan and, especially, F. Delduc and
S. Krivonos for many useful
and clarifying discussions.
We are grateful to V. Ogievetsky for his interest in this work.
Two of us (E.I. \& A.S.) acknowledge a partial support from the
Russian Foundation of Fundamental Research, grant 93-02-03821,
and the International Science Foundation, grant M9T300.

\vspace{1cm}

{\Large \bf Appendix A: Notations for $sl(2|1)$ superalgebra}
\setcounter{equation}{0}
\def\theequation{A.\arabic{equation}}
\vspace{0.5cm}

The generators of $sl(2|1)$ superalgebra in the complex basis introduced
in Section $2$ for the fundamental representation are given by
\bea
t_{1}=
\left( \begin{array}{ccc}
0 & 1 & 0   \\
0 & 0 & 0  \\
0  & 0 & 0   \\
\end{array} \right) \;,
t_{2}=
\left( \begin{array}{ccc}
0 & 0 & 0   \\
0 & 1 & 0  \\
0  & 0 & 1   \\
\end{array} \right) \;,
t_{3}=
\left( \begin{array}{ccc}
0 & 0 & 1  \\
0 & 0 & 0  \\
0  & 0 & 0   \\
\end{array} \right) \;,
t_{4}=
\left( \begin{array}{ccc}
0 & 0 & 0   \\
0 & 0 & 1  \\
0  & 0 & 0   \\
\end{array} \right) \;, \nn \\
t_{\bar{1}}=
\left( \begin{array}{ccc}
0 & 0 & 0   \\
1 & 0 & 0  \\
0  & 0 & 0   \\
\end{array} \right) \;,
t_{\bar{2}}=
\left( \begin{array}{ccc}
1 & 0 & 0   \\
0 & 0 & 0  \\
0  & 0 & 1   \\
\end{array} \right) \;,
t_{\bar{3}}=
\left( \begin{array}{ccc}
0 & 0 & 0   \\
0 & 0 & 0  \\
1 & 0 & 0   \\
\end{array} \right) \;,
t_{\bar{4}}=
\left( \begin{array}{ccc}
0 & 0 & 0   \\
0 & 0 & 0  \\
0  & 1 & 0   \\
\end{array} \right),
\eea
where Cartan generators $t_{2}, t_{\bar{2}}$ together with $t_{1}, t_{
\bar{1}}$ form the bosonic subalgebra $sl(2) \oplus u(1)$, while the
generators
$t_{3}, t_{\bar{3}}, t_{4}, t_{\bar{4}}$ are fermionic roots.
In this basis the structure constants of $sl(2|1)$ are
\bea
{f_{2,1}}^{1} & = & -1,\;
{f_{2,\bar{1}}}^{\bar{1}}=1,\;
{f_{2,3}}^{3}=-1,\;
{f_{2,\bar{3}}}^{\bar{3}}=1, \nn \\
{f_{\bar{2},1}}^{1} & = & 1,\;
{f_{\bar{2},\bar{1}}}^{\bar{1}}=-1,\;
{f_{\bar{2},\bar{4}}}^{\bar{4}}=1,\;
{f_{\bar{2},4}}^{4}=-1, \nn \\
{f_{1,\bar{1}}}^{2}& = & -1, \;
{f_{1,\bar{1}}}^{\bar{2}}=1,\;
{f_{1,\bar{3}}}^{\bar{4}}=-1,\;
{f_{1,4}}^{3}=1,\; \nn \\
{f_{\bar{1},\bar{4}}}^{\bar{3}} & = & -1,
{f_{\bar{1},3}}^{4}=-1,\;
{f_{\bar{1},3}}^{4}=1,\; \nn \\
{f_{3,\bar{3}}}^{1} & = & 1,\;
{f_{3,\bar{4}}}^{1}=1 ,\;
{f_{4,\bar{3}}}^{\bar{1}}=1, \;
{f_{4,\bar{4}}}^{2}=1,\;
\eea
and nonzero elements of Killing metric are given by
\bea
g_{1 \bar{1}}=-g_{2 \bar{2}}=g_{3 \bar{3}}=g_{4 \bar{4}}=1.
\eea
The explicit relations between affine supercurrents
$\cJ_{a}, \cJ_{\bar{a}}$
in this basis and the entries $\cJ_{mn}$ of the $sl(2|1)$ superlagebra
valued affine supercurrent
introduced in Section $3$ are as follows
\bea
\cJ_{1}=\cJ_{12}, \;\; \cJ_{2} \equiv \cH_{1}=\cJ_{11},
\;\; \cJ_{3}=\cJ_{13}, \;\;
\cJ_{4}=\cJ_{23}, \nn \\
\cJ_{\bar{1}}=\cJ_{21}, \;\; \cJ_{\bar{2}} \equiv \cH_{\bar{1}}=\cJ_{22}, \;\;
\cJ_{\bar{3}}=\cJ_{31}, \;\; \cJ_{\bar{4}}=\cJ_{32}.
\eea

\vspace{1cm}

{\Large \bf Appendix B: Notations for $sl(3|2)$ superalgebra}
\setcounter{equation}{0}
\def\theequation{B.\arabic{equation}}
\vspace{0.5cm}

We choose four Cartan generators
$t_{1}, t_{\bar{1}}, t_{2}, t_{\bar{2}}$ of $sl(3|2)$
superalgebra in the fundamental representation in the following form
\bea
t_{1}=
\left( \begin{array}{ccccc}
0  & 0 & 0  & 0 & 0  \\
0  & 1 & 0  & 0 &  0  \\
0  & 0 & 0  & 0 & 0   \\
0  & 0 & 0  & 1 & 0  \\
0  & 0 & 0  & 0 & 0  \\
\end{array} \right), \;\;
t_{\bar{1}}=
\left( \begin{array}{ccccc}
1  & 0 & 0  & 0 & 0  \\
0  & 0 & 0  & 0 &  0  \\
0  & 0 & 0  & 0 & 0  \\
0  & 0 & 0  & 1 & 0  \\
0  & 0 & 0  & 0 & 0  \\
\end{array} \right), \nn \\
t_{2}=
\left( \begin{array}{ccccc}
0  & 0 & 0  & 0 & 0  \\
0  & 0 & 0  & 0 &  0  \\
0  & 0 & 1  & 0 & 0   \\
0  & 0 & 0  & 0 & 0  \\
0  & 0 & 0  & 0 & 1  \\
\end{array} \right), \;\;
t_{\bar{2}}=
\left( \begin{array}{ccccc}
0  & 0 & 0  & 0 & 0  \\
0  & 1 & 0  & 0 &  0  \\
0  & 0 & 0  & 0 & 0   \\
0  & 0 & 0  & 0 & 0  \\
0  & 0 & 0  & 0 & 1  \\
\end{array} \right).
\label{eq:Cartan}
\eea
Each of the~ remaining $10$ unbarred~ generator
$t_{a}, a=3, 4, \dots , 12$ is represented~ by a $5 \times 5$
supermatrix~ with the only~ non-zero entry $1$ on the
intersection of $m$-th line and $n$-th row
$(m=1, 2, 3, 4, n > m)$. The barred generators $t_{\bar{a}}$ have their
nonzero entries $1$ in the bottom triangular part. Using the explicit
form of $sl(3|2)$ generators we can find their
(anti)commutators, taking into account their statistics (the generators
with non-zero entries inside the diagonal $3 \times 3$ and $2 \times 2$ blocks
are bosonic, all others are fermionic). In the complex basis, they
satisfy the following graded commutators
\bea
[ t_{a}, t_{b} \}= {F_{a b}}^{c} t_{c}, \;\;
[ t_{\bar{a}}, t_{\bar{b}} \} = {F_{\bar{a} \bar{b}}}^{\bar{c}}
t_{\bar{c}}, \;\;
[ t_{a}, t_{\bar{b}} \} ={F_{a \bar{b}}}^{c} t_{c} +
{F_{a \bar{b}}}^{\bar{c}} t_{\bar{c}}
\eea
  From this we can read off all the structure constants ${F_{AB}}^{C}$ which
are $1$ or $-1$ (remember that ${f_{AB}}^{C}=(-1)^{(d_{A}+1) d_{B}}
{F_{AB}}^{C}$). Killing metric $g_{a \bar{b}}$ is given by
$Str(t_{a} t_{\bar{b}})$ where we take usual convention for
supertrace.
Just as an example, we write down nonzero elements of
$g_{a \bar{b}}$ for the subset (\ref{eq:Cartan})
\bea
-g_{1 \bar{1}}= g_{1 \bar{2}}= -g_{2 \bar{2}}=1.
\eea
\vspace{1cm}

{\Large \bf Appendix C: $u(m|n)$ SCAs \cite{DTH} in terms of currents}
\setcounter{equation}{0}
\def\theequation{C.\arabic{equation}}
\vspace{0.5cm}

This algebra includes Virasoro stress tensor $T$, $2n$ spin
$3/2$ bosonic currents, $G^{a}, {\bar{G}}_{a}$, $2 m$
spin $3/2$ fermionic currents, $G^{b}, {\bar{G}}_{b}, a=m+1, m+2, \dots, m+n,
b=1, 2, \dots,
m $, $(m^2+n^2)$ spin $1$ bosonic currents, ${J^{c}}_{d}, c,d=1,2, \dots,
m, {J^{e}}_{f}, e=m+1, m+2, \dots, m+n, f=n+1, n+2, \dots, m+n,$
and $2mn$ spin $1$ fermionic ones,
${J^{g}}_{h}, {J^{i}}_{j}, g=m+1,m+2, \dots, m+n,
h=1, 2, \dots, n, i=n+1, n+2, \dots, m+n, j=1, 2, \dots, m $.
The total set of $(m+n)^2$ spin $1$ currents forms $u(m|n)$
current algebra. Spin $3/2$ currents transform under fundamental
and conjugate representaions of $u(m|n)$, for upper and lower
positions of the indices, respectively.

These currents satisify the following OPEs
\bea
T(z) T(w) & = &
\frac{1}{(z-w)^4} 3 \kappa +\frac{1}{(z-w)^2} 2 T+
\frac{1}{(z-w)} T', \nn \\
T(z) {J^{a}}_{b}(w) & = &
\frac{1}{(z-w)^2} {J^{a}}_{b}+
\frac{1}{(z-w)} {J^{a}}_{b}', \nn \\
T(z) {G}^{a}(w) & = &
\frac{1}{(z-w)^2} \frac{3}{2} {G}^{a}+
\frac{1}{(z-w)} {{G}^{a}}', \nn \\
T(z) {\bar{G}}_{a}(w) & = &
\frac{1}{(z-w)^2} \frac{3}{2} {\bar{G}}_{a}+
\frac{1}{(z-w)} {\bar{G}}_{a}', \nn \\
{J^{a}}_{b} (z) {J^{c}}_{d} (w) & = &
\frac{1}{(z-w)^2} \left[ \frac{1}{2-(m-n)} (-1) ^{(d_{a}+
1)(d_{b}+1)+(d_{c}+1)(d_{d}+1)}
\delta^{a}_{b} \delta^{c}_{d} + \right. \nn \\
& & \left. (-1)^{(d_{a}+d_{b}+d_{c}+1)(d_{d}+1)}
\delta^{a}_{d} \delta^{c}_{b} \right] \kappa
+
\frac{1}{(z-w)} \left[ (-1)^{(d_{a}+1)(d_{b}+d_{c})}
\delta^{c}_{b} {J^{a}}_{d} - \right. \nn \\
& & \left. (-1)^{(d_{a}+1)(d_{d}+1)+(d_{b}+1)(d_{c}+1)+
(d_{d}+1)(d_{b}+1)+(d_{d}+1)(d_{c}+1)} \delta^{a}_{d} {J^{c}}_{b}
\right], \nn \\
{J^{a}}_{b} (z) G^{c} (w) & = & \frac{1}{(z-w)} \delta^{c}_{b} G^{a},
 \nn \\
{J^{a}}_{b} (z) {\bar{G}}_{c} (w) & = & -\frac{1}{(z-w)} \delta^{a}_{c}
 (-1)^{(d_{b}+1)d_{c}} {\bar{G}}_{b}, \nn \\
G^{a} (z) {\bar{G}}_{b} (w) & = &-\frac{1}{(z-w)^3} (-1)^{(d_{a}+1)(
d_{b}+1)} \delta^{a}_{b} 4 \kappa + \nn \\
& & \frac{1
}{(z-w)^2} \left[2 (-1)^{(d_{a}+1)(d_{b}+1)} \delta^{a}_{b} {J^{c}}_{c}-
4 {J^{a}}_{b} \right] + \nn \\
& &
\frac{1}{(z-w)} \left[(-1)^{(d_{a}+1)(d_{b}+1)} \delta^{a}_{b} {J^{c}}_{c}'-
2 {J^{a}}_{b}'-
2 (-1)^{(d_{a}+1)(d_{b}+1)} \delta^{a}_{b} T - \right. \nn \\
& & \frac{1}{\kappa} (-1)^{(
d_{a}+1)(d_{b}+1)} \delta^{a}_{b} {J^{c}}_{c} {J^{d}}_{d}+
\frac{2}{\kappa} {J^{c}}_{c} {J^{a}}_{b}  + \nn \\
& &
\left(
-\frac{1}{\kappa} \delta^{a}_{c} \delta^{d}_{b}+\frac{1}{2 \kappa}
(-1)^{(d_{a}+1)(d_{b}+1)} \delta^{a}_{b}
\delta^{c}_{d} \right) \times \nn \\
& & \left.  \left( (-1)^{d_{e}+1} {J^{c}}_{e} {J^{e}}_{d}+
(-1)^{(d_{c}+1)(d_{d}+d_{e})+(d_{d}+1)(d_{e}+1)} {J^{e}}_{d} {J^{c}}_{e}
\right)\right],
\label{eq:comp}
\eea
\vspace{1cm}

{\Large \bf Appendix D: A Different Realization of $sl(n|n-1)$}
\setcounter{equation}{0}
\def\theequation{D.\arabic{equation}}
\vspace{0.5cm}

We can realize superalgebra $sl(n|n-1)$ in a different, though
equivalent way by $(2n-1) \times (2n-1)$ supermatrix whose entries
$\cJ_{\tilde k \tilde l}$ are related
to those $\cJ_{kl}$ in the standard realization according to the
following rule \cite{BLNW}
\bea
\widetilde{k} & = & 2k-1, \widetilde{l} =2l-1 \;\; \mbox{if} \;\;
1 \leq k, l \leq n \nn \\
\widetilde{k} & = & 2(k-n), \widetilde{l} =2(l-n) \;\; \mbox{if} \;\;
n < k, l \leq 2n-1 \;.
\eea
This parametrization corresponds to choosing the system of purely
fermionic simple roots in $sl(n|n-1)$.
It is very convenient when studying embeddings
of $sl(2|1)$ into $sl(n|n-1)$: the former is identified with proper
$3 \times 3$ blocks in the $sl(n|n-1)$ supermatrix ${}^{\ddagger\ddagger}$.

\footnotetext{We are grateful to
F. Delduc for explaining us the merits of this realization.}

Using this convention, the set of hamiltonian reduction constraints
we dealt with in the $sl(2|1)$ case can be rewritten in the following
suggestive way
\bea
N=2 \;SCA: \;\; {\cJ_{mn}}^{\mbox{constr}}=
\left( \begin{array}{ccc}
\ast & 0 & 1 \\
\ast & \ast & \ast \\
\ast & 1 & \ast
\end{array} \right)
(\ref{eq:sl21constr})
\Longrightarrow
\left( \begin{array}{ccc}
\ast & 1 & 0 \\
\ast & \ast & 1 \\
\ast & \ast & \ast
\end{array} \right).
\label{eq:diasl21}
\eea
This picture shows that the constraints are concentrated in the upper
triangular part of the supercurrent matrix, and this is true as well
for the $sl(3|2)$ constraints except for (\ref{eq:conso5}). We first
present the matrices of constraints for the cases of $N=2$ $W_3$, $N=2$
$W_3^{(2)}$, $N=2$ $u(2|1)$ and $N=2$ $u(3)$ SCAs
\bea
N=2\; W_3: \;\;\;\cJ_{mn}^{\mbox{constr}}=
\left( \begin{array}{ccccc}
\ast & 0    & 0  & 1    & 0  \\
\ast & \ast & 0  & \ast & 1 \\
\ast & \ast & \ast & \ast & \ast \\
\ast & 1 & 0 & \ast & 0 \\
\ast & \ast & 1 & \ast & \ast
\end{array} \right)
(\ref{eq:w3cons})
\Longrightarrow
\left( \begin{array}{ccccc}
\ast & 1    & 0  & 0    & 0  \\
\ast & \ast & 1  & 0 & 0 \\
\ast & \ast & \ast & 1 & 0 \\
\ast & \ast & \ast & \ast & 1 \\
\ast & \ast & \ast & \ast & \ast
\end{array} \right) \label{w3}
\eea
\bea
N=2\; W_3^{(2)}: \;\;\cJ_{mn}^{\mbox{constr}}=
\left( \begin{array}{ccccc}
\ast & 0    & 0  & 1    & 0  \\
\ast & \ast & 0  & \ast & 1 \\
\ast & \ast & \ast & \ast & \ast \\
\ast & 1 & 0 & \ast & 0 \\
\ast & \ast & \ast & \ast & \ast
\end{array} \right)
(\ref{eq:consw32})
\Longrightarrow
\left( \begin{array}{ccccc}
\ast & 1    & 0  & 0    & 0  \\
\ast & \ast & 1  & 0 & 0 \\
\ast & \ast & \ast & 1 & 0 \\
\ast & \ast & \ast & \ast & \ast \\
\ast & \ast & \ast & \ast & \ast
\end{array} \right)  \label{w32}
\eea
\bea
N=2\;u(2|1):\;\;\cJ_{mn}^{\mbox{constr}}=
\left( \begin{array}{ccccc}
\ast & 0    & 0  & 1    & 0  \\
\ast & \ast & \ast  & \ast & \ast \\
\ast & \ast & \ast & \ast & \ast \\
\ast & 1 & 0 & \ast & 0 \\
\ast & \ast & \ast & \ast & \ast
\end{array} \right)
(\ref{eq:consu2})
\Longrightarrow
\left( \begin{array}{ccccc}
\ast & 1    & 0  & 0    & 0  \\
\ast & \ast & 1  & 0 & 0 \\
\ast & \ast & \ast & \ast & \ast \\
\ast & \ast & \ast & \ast & \ast \\
\ast & \ast & \ast & \ast & \ast
\end{array} \right)
\label{eq:diau21}
\eea
\bea
N=2\;u(3):\;\;\cJ_{mn}^{\mbox{constr}}=
\left( \begin{array}{ccccc}
\ast & 0    & \ast  & \ast    & 0  \\
\ast & \ast & 0  & \ast & 1 \\
\ast & \ast & \ast & \ast & \ast \\
\ast & 1 & 0 & \ast & 0 \\
\ast & \ast & \ast & \ast & \ast
\end{array} \right)
(\ref{eq:cou3})
\Longrightarrow
\left( \begin{array}{ccccc}
\ast & \ast    & 0  & 0   & \ast  \\
\ast & \ast & 1  & 0 & 0 \\
\ast & \ast & \ast & 1 & 0 \\
\ast & \ast & \ast & \ast & \ast \\
\ast & \ast & \ast & \ast & \ast
\end{array} \right) \;.
\label{eq:diau3}
\eea
The supermatrices of constraints for two ``noncanonical'' cases
described in Subsection 5.3, respectively with
the constrained $N=2$ stress tensor and/or spin $0$
supercurrents present, are given by
\bea
\cJ_{mn}^{\mbox{constr}}=
\left( \begin{array}{ccccc}
\ast & \ast    & \ast  & 1    & \ast  \\
\ast & \ast & \ast  & 0 & \ast \\
\ast & \ast & \ast & 0 & \ast \\
\ast & \ast & \ast & \ast & \ast \\
\ast & \ast & 1 & 0 & \ast
\end{array} \right)
(\ref{eq:conso5})
\Longrightarrow
\left( \begin{array}{ccccc}
\ast & 1    & \ast  & \ast    & \ast  \\
\ast & \ast & \ast  & \ast & \ast \\
\ast & 0 & \ast & \ast & \ast \\
\ast & 0 & \ast & \ast & 1 \\
\ast & 0 & \ast & \ast & \ast
\end{array} \right)  \label{1uncan}
\eea
\bea
\cJ_{mn}^{\mbox{constr}}=
\left( \begin{array}{ccccc}
\ast & 0    & 0  & 1    & \ast  \\
\ast & \ast & \ast  & \ast & \ast \\
\ast & \ast & \ast & \ast & \ast \\
\ast & 1 & 0 & \ast & \ast \\
\ast & \ast & \ast & \ast & \ast
\end{array} \right)
(\ref{eq:consn5})
\Longrightarrow
\left( \begin{array}{ccccc}
\ast & 1    & 0  & \ast   & 0  \\
\ast & \ast & 1  & \ast & 0 \\
\ast & \ast & \ast & \ast & \ast \\
\ast & \ast & \ast & \ast & \ast \\
\ast & \ast & \ast & \ast & \ast
\end{array} \right)\;.  \label{2uncan}
\eea

These pictures clearly demonstrate the relations between different reductons
in accord with the diagram of Section 6. Also it is seen from them
that it is natural to treat all the considered cases in the language
of $sl(2|1)$ embeddings. The case of $N=2$ $W_3$ corresponds to the
principal embedding of $sl(2|1)$ into $sl(3|2)$ while $N=2$ $u(2|1)$ and
$N=2$ $u(3)$ ones to two inequivalent non-principal embeddings. It would be
interesting to understand from an analogous point of view the cases \p{w32},
\p{1uncan}, \p{2uncan}. It seems that in this way one could
explain some peculiar features of them (lacking of the superprimary basis
in the $N=2$ $W_3^{(2)}$ case, the presence
of constrained $N=2$ stress tensor and/or spin $0$ supercurrents in two
remaining cases). Note that the complete classification of
$sl(2|1)$ embeddings, at the component level and in the string theory
context, is undertaken in \cite{RSS}.

\end{document}